\documentclass[10pt]{article}
\usepackage{a4}
\usepackage{color}
\usepackage{cite}
\begin{document}
\title{State-space Manifold and Rotating Black Holes}
\author{}
\date{ Stefano Bellucci$^{a}$  \thanks{\noindent bellucci@lnf.infn.it},
 Bhupendra Nath Tiwari $^{a}$\thanks{\noindent tiwari@lnf.infn.it}\\
$^{a}$ INFN-Laboratori Nazionali di Frascati\\
Via E. Fermi 40, 00044 Frascati, Italy.}
\maketitle \abstract{ We study a class of fluctuating higher
dimensional black hole configurations obtained in string theory/
$M$-theory compactifications. We explore the intrinsic Riemannian
geometric nature of Gaussian fluctuations arising from the Hessian
of the coarse graining entropy, defined over an ensemble of brane
microstates. It has been shown that the state-space geometry
spanned by the set of invariant parameters is non-degenerate,
regular and has a negative scalar curvature for the rotating
Myers-Perry black holes, Kaluza-Klein black holes, supersymmetric
$AdS_5$ black holes, $D_1$-$D_5$ configurations and the associated
BMPV black holes. Interestingly, these solutions demonstrate that
the principal components of the state-space metric tensor admit a
positive definite form, while the off diagonal components do not.
Furthermore, the ratio of diagonal components weakens relatively
faster than the off diagonal components, and thus they swiftly
come into an equilibrium statistical configuration. Novel aspects
of the scaling property suggest that the brane-brane statistical
pair correlation functions divulge an asymmetric nature, in
comparison with the others. This approach indicates that all above
configurations are effectively attractive and stable, on an
arbitrary hyper-surface of the state-space manifolds. It is
nevertheless noticed that there exists an intriguing relationship
between non-ideal inter-brane statistical interactions and phase
transitions. The ramifications thus described are consistent with
the existing picture of the microscopic CFTs. We conclude with an
extended discussion of the implications of this work for the
physics of black holes in string theory. }

\vspace{1.0cm}

\textit{PACS numbers: 04.70.-s Physics of black holes; 04.70.Bw
Classical black holes; 04.70.Dy Quantum aspects of black holes,
evaporation, thermodynamics; 04.50.Gh Higher-dimensional black
holes, black strings, and related objects.}

%
\newpage
\begin{Large} \textbf{Contents:} \end{Large}\\
\begin{enumerate}
\item{Introduction.}
\item{State-space Geometry.}
\item{Rotating Black Holes:}
\subitem{3.1 \ \ Myers-Perry Black Holes.}
\subitem{3.2 \ \ Kaluza-Klein Black Holes.}
\subitem{3.3 \ \ Supersymmetric $AdS_5$ Black Holes.}
\subitem{3.4 \ \ $D_1$-$D_5$ Configurations.}
\subitem{3.5 \ \ BMPV Black Holes.}
\item{Conclusion and Discussion.}
\end{enumerate}
\section{Introduction}
Motivated from the physics and mathematics of higher dimensional
brane configurations and their microscopic descriptions, we focus
on the possible investigations of covariant intrinsic
thermodynamic Riemannian geometries for a class of rotating black
hole solutions obtained in either string theory or $M$-theory
compactifications. Our study is prompted in part by the recent
insight perceived from the string theory solutions, which may in
turn impel significant progress towards an understanding of the
nature of microscopic data, in the limiting case of extremal and
near-extremal black brane configurations. In particular, it is
well-known that one can count the microstates for certain rotating
and non-rotating black brane configurations carrying the same
charges, angular momenta and energy, with a microscopic entropy
$S_{micro} $ like that of the corresponding macroscopic
configurations having Bekenstein-Hawking entropy $S_{BH}=
\frac{A}{4G}$, see for instance \cite{9601029v2,
9602043v2,0706.3884v1}.

To refine this issue, one should first note that the concept of
AdS/CFT correspondence \cite{9711200v3,0002092v1, 0206126,
0212204} suggests that the $S_{micro}$ counts the microstates of
the black brane configuration that belong to the boundary CFT,
which is dual to the gravitational description. This has been very
useful in understanding the black brane entropy, which is based on
the coarse graining of an ensemble of brane microstates
parameterized by a finite number of asymptotic parameters. We
believe that the underlying space-time geometries may be capturing
generic CFT states of the black holes and black branes with
non-zero microscopic degeneracy, and thus capturing enough of them
to account for their fluctuating horizon area. Here, we will use
this result as a crucial ingredient in the realization of possible
state-space fluctuations around an equilibrium statistical
configuration of the higher dimensional rotating black hole
solutions.

The fact that we are only looking at a special class of Gaussian
fluctuations means that we are necessarily restricting the degrees
of freedom of the interested configurations. Thus, the crucial
idea would be first to develop an intrinsic Riemannian geometric
notion, for synthesizing the statistical correlations among the
underlying microstates of the various rotating black hole
solutions. To fully support this construction, we shall point out
an adequate ground for the microscopic perspective ratifying the
state-space geometry. The metric tensor arises from the negative
Hessian matrix of the corresponding coarse graining entropy
defined over a large number of brane microstates characterizing
the chosen black hole macroscopic configuration. Here, the
important progress is thus to appreciate how the microstate
geometries may capture enough entropy to account for statistical
correlations and their dependence on finitely many conserved
electric-magnetic charges, angular momenta and other possible
physical charges, if any.

As in the classical theory of four dimensional black holes, the
uniqueness theorems assert that the parameters of a black hole are
precisely those which correspond to certain conserved quantities,
{\it viz.}, the mass, angular momenta, and promising conserved
charges, if any, associated with local gauge symmetries. Moreover,
we know that the possible black hole solutions of the
four-dimensional Einstein-Maxwell theory are only the Kerr-Newman
black holes \cite{Hawking, Robinson, Mazur}. More strongly, this
result precludes the possibility that the black hole possesses
higher multipole moments (e.g., a mass quadrupole or a charge
dipole) that are not completely fixed by the values of conserved
charges. It is however worth to notice that the higher dimensional
rotating black branes may in general violate the black hole
uniqueness theorems, which exist for the ordinary four dimensional
configurations \cite{mathur1,mathur2}. Another consequence is that
the microscopic states, that are responsible for the large
degeneracy implied by the Bekenstein-Hawking entropy of a rotating
configuration, are however invisible at the level of the classical
gravitational theory, but their radiative effects may be observed
in the corresponding state-space geometry.

There is indeed a significant body of evidences that supports the
idea of the state-space geometry for the Myers-Perry black holes,
which are the simplest rotating black hole configurations in five
dimensional space-times and thus capture the basic notion of our
geometric investigations. Here, we will  focus on the recent
discovery that the five-dimensional black holes exhibit
qualitatively new properties, which are rather not shared by their
four-dimensional siblings. We would like to investigate this from
the perspective of state-space manifolds associated with spherical
horizon topology black holes, where the conventional notions of
uniqueness theorems do apply \cite{0110260v2}. In short, we would
investigate on certain possible developments, which may notably be
appreciated in the higher dimensional versions of the Kerr
solutions constructed by the Myers and Perry \cite{MP} and in the
other associated brane configurations.

The statistical importance of charged supersymmetric black holes
in string theory has originally been motivated by the study of
Strominger and Vafa \cite{9601029v2}. They have shown that the
microscopic understanding of black holes in string theory may be
characterized by the parameters of the solutions, and in
particular, the charges of supersymmetric black holes serve as the
tag marks to identify their microscopic constituents in the string
theory. Moreover, the phase-space of the supersymmetric black hole
solutions turns out to be drastically constrained and is subject
to various powerful non-renormalization theorems and
state-counting techniques. However, it is known that the neutral
black holes in $M$-theory carry a minimal set of quantum numbers,
{\it viz.}, mass and angular momentum \cite{MP}. Therefore, it is
hard to restrict the phase-space into a sector in which it is
simple enough to count the microstates. Besides the microscopic
deficiency, we find for these solutions that it is not difficult
to give an intrinsic state-space meaning to the associated
statistical fluctuations, under the Gaussian approximation.

There exist vacuum black holes in $M$-theory that can be mapped to
certain well-defined bound states of $D$-brane configurations in
string theory, which in an intriguing limit become asymptotically
flat. In particular, we shall restrict our attention on the
state-space manifolds of five-dimensional dyonic black hole
solutions described in Kaluza-Klein theory \cite{gw}. It has been
shown in Ref. \cite{itz} that, in an ascription akin to the
decoupling limit in AdS/CFT, such black holes include the
five-dimensional asymptotically flat neutral rotating solutions of
Myers and Perry \cite{MP}. Furthermore, one can view these black
holes as the Myers-Perry black holes placed at the tip of the
Taub-NUT geometry. Our purpose would here be to study and compare
the nature of their state-space interactions, {\it viz.}, for the
Kaluza-Klein black holes and that of the Myers-Perry black holes
and thus one may simultaneously analyze whether they both pertain
a well-defined state-space.

The case of the Kaluza-Klein configuration implies that an
addition of electric charge or energy to the extremality condition
leads to a finite horizon black hole at the tip of the cigar
solution, which is being fibered with the spherical horizon of the
corresponding four-dimensional black hole. The corresponding
magnetically charged Kaluza-Klein black holes may thus be viewed
as the five-dimensional black holes with $S^3$ horizon topology,
which may in turn be localized inside the Taub-NUT geometry. On
the other hand, it is much more illuminating to note that both
descriptions of this model are considerably interesting for our
geometric analysis. In particular, it is worth to recall that the
electric charges do not correspond to a boost, but to the
rotations of the five dimensional black holes, which are aligned
along the Kaluza-Klein circle \cite{itz}. We find that these
rotations may be analyzed from the perspective of general
co-ordinate transformations on the concerned state-space manifold.
This also follows directly from the fact that the component of the
rotation that is not aligned with the fiber of the five
dimensional solution determines the four-dimensional rotation,
which describes the usual four dimensional Kerr-Newman black
holes.

Microscopically, we can count the microstates of Kaluza-Klein
configurations by taking the tensor product of $T^6$ with the
$S^1$ and view it as a vacuum solution in the $M$-theory, with the
Kaluza-Klein circle being the $M$-theory compactification circle.
The starting point of our discussion may thus be concerned with
the consideration of usual duality relations between the
$M$-theory and type-IIA string theory, which comprise certain
electric and magnetic charges corresponding to the given number of
constituent $D_0$ and $D_6$-branes \cite{itz}. In this case, it
turns out that the quantization conditions exist for the
individual electric and magnetic charges, which may simply be
written in terms of the net number of corresponding $D$-branes.
Furthermore, this yields that the net degeneracy of the $D_0$ and
$D_6$-brane microstates may be given by the central charge of the
conformal field theory. More precisely, we would like to appraise
how our state-space geometry of an extremal Kaluza-Klein black
hole entangles with its degeneracy of CFT microstates, and it
describes the corresponding $D_0$-$D_6$ microscopic configuration.

This is positive and compelling evidence for the identification of
black hole entropy formulas, and thus of the derived state-space
quantities, which may indeed be done so, just by taking the T-dual
configurations in which the microscopic descriptions become more
transparent. For simplicity, let us consider the case of non
rotating four-dimensional solutions. Then the supersymmetric
four-charge black hole configurations in Type II string theory
compactified on $T^6$ have many possible choices for the charges,
which in general are related by U-duality. It has been shown in
\cite{Balasubramanian:1996rx} that one may explain the microscopic
origin of the Kaluza-Klein black holes, just by considering four
stacks of $D_3$-branes. Among these, any two stacks intersect over
a line, while the rest four intersect at a point. Moreover, the
orientation of the first three stacks can be chosen arbitrarily,
but the orientation of the last set of $D_3$-branes is required to
be fixed, in order to preserve the $\mathcal N=2$ supersymmetry.
In the case when the number of branes in each stack remains same,
then one finds that the torus moduli remain constant. In
particular, the corresponding solution reduces to the extremal
Reissner-Nordstr\"{o}m black hole, which has a non-interacting
state-space geometry \cite{0801.4087v1}. Here, one educes that the
covariant intrinsic structures may easily be determined from the
entropy (or mass) which grows with the number of $D_3$-branes
wrapped around the torus.

Investigations of the relationship between the type IIB
superstring theory and Yang-Mills theory have further motivated us
to analyze the state-space structure of the supersymmetric $AdS_5$
black holes. In particular, the spectrum of the microstates in
$\frac{1}{2}$-BPS sector has extensively been studied in both the
string theory and $\mathcal N=4$ Yang-Mills theory \cite{llm,
ber-1}. It has been known that the spectrum of type IIB
superstring theory contains a large class of BPS states, when
compactified on $AdS_5\times S^5$. Thus, it would be interesting
to describe the nature of possible statistical correlations for
these black hole configurations, under the general consideration
of microscopic superconformal field theories. In turn, we identify
that the nonzero $U(1)\subset SO(6)$ momenta or certain R-charges
associated with the possible fractional BPS states serve as the
co-ordinates for our state-space geometry of these configurations.
More recently, some pioneering works have also been done in the
supergravity configurations and in super Yang-Mills theories,
which provide an extent of confirmation for our investigations,
see for example \cite{surya,sh-to,mandal,gmmpr,dms}.

It would thus be interesting to consider the state-space geometry
in a certain specific supersymmetric sector. In particular, the
fact that the $\frac{1}{16}$-BPS states preserve 2 real
supersymmetries provides viable support for our geometric
examination. Our analysis shows that the underlying microscopic
configurations should have a finite and regular correlation
volume. One such motivation for the concerned behavior of the
supersymmetric configurations comes from the study of the $AdS_5$
black holes, which have been rather recently obtained from the
five dimensional gauged supergravity theories
\cite{gu-re-1,gu-re-2,cclp1,cclp2,klr}. It turns out that these
black holes carry three $U(1)^3\subset SO(6)$ momenta in the
$S^5$, as well as two $U(1)^2\subset SO(4)$ in the $AdS_5$, which
define the co-ordinate charts for the underlying state-space
geometry. In order to have a regular horizon black hole, it may
however be urged that the angular momenta in the $AdS_5$ should be
nonzero: otherwise the  black hole solutions of interest would
develop certain naked singularities \cite{my-ta,bcs}. In
particular, the presence of angular momenta in $AdS_5$ forces them
to preserve a symmetry not larger than the
$\frac{1}{16}$-supersymmetry. We observe in the present case that
the associated state-space turns out to be an intrinsic real
Riemannian geometry, which effectively implies an attractive
statistical configuration.

To be concrete, we shall be interested here in the $AdS_5\times
S^5$ black hole configurations. In particular, we shall consider
five dimensional $\mathcal N=1$ gauged supergravity coupled with
$n$ abelian vector multiplets, which turns out to be more
convenient, in order to deal with these solutions, including
certain general supergravity theories of interest. For simplicity,
we consider the supergravity black holes in the symmetric spaces,
which involve a self-dual angular momentum in $AdS_5$ described in
\cite{gu-re-2}. Before commencing further details, we wish to
recall that the physical quantities which define our state-space
geometry are only the $U(1)^n$ charges $Q_i$ and the self-dual
angular momentum $ J= \frac{J_1+J_2}{2} $, where the $J_1$, $J_2$
are the Cartan generators of $SO(4)$ rotations of the $AdS_5$.
Note, further, that the Bekenstein-Hawking entropy of such black
holes may easily be given by the horizon area of a squashed
3-sphere. Our computation thus allows for compact expressions for
the state-space quantities. We also notice that there exists a set
of interesting scaling properties for these underlying
configurations.

Another consequence is that there are $n+1$ independent charges
carried by these black holes, {\it viz.}, $n$ electric charges
$Q_i$'s and a self-dual angular momentum $J$, but there are only
$n$ independent physical parameters $q_i$ for the genuine
solutions. This is due to the fact that there exists an intriguing
relationship between the charges, In particular, if one tries to
express the macroscopic entropy in terms of the charges $Q_i$ and
the rotation $J$, then there exists an ambiguity in its
expression, see for more details \cite{kmmr,brs}. Thus, one may
take advantage of this ambiguity and write the macroscopic entropy
in terms of the electric charges accompanied by these black hole
configurations. For a symmetric space, Kim et. al. found that
there exists an appropriate combination: $C^{ijk}\bar{X}_iQ_j Q_k$
of the charges $Q_i$ and vacuum scalars $\bar{X}_i$, which defines
an illuminating formula for the concerned entropy, $
S_{BH}=2\pi\sqrt{D^{ij}Q_j Q_k- 4cJ} $, see for the details
\cite{hep-th/0607085}. Here, the constant $c$ actually defines the
central charge of a 4 dimensional holographic superconformal field
theory living on the boundary, and may as well be computed from
the dual gravity data. In particular, for the choice of
${\mathcal{N}}=4$ $SU(N)$ Yang-Mills theory, the central charge
$c$ may be normalized to be $c=N^2/4$. Our analysis thus
determines the allowed physical domain for the statistical
correlations and possible invariants of underlying state-space
geometry. We find, specifically, that the $AdS_5$ black holes are
stable under the Gaussian fluctuations, if at least one of the
parameter remains fixed.

Notice that the mentioned premises may easily be analyzed from the
microscopic duality transformations on the charges, which in the
large charge limit imply the existence of a correspondence with
the general coordinate transformations on the associated
state-space manifold. In fact, the state-space geometry of such a
general black brane configuration may be achieved from the
parameters of an effective space-time geometry, describing an
ensemble of equilibrium microstates. At this juncture, we may
demonstrate that the state-space geometry, thus characterized in
terms of physical parameters of the black brane configurations,
may always be defined as an intrinsic real Riemannian manifold. Up
to a certain quotient, corresponding to the flat directions, we
nevertheless encounter that most of the known higher dimensional
solutions display certain regular, negatively curved state-space
manifolds. Accordingly, it turns out, in the present prescription,
that the scalar curvature signifies nothing more than the possible
attractions, in the underlying statistical configurations.

Furthermore, we wish to focus our attention on the statistical
pair correlation functions. In particular, we would like to
analyze the consequences emerging from the existing macroscopic
considerations and intend to indicate the nature of the
microscopic configuration. Typically, the approach to our
intrinsic geometric study of the black hole thermodynamic starts
with the Gaussian fluctuations of the black hole entropy. In
particular, the components of the intrinsic Riemannian geometry,
thus invoked, may be obtained as the negative Hessian matrix of
its macroscopic entropy, computed from the Bekenstein-Hawking-Wald
area law. Thus, one may easily analyze possible geometric
structures arising from the Gaussian fluctuations and there by try
to understand underlying statistical configurations of the higher
dimensional black holes.

This article has consequently been organized into the following
sections and their respective subsections. The very first section
has in turn illuminated several intriguing motivations to examine
the statistical fluctuations and thereby initiates why to study
our state-space geometry. In section 2, we shall briefly explain
what is the state-space geometry based on a large number of
equilibrium parameters, characterizing the black hole solution. In
section 3, we investigate the state-space geometry for the neutral
Myers-Perry black holes, and anticipate two and three charged
rotating configurations, in higher dimensional space-times.
Section 4 consists of some concluding notes and a set of remarks
arising from the state-space geometry. The general implications
thus obtained may be divulged, for various possible black holes in
string theory and $M$-theory.
\section{Motivations from microstates counting}
In order to see the key points of our state-space investigations,
it is worth to note that Mathur has proposed a class of fuzzball
solutions by analyzing certain implications of the AdS/CFT
correspondence \cite{mathur1,mathur2,LMM,fuzzball}. In particular,
Mathur has shown, for the physics of a $D_1$-$D_5$-system, that
each CFT vacuum of this configuration is dual to the smooth bulk
solution, which neither has a horizon nor the loss of information
\cite{fuzzball}. Thus, we may take account of the fuzzball
geometry into our state-space quantities, and our geometric
observations thus made are based on the well-known fact that the
$D_1$-$D_5$-configuration acquires a large microscopic degeneracy.
The success of this endeavor has in turn led us to a realization
that one may similarly procure an appropriate state-space notion
for the $D_1$-$D_5$-$P$ fuzzball solutions, which receive
correction form the corresponding Kaluza-Klein-momentum
\cite{0706.3884v1}. Moreover, this comprehension leads one to
accept through the AdS/CFT correspondence that the $D_1$-$D_5$-$P$
black holes may be thought of as an ensemble of fuzzball geometry
\cite{LMM}.

Thus, one may clearly see how our intrinsic geometric results open
a new fascinating window into the understanding of statistical
correlation functions for the black holes in string theory and
$M$-theory. The main line of hypothesis has already provided a
parallel motivation for much of the striking task on black brane
state-space manifolds \cite{BNTBull}. In particular, many other
related investigations including the proposition of Mathur's
fuzzballs \cite{fuzzball} can be also introduced for the
fundamental state-space structures of a class of higher
dimensional black brane configurations. To be more specific, we
notice an interesting agreement with the Mathur's proposal that
there should exist some U-duality frame, in which the
non-singular, horizon-free supergravity black hole microstates
admit a coarse grained space-time geometric description for these
solutions.

In a similar spirit, the intrinsic Riemannian geometry
characterizes possible statistical correlations and shows that the
underlying $D_1$-$D_5$ or $D_1$-$D_5$-$P$  black holes entail an
attractive regular state-space configuration. In order to resolve
the black hole information paradox, the Mathur's programme has
inspired a search for gravitating microstates which are horizon
free and non-singular, with the same brane charges as that of the
considered black hole \cite{0706.3884v1,fuzzball}. Various
computations in string theory have suggested that the usual
picture of a black hole arising with an event horizon could be an
emergent phenomenon of the empty space with a central singularity,
when we coarse grain the configuration over an ensemble of
microstates.

Recently, Lunin and Mathur have made a stronger conjecture
\cite{lu-ma} which connotes that the black hole microstates may be
characterized by string theory backgrounds with no horizons. These
solutions saturate an exact bound on their angular momentum and
have the same set of asymptotic charges, as that of the respective
black hole, whose horizon area classically vanishes. Some finite
temperature (non-supersymmetric) generalizations appear as well,
which correspond to definite black holes with finite size
horizons; we may thus be interested in finding state-spaces for
the BPS black holes which are already known to have a finite size
horizon at zero temperature \cite{mathur1,mathur2}. In five
space-time dimensions, the complete class of two charge
supersymmetric solutions is in fact well-known since the invention
of \cite{LMM}. Thus, it would be inspiring to associate to them
the corresponding state-space manifold. Since the number of
parameters for such configurations is the same as for the
macroscopic supergravity solutions, this might indicate the
behaviour of the underlying CFT configurations.

In the case of two charge configurations, the microscopic
solutions do appear surprisingly like a black-hole
\cite{fuzzball}. The higher-derivative corrections cause the
heterotic string solutions to have well defined two charge small
black hole configurations. Both of them have well defined
state-space correlations \cite{BNTBull}. This is because they
exhibit throat regions, which close off at the compactification
radius. The latter may be fixed in terms of the charges of the theory,
by employing an associated attractor mechanism. Now, if one
computes the Bekenstein-Hawking entropy associated with this
radius of capped throat space-time geometry, then one may easily
define the contraventions associated with state-space correlations
and thus examine them in contrast to the existing microscopic
configurations. In particular for the two charge rotating small
black branes, it has been demonstrated in \cite{BNTBull} that
these notions indeed agree with those anticipated in terms of the
parameters of underlying statistical configurations.

Supersymmetric five dimensional black holes are of considerable
interest in string theory, because they admit simple microscopic
descriptions in terms of the charges of underlying $D$-branes
\cite{9601029v2}. Here, the largest known family of such black
holes is the family of four-parameter BMPV solutions
\cite{9603078}. The present work shows that these configurations
have an interesting state-space structure, in general. In
particular, we discover that the statistical pair correlations
satisfy a definite set of scaling properties, with respect to each
other, and the underlying statistical systems are effectively an
attractive configuration under the Gaussian fluctuations. Clearly
the challenge is now to obtain a microscopic description of their
entropy, that correctly accounts for the statistical fluctuations
into these solutions. What appears to be an obstacle may actually
be a very useful ingredient, towards a more complete understanding
of the $D_1$-$D_5$-$P$ system.

In this direction, a remarkable aspect of the supergravity
solutions is the way in which they capture highly nontrivial
constraints between the parameters that arise in their possible
microscopic descriptions. As for the BMPV black holes, it has
already been known that the condition that the closed casual
curves be absent yields appropriate upper bounds on the angular
momentum. Such bounds are also supported from the microscopic CFT
or worldvolume theory. Similar constraints do arise from our
state-space construction, that the supergravity solutions may
alternatively be viewed from the requirement that they should
admit a thermal deformation. It would indeed be very surprising if
string theory could not account for these solutions, which in
their low energy supergravity limit seem completely pathology-free
and describe an absolute picture for our state-space geometry.

From the viewpoint of thermodynamics, we realize as well that the
solutions, which are characterized by the same asymptotic charges
as that of the black holes, may be regarded as being only locally
stable and correspond to an attractive configuration. However, if
we partially maximize the entropy by varying one or more
parameters, then we discover the existence of a unique, globally
stable, thermodynamic configuration. It is worth to note, in
general, that the state-space constraints thus derived, by
requiring hyper-planar stabilities, do not merely fix the angular
momentum for the given charges, but instead they impose an upper
bound on it. This is indeed in an accordance with certain
supergravity constraints, which actually require the absence of
causal anomalies and similarly impose an upper bound on the
allowed angular momentum \cite{0003063}. For the purpose of future
analysis, we thus note that the $D_1$-$D_5$-$P$ black holes of
interest may be obtained from the solution of eleven-dimensional
supergravity, which may easily be Kaluza-Klein reduced to a
solution of type IIA supergravity and then dualized to the type
IIB solution, with net $D_1$, $D_5$ charges and Kaluza-Klein
momentum.

We here study the state-space properties of these black brane
solutions and specifically analyze them for the limiting case of
spherical horizon configurations. In particular, it has been
realized, during the study of stationary, asymptotically flat
black hole solutions, that the $D=5$ vacuum Einstein equations
admit BMPV black holes, with an event horizon of $S^3$ topology
which exclusively describes intriguing higher dimensional rotating
black holes \cite{0110260v2}. It turns out that our state-space
geometry determining the nature of statistical configuration only
involves the asymptotic charges of underlying black hole
solutions. This lies on the fact that these black holes may
uniquely be characterized by their asymptotic parameters, {\it
viz.}, conserved charges, mass and angular momentum. Moreover,
these charges do not distinguish the black holes of spherical
topology from other possible black brane configurations. See for
example how the solutions constructed in \cite{HE,EE} describe the
case of charged black rings. Furthermore, it may be interesting to
determine the possible mutual relevance of the state-space
quantities, among the various horizon changing solutions.

Nevertheless, one could equivalently work by constructing the
associated CFT microstates, whose degeneracy formula gives rise to
the microscopic entropy. Thus the statistical pair correlation
functions and correlation volume may also be observed for the
underlying phase-space configurations. Therefore, it seems
feasible, via an appropriate AdS/CFT-map, that one may even
predict certain equivalences among the parameters describing the
CFT correlators and the co-ordinates of state-space manifolds, for
the specific black brane configurations. It would indeed be
interesting to ponder, whether this kind of approach can be pushed
further, for example to examine certain properties of the
correlations in the above possible descriptions, for the various
new black branes in $D>4$ space-time solutions. In general, we
contemplate that such predictions are not of much help in
realizing the statistical pair correlation functions and
correlation volume of an arbitrary black brane configuration
characterized by a large number of charges, angular momenta, ADM
mass and possibly other local charges. In particular, one may
easily perceive the order of the difficulty, in the analysis of
the concerned statistical configuration, just by knowing the
dimension of the state-space manifold. However, we notice that
there exist certain special configurations, where some amount of
supersymmetry is preserved, which turn out to be more tractable
than the general configurations.

Furthermore, it may be observed that the string theory/ $M$-theory
descriptions play a central role in understanding certain
state-space properties of the rotating black holes, or in general
black brane configurations, see for instance \cite{9601029v2,
0110260v2, 9508072v3, 9602111v3, 9607235, MSW, EEMR2}. In
particular, our analysis might deploy the AdS/CFT correspondence,
involving the equality of the partition function of a string
theory (or $M$-theory) compactified on a space $K$, with
asymptotics of form $AdS_n \times K$, and that of the associated
holographic dual conformal field theory, described on the
conformal boundary of the $AdS_n$, see \cite{9905111v3,MOOREICTP,
0805.4216v1,0805.0095v3, 0806.0053v2, 0707.3437v1}, for some
admissible compact spaces and Calabi-Yau compactifications.
Moreover, a conjecture exists for type IIA extremal p-dimensional
black brane attractors, i.e. that they correspond to
$Ads_{p+2}\times S^{8-2n-p}$ horizon geometries \cite{0809.1114}.

Having thus provided a brief account of the higher dimensional
black hole configurations, we find interesting to note that there
exists an asymptotic expansion of the exact spectrum of the
degeneracy formula, for large charges and angular momenta, which
not only reproduces the entropy of corresponding black branes in
the leading order, but also to the first few subleading orders,
which may be given as an expansion in the inverse power of the
charges \cite{MOOREICTP,0412287}, and particularly see \cite{ASen}
for the spectrum of half-BPS states in $\mathcal N=4$
supersymmetric string theory. Given this correspondence between
the microscopic spectrum and the macroscopic entropy of a class of
black brane configurations, a natural question would be to
assimilate a microscopic CFT origin of the state-space geometric
interactions, arising from the degeneracy formula and, in
specific, what does this correspondence mean for rotating black
holes? However, we find this question beyond the scope of the
present interest and would relegate it for the future analysis.

Here, we focus our attention on the asymptotic parameters defining
these black hole solutions, which divulge certain associated
macroscopic and microscopic acquisitions as well. There by one may
critically examine the physical implications arising from the
Gaussian fluctuations of the entropy, defining the intrinsic
state-space geometry. The present paper thus concentrates on the
macroscopic-microscopic descriptions of consistent rotating black
brane solutions and discuss the underlying state-space geometry,
thus anticipated as certain bound states of the possible
configurations arising, either from the $D$-branes, or dual
$M$-branes. Therefore, our study of the equilibrium state-space
geometry, characterized by certain electric-magnetic charges,
other invariant quantities, such as ADM mass, and angular momenta,
is well suited for an intriguing application of the AdS/CFT
correspondence and vacuum phase-transitions, if any, in the
considered black brane configurations.

Moreover, the state-space geometry has been one of the pivotal
intervention in the study of thermodynamic structures of a large
class of condensed matter systems, black holes or higher
dimensional rotating black branes, see for instance
\cite{Weinhold1, Weinhold2, RuppeinerRMP, RuppeinerA20,
RuppeinerPRL, RuppeinerA27, RuppeinerA41,
grqc0512035v1,grqc0601119v1,grqc0304015v1,0510139v3,0606084v1,0508023v2,SST}.
In particular, the article \cite{BNTBull} describes the nature of
the state-space configuration for various extremal and
non-extremal black branes, including the multi-centered branes,
fractional branes, fuzzy rings and $M$-theory bubbling solutions.
It may be observed that our state-space geometry analyzes the most
probable interactions between the microstates of the black branes.
Specifically, it determines the nature of statistical pair
correlation functions, correlation volume and the possible
stability criteria under the Gaussian fluctuations, in the
underlying equilibrium configurations. Our analysis thus provides
an appropriate ground for the implications arising from the
physics of black holes in string theory/ $M$-theory.

The applications thus considered are analogous to those mimicked
in \cite{BNTBull}, which give an appropriate notion of statistical
interactions, for the bubbling $AdS$ geometry and other well-known
configurations. Here, we shall explicitly construct the general
features of the state-space configurations, which come from the
underlying black brane entropy. It may thus naively be possible to
compare the concerned geometric quantities, with the fundamental
characteristics of the underlying microscopic configurations. In
particular, the state-space, thus constructed, primarily out-lines
the stability criteria of the state-space manifold, defining the
fluctuating higher dimensional black holes, with or without
rotations. Indeed, we have shown, for a certain range of the
parameters, that our intrinsic state-space geometric implications
provide a unified framework to discuss both the statistical
correlations, as well as the associated singularities and
equilibrium configurations, which in general live on an arbitrary
finitely curved intrinsic Riemannian manifold.

More precisely, we have analyzed a series of interesting examples,
in order to illustrate the state-space geometry, which, as an
intrinsic Riemannian geometry, may easily be defined from the
negative Hessian matrix of the counting entropy of the associated
rotating black hole configurations. In the concerned solutions, it
has been shown that there exists an interesting set of scaling
properties for the pair correlation functions. Thereby we have
analyzed the stability on possible hyper-surfaces and related
global properties of the state-space manifold of string theory and
$M$-theory black holes. In particular, we have indicated that our
ideas concerning macroscopic/ microscopic descriptions may easily
examine the phase-space stability of underlying brane
configurations, under the Gaussian fluctuations. It is thus
possible to discuss the statistical implications, associated with
an ensemble of boundary CFT microstates describing the particular
black hole configurations, in chosen microscopic descriptions.

An important ingredient in the discussion of the state-space
geometry turns out to be the fact that there exists a number of
extremal black branes, non-extremal black branes, multi-centered
black branes and the various other possible black brane
configurations, inspiring certain general physical ramifications.
Furthermore, it may be argued that the nature of state-space
configurations remains the same under the possible subleading
corrections. In fact, we find that the stability and critical
properties of the considered black hole solutions remain intact in
the large charge limit. The state-space geometry thus implied may
have an interesting inter-relevance with the AdS/CFT
correspondence. In particular one may try to come up with an
adequate account for the correlated microstates and associated
degeneracy of the fluctuating black hole configurations.

A parallel motivation may also be observed with certain logical inference of the string duality symmetries
involving the parameters of the brane configurations and in particular, it turns out that the corresponding
microscopic duality relations may easily be divulged by a set of decisive general coordinate transformations
on the associated state-space manifold of the typical macroscopic solutions.
Moreover, in the lieu of \textit{``thermodynamical/ statistical''} behavior of the associated configurations,
it appears that the significant parameters involving \textit{``macroscopic/ microscopic''} relation(s) thus
indicated may be in the close connection with those defining the \textit{``AdS/ CFT correspondence''}.
Such an inspiration for the duality may in fact be envisaged via the associated degeneracy formula of
underlying boundary conformal field theory, and thus it entangles only with the parameters of the configuration.
Incrementally, there exists a series of interesting circumstances which support the fact that our study
offers a geometrical approach to analyze the microscopic configuration of an ensemble of microstates
which describe an equilibrium thermodynamic configuration, and thus one may actually apply it to divulge
certain important physical and chemical behavior of the string theory/ M-theory black brane configurations.

In this paper, we specifically wish to consider a general
situation of the higher dimensional configurations, where the
state-space geometry determines the statistical pair correlation
functions and correlation volume which may disclose certain
critical nature of associated dual conformal field theory living
on the boundary. Thus, one may exhibit distinguishing features of
the microscopic duality acquisitions well-complied with the usual
understanding of the associated parameters defining the concerned
macroscopic configuration. In fact, it is well-known that both
descriptions thus apprised may further be elucidated from the very
possible application of the AdS/CFT correspondence \cite{LMM}. In
the present consideration, we find that the state-space scalar
curvature, as a function of the parameters, entails the central
nature of the stabilities in the considered black hole solutions,
while the determinant of the state-space geometry thus analyzed
entangles with the notion of degeneracy of brane microstates
constituting possible equilibrium statistical configurations.

Here, it is worth to mention that our intrinsic geometric formulation tacitly involves an underlying
statistical basis, which for the rotating black hole configurations requires to chose an ensemble of
CFT microstates in the thermodynamic limit and thus the interested formalization accommodates only the
invariant parameters concerning the string theory or $M$-theory compactifications.
In precise, this will imply that the state-space scalar curvature of rotating black brane
configurations apprehends the nature of the correlation volume and thus the largest possible
correlation being present in the underlying equilibrium microscopic systems.
This strongly suggests that in the context of state-space manifold of the rotating black holes
arising from a consistent string theory or $M$-theory solution when treated as closed systems,
one may realize in both cases that the non-zero scalar curvature provides valuable information
in analyzing the nature of quantum statistical interactions existing among the microstates
of the considered brane configurations.
Microscopically, as there exists many dual CFT descriptions, thus such brane configurations
of interest may be anticipated as an ensemble of the bound states whose vacuum degeneracy takes
an account of a large number of $D$-branes or an appropriate combination of dual $M$-branes.
\section{State-space Geometry}
In this section, we provide a brief account of the state-space
geometry, which as an intrinsic Riemannian manifold $(M, g)$
serves our purpose, for describing the nature of statistical
fluctuations in higher dimensional rotating black holes. Before
focusing our attention on specific configurations, it is
worthwhile reviewing some of the basics of intrinsic geometries,
and then in general finding the state-space meanings, emerging
from the asymptotic parameters of an ensemble of degenerate CFT
microstates. For more details, see \cite{Weinhold1, Weinhold2,
RuppeinerRMP, RuppeinerA20, RuppeinerPRL, RuppeinerA27,
RuppeinerA41, grqc0512035v1, grqc0601119v1, grqc0304015v1,
0510139v3, 0606084v1, 0508023v2} for certain black holes in
general relativity, \cite{SST} in string theory, \cite{BNTBull}
for the recent account of the concerned notions in string theory
and $M$-theory, and \cite{BNTSBVC,12p1,12p2,12p3} for the chemical
correlations and associated quark number susceptibilities in the
$2$- and $3$-flavor hot QCD configurations.

In the present investigation, we shall provide detailed
explanations for the state-space geometry, which analyzes
fluctuations about an equilibrium ensemble of brane microstates,
and there by describes the possible statistical configurations of
the (rotating) black brane solutions. In particular, one may
realize that our explorations shed light on the essential
characteristics of the statistical correlations, quantum phase
transitions and underlying critical behavior, if any, extant in
the higher dimensional black brane configurations. The intrinsic
Riemannian geometry, thus, turns out to be one of the main
appliance, in order to study the state-space structures of the
large class of black holes or black brane configurations being
described in general relativity, as well as in the framework of
fundamental string theories and $M$-theory.

First of all, let us recall that the usual probability distributions of thermodynamic fluctuations
about an equilibrium statistical configuration characterize an invariant line interval for the corresponding
intrinsic state-space geometry. Thus, by expanding the entropy about an equilibrium configuration,
it turns out that the underlying probability distribution in the Gaussian approximation reads
\begin{equation}
\Omega(x)= A \ \ exp [-\frac {1}{2} g_{ij}(x)dx^i \otimes dx^j],
\end{equation}
where the pre-factor $A$ is the normalization constant of the Gaussian distribution under consideration,
see \cite{LandauLifshitz} for the preliminary introduction and concerned physical motivations.
In this case, one may easily deduce that the second moments of the fluctuations turn out to be defined by,
\begin{equation}
\langle x_i x_j \rangle= \int  x_i x_j \Omega(x) \sqrt{\Vert g \Vert} \prod_i dx_i, \ \forall \ i= 0,1,..n
\end{equation}
Therefore, it may easily be perceived in the critical set-up of
\cite{RuppeinerRMP} that the components of the inverse metric
tensor are the second moments of the fluctuations, or the
contravariant statistical pair correlation functions. Here, it
turns out that the second moments of the concerned statistical
fluctuations may in turn be given by,
\begin{equation}
\langle x_i x_j \rangle= g^{ij}= < X^i X^j>,
\end{equation}
where the state-space co-ordinates $\lbrace X^i \rbrace $ are conjugate to the intensive variables
$\lbrace x_i \rbrace$, which may in particular be defined as,
\begin{equation}
X^i:= \frac{\partial S(x)}{\partial x_i}
\end{equation}
Furthermore, one obtains that the underlying Gaussian distribution
$\Omega(x)$ may be represented in terms of the conserved extensive
charges $\lbrace X^i \rbrace $. In particular, it is easy observe
that the Gaussian distribution describing our state-space geometry
may be expressed as,
\begin{equation}
P(X)= A \ \ exp [ -\frac {1}{2} g_{ij}(X)dX^i \otimes dX^j],
\end{equation}
Following \cite{RuppeinerRMP, 0510139v3, 0606084v1}, it turns out that the natural inner product
just mentioned may easily be ascertained, for an arbitrary $n$ parameter black brane configuration,
and the concerned state-space turns out to be a n-dimensional intrinsic Riemannian manifold $ M_n $.
In particular, the associated entropy, as an embedding function, defines the covariant components of
the metric tensor of the thermodynamic state-space geometry, which has originally been anticipated by
Ruppeiner in the related articles \cite{RuppeinerRMP, RuppeinerA20, RuppeinerPRL, RuppeinerA27, RuppeinerA41}.
Here, we shall take this representation of the intrinsic geometry, and thus note that the covariant
components of state-space metric tensor may be defined to be,
\begin{equation}
g_{ij}:=-\frac{\partial^2 S(\vec{X})}{\partial X^i \partial X^j}
\end{equation}
It is thus entirely clear from our outset, how this can be employed
to describe the state-space geometry of black holes. However, if we
restrict ourselves to the co-ordinates, which are extensive
parameters, then one may be able to draw certain properties of the
underlying fluctuations about an equilibrium statistical
configuration. We may further urge that the variables $\lbrace X^i
\rbrace $, associated with a set of asymptotic parameters carried
by the higher dimensional black hole, define a suitable co-ordinate
chart on the state-space manifold. In this representation, the
coordinate $\vec{X}$ turns out to be a finite collection of the
mass $M$, electric-magnetic charges $(P^i,Q_i)$, angular momenta
$\lbrace J_i \rbrace$ and certain concerned local invariant
charges $\lbrace q_i \rbrace$, if any. Thus, a most general
co-ordinate on an arbitrary state-space manifold $(M, g)$ may be
expressed as, $\vec{X}= (M, P^i,Q_i, J_i, q_i) \in M_n $.

With the the considerations developed in Appendix A, it is worth
to perceive that the local stability of the underlying statistical
configuration requires that the principal components of
state-space metric tensor $g_{ii}$ signifying heat capacities
should be positive definite
\begin{eqnarray}
g_{QQ} &>& 0 \nonumber  \\
g_{JJ} &>& 0
\end{eqnarray}
Moreover, we may inspect that the metric tensor for three
parameters $P,Q,J$ comprises Gaussian fluctuations into the
entropy as an embedding function and thus defines an intrinsic
state-space manifold for the two charged rotating black brane
configurations. This is because the components of the state-space
metric tensor are related to the statistical pair correlation
functions, which may be defined in terms of the three invariant
parameters describing the microscopic conformal field theory on
the boundary. Here, the simplest such examples which concern to
the present analysis are the usual $D_1$-$D_5$ CFT configurations
or the vacuum $M$-theory Kaluza-Klein black holes involving
$D_0$-$D_6$-branes. Furthermore, it is worth to note that the
local stability of the underlying statistical configurations
requires that the principal components of the state-space metric
tensor $g_{ii}$, which signify heat capacities, must be positive
definite,
\begin{eqnarray}
g_{QQ} &>& 0 \nonumber  \\
g_{PP} &>& 0 \nonumber  \\
g_{JJ} &>& 0
\end{eqnarray}
In this case, the stability in the hyper-planes of the statistical
configuration requires that all the off-diagonal fluctuations must
vanish, as well. This implies that the consequent principal minors
$\lbrace p_2, p_3 \rbrace $ of the metric element should be
strictly positive definite quantities on the entire state-space
$(M_3,g)$. One thus finds that the hyper-planes stability
criterion may rather be expressed as
\begin{eqnarray}
p_2:=  \left \vert\begin{array}{rr}
    g_{QQ} & g_{QP}  \\
     g_{QP} & g_{PP}  \\
\end{array} \right \vert > 0
\end{eqnarray}
Moreover, the existence of positivity of the state-space volume form imposes a stability condition on
the Gaussian fluctuations that for the underlying statistical configurations one must have $p_3>0$.
This in particular requires that the  determinant of the state-space metric tensor must be positive
definite as a function of electric-magnetic charges and angular momentum. In fact, a straightforward
computation yields that the determinant of the metric tensor may be given to be,
\begin{eqnarray}
\Vert g \Vert &= &
\frac{\partial^2 S}{\partial P^2}
  (\frac{\partial^2 S}{\partial Q \partial J})^2
+(\frac{\partial^2 S}{\partial Q \partial P})^2
  \frac{\partial^2 S}{\partial J^2}
+ (\frac{\partial^2 S}{\partial P \partial J})^2
  \frac{\partial^2 S}{\partial Q^2} \nonumber \\
&-& \frac{\partial^2 S}{\partial P^2}
 \frac{\partial^2 S}{\partial Q^2}
 \frac{\partial^2 S}{\partial J^2}
-2 \frac{\partial^2 S}{\partial Q \partial P}
  \frac{\partial^2 S}{\partial P \partial J}
  \frac{\partial^2 S}{\partial Q \partial J}
\end{eqnarray}
Consequently, it may be noticed that the scalar curvature
corresponding to this state-space geometry elucidates the typical
feature of Gaussian fluctuations about an equilibrium $D$-brane
(or dual $M$-brane) microstates for possible two charge rotating
configurations. Furthermore, we procure that it is easy to
determine a global invariant on the three variable state-space
manifolds. In this case as well, we nevertheless notice that one
may actually explicate the nature of the underlying microscopic
configurations. Just an interesting invariant which accompanies
the information of correlation volume of underlying statistical
systems turns out to be the intrinsic scalar curvature. A detailed
analysis shows that the concerned scalar curvature may easily be
determined
\begin{equation}
 R= \frac{1}{2 \Vert g \Vert^{2}} f(Q,P,J)
\end{equation}
The exact expression for the $f(Q,P,J)$ is rather involved and we relegate it to the Appendix B.
Furthermore, a systematic examination demonstrates that the state-space geometry with more
charges and angular momenta can similarly be defined, and in turn,
we must remark that one may easily deduce the statistical correlations and stability in
terms of the corresponding ensemble of microstates which delicately characterize an
elementary microscopic statistical basis against the macroscopic black brane configurations.
Thence, it is not difficult to ponder from the theory of Gaussian distribution that the components
of covariant state-space metric tensor may further be defined as the negative Hessian matrix
of underlying entropy with respect to the mass, invariant charges and angular momenta,
as well as certain possible exotic charges, if any, carried by the black brane configuration.
It is worth to note in the case of present interest that the black holes, when viewed from out-side,
may only be characterized by their mass, angular momenta, and electric-magnetic charges.
Detailed internal structure and history of their formation are thus irrelevant in this ``no hair'' property.
Such a drastic reduction of complexity is thus an important characteristic of the state-space investigations.

In the present description, the local stability condition of the
underlying statistical configuration under the Gaussian
fluctuations requires that all principal components of the
fluctuations be positive definite, i.e. for given set of
state-space variables $\lbrace X^1, X^2, \ldots, X^n \rbrace$ one
must demand that $\lbrace g_{ii}(X^i) > 0; \ \forall i= 1, 2,
\ldots, n \rbrace$. In particular, it is important to note that
this condition is not sufficient to insure the global stability of
the chosen configuration and thus one may only accomplish a
locally equilibrium configuration. It is however known that the
complete stability condition requires that all
principal components of the Gaussian fluctuations should be
positive definite and the other components of the fluctuations
should vanish \cite{RuppeinerRMP}. In order to ensure this
condition, one appraises that all the principal components and all
the principal minors of the metric tensor must be strictly
positive definite. This implies that the global stability
condition constrains the allowed domain of the parameters of black
hole configurations, which may interestingly be expressed by the
following set of simultaneous equations:
\begin{eqnarray}
p_0&:=& 1,  \nonumber \\
p_1&:=& g_{11} > 0, \nonumber \\
p_2&:=&  \left \vert\begin{array}{rr}
    g_{11} & g_{12}  \\
     g_{12} & g_{22}  \\
\end{array} \right \vert > 0, \nonumber \\
p_3&:=&  \left \vert\begin{array}{rrr}
    g_{11} & g_{12} & g_{13} \\
     g_{12} & g_{22} & g_{23} \\
     g_{13} & g_{23} & g_{33} \\
\end{array} \right \vert > 0, \nonumber \\
\vdots \nonumber \\
p_n&:=& \Vert g \Vert > 0
\end{eqnarray}
Strictly speaking, the determinant condition $p_n> 0 $ implies that there should exist a positive definite
volume form on $(M_n,g)$. However, it is worth to notice that the classical thermodynamic fluctuation theory
may not remain viable for the general black hole configurations, see for detail \cite{RuppeinerRMP}.
This is because a black hole cannot come into an equilibrium with any extensive, infinite environment,
and in particular, for all asymptotic parameters the full Hessian determinant of the black hole entropy
$S:= S(X^i)$ turns out to be negative definite, see for the related motivation \cite{Landsberg}.
This is in fact required, in order to produce a local maximum in the total entropy of the configuration.
Thus, the non-existence of the positive definite volume form on an intrinsic state-space manifold $(M_n,g)$
puts the full problem beyond the control of standard equilibrium fluctuation formalism, when all the state-space
parameters $\lbrace X^i:= (M, J_i, P^i, Q_i) \rbrace$ are fluctuating.

Physically, the brane dynamics may ultimately favor extreme situations where the black hole may
either completely evaporate or grow without limit. Moreover, the black hole will presumably create
whatever particles or branes it likes near its event horizon, and an equilibrium is quite unlikely,
until the environment is populated by the particles or branes in the same proportion to those created.
However, in a limited domain of the parameters, when at least one of the parameter may slowly be changing in
time, in comparison to the remaining ones, it obliges a class of stable configurations under the Gaussian fluctuations.
This investigation thus implies a certain restricted class of stable configurations with arbitrary $m$ number of
fluctuating parameters and the remaining $n-m$ parameters being effectively fixed or drifting very slowly
out of the equilibrium configuration with the environment, where $1\leq m \leq n $.

Interesting discussions of the phase transitions exist in
literature which involve some change of the black hole horizon
topology, \cite{Arcioni, Aman}. However, there exists a large class
of black holes with a spherical topology in the physical
regime, which are quite energizing in their own, as we shall
exhibit in the next section. However, we shall now analyze
the behavior of collective state-space correlations, and in
particular, we wish to illustrate certain peculiarities of the
state-space geometry which in the large charge limit provide the
correlation length for the associated microscopic configurations.
Additionally, Ruppeiner has revived that the state-space scalar
curvature remains proportional to the correlation volume $\xi^d$,
where $d$ is the system's spatial dimensionality and $\xi$ is its
correlation length which reveals related information residing in
the microscopic models \cite{RuppeinerA20}. Furthermore, the
state-space curvature for higher dimensional black holes may be
intertwined with vacuum phase transitions existing in the possible
brane configurations and thus one may revel certain intriguing
information of the possible microscopic CFTs.

Ruppenier has conjointly interpreted with the assumption ``that
all the statistical degrees of freedom of a black hole live on the
black hole event horizon'' and thus the scalar curvature signifies
the average number of correlated Planck areas on the event horizon
of the black hole \cite{RuppeinerPRD78}. In particular, zero
scalar curvature indicates certain bits of information on the
event horizon fluctuating independently of each other, while the
diverging scalar curvature signals a phase transition indicating
highly correlated pixels of the informations. Moreover, Bekenstein
has introduced an elegant picture for the quantization of the area of
the event horizon being defined in terms of Planck areas
\cite{Bekenstein}, and in this concern it appears from our
analysis that the microscopic resolution would possibly involve
the framework of well-celebrated Mathur's fuzzballs
\cite{fuzzball}. Thus, our perspective discloses ground with the
statements that the state-space scalar curvature of interest has
not only exiguous microscopic knowledge of the black hole
configuration, but in particular it has nice intrinsic geometric
structures. Specifically, we find that the configurations under
present analysis are effectively attractive in general, while they
are stable only if at least one of the parameters remain fixed. Our
hope is that finding statistical mechanical models with like
behavior might yield further insight into the microscopic
properties of black holes and conclusive physical interpretation
of their state-space curvatures and related intrinsic geometric
invariants.

With this general geometric prelude to the physics of a large class of higher dimensional rotating solutions,
we shall now proceed to systematically analyze the state-space geometric structures characterized by a
finite number of invariant electric-magnetic charges, angular momenta, mass, and other veritable parameters,
if any, defining the microstates of the black hole configurations.
\section{Rotating Black Hole Configurations}
In the present section, we shall first apply the notion of
state-space geometry thus developed to the rotating Myers-Perry
configurations. There after, we move on to investigate the related
well-known two, three, charged rotating black hole configurations.
We shall devote the present section to the possible descriptions
and their interpretations against our intrinsic Riemannian
geometry, which describes the vitality of an ensemble of
equilibrium microstates of the extremely simple black holes. In
particular, in the very first subsection we shall approach our
study by analyzing the state-space manifold of the rotating black
holes, which only comprises the mass and an angular momentum as
invariant parameters, and then we shall proceed into the next
subsections, which expose that the very close conclusions hold as
well for the various higher charged rotating black hole
configurations in the vacuum $M$-theory, string theory and
possible supergravity theories.

In this spirit, we shall explain how the two point statistical correlation functions,
as the function of parameters characterizing an ensemble of equilibrium microstates of the
rotating black hole configurations, are well-structured in the framework of our state-space geometry.
In turn, it shall be interesting to notice whether the correlation length of higher dimensional rotating
configurations may be described as a regular or singular function of the charges and angular momentum
on the concerned state-space manifolds. In both cases, we recognize from the perspective of state-space
manifolds that it is not difficult to analyze the nature of the singularities, if any, present in the
corresponding scalar curvature for the underlying microscopic configurations.
In particular, we shall show that there may exist certain critical point(s), or critical line(s),
or critical (hyper)-surface(s) on the state-space manifolds of the definite rotating black brane
configurations on which the associated state-space scalar curvatures diverge, as a function of
the parameters being specified by the considered configurations.

A main difficulty in assimilating the black hole solutions, however,
appears to be somewhat anomalous. The concerned problem may be
taken into account by the coordinate systems of the state-space
manifold, in which the coordinates illuminating the abovementioned
configuration take the simplest known form. Therefore, our
conclusions thus accomplished would also be useful for obtaining
adapted space-time parameters in the other possible settings. In
the following, we shall study how the uniqueness theorems play a
central role in the very possible study of generalized state-space
manifolds encompassing a definite set of brane charges, concerning
the higher dimensional rotating black hole configurations.
\subsection{Myers-Perry Black Holes}
This subsection, as the first exercise, elucidates the state-space
geometry of five dimensional neutral black holes in the vacuum
$M$-theory. We notice that these solutions are natural to be analyzed
in the type-$II$ string theory description. In particular, it is
known that these neutral black holes carry a minimal set of
quantum numbers, {\it viz.}, mass and angular momentum which
characterize the vacuum $M$-theory configurations. From the
microscopic perspective, the intriguing idea emerges from the fact
that these black holes may be appreciated by taking the product of
$S^1$ with flat $T^6$, and then one may obtain the distinguishing
$M$-theory solutions, whose type-$IIA$ reduction has $D_0$ and
$D_6$ charges. Moreover, it has actually been shown in \cite{itz}
that in a certain limit, akin to the decoupling limit in AdS/CFT,
one may acquire the asymptotically flat neutral Myers and Perry
black hole solutions \cite{MP}. However, it is worth to mention
from the type $IIA$ dual standpoints that there are
non-supersymmetric quadratically stable $D_0$-$D_6$ bound states
\cite{wati}, which serve as the basis for the microscopic picture of
the statistical configurations.

Although supersymmetry turns out to be completely broken in this
limit \cite{KO}, nonetheless one may prove that, in the same
limit, there are certain supersymmetric bound states of the
constituent $D_0$ and $D_6$ branes \cite{Horowitz:1996ay}. Indeed,
it is stimulating to note that we are obliged to use two different
sets of branes, and then, from the standpoints of type-$IIA$
string theory, the number of brane charges is doubled. In
particular, this is reminiscent of the earlier suggestions that
the neutral black holes should be viewed as a collection of branes
and anti-branes \cite{Horowitz:1996ay}. Moreover, there exists a
mysterious duality invariant formula which, including the
Schwarzschild black holes, reproduces the entropy of all known
nonextremal black holes ascribed in terms of charges carried by the
constituent branes and antibranes, see for details
\cite{Horowitz:1996ay,ddbar}. In both cases, there exist simple
microscopic $D$-brane descriptions which precisely reproduce the
Bekenstein Hawking entropy, which may thence be expressed in terms
of the mass $M$ and angular momentum $J$ possessed by these
extremal rotating black hole configurations in the vacuun
$M$-theory. Specifically, one finds an illuminating expression for
the Bekenstein Hawking entropy of the extremal Myers-Perry black
holes, which may be expressed as
\begin{equation}
S(M,J):= \frac{8}{3G}\sqrt{\frac{2\pi}{3}\bigg(G^3M^3-\frac{27}{32}\pi G^2J^2\bigg)}
\end{equation}
where $G$ shall henceforth denote the Newton coupling constant,
see for details \cite{MP}. In order to take a closer look at the
state-space geometry arising from the equilibrium microstates of
underlying $D$-brane configuration, we focus our attention on the
negative Hessian matrix of the entropy of these simplest rotating
Myers-Perry neutral black hole configurations. Here, it is worth
to note that the mass and angular momentum are the unique
parameters which characterize the Myers-Perry configurations, and
consequently they form the co-ordinate chart for the underlying
state-space geometry. Employing the previously proclaimed
formulation, one may easily read off the components of the covariant
state-space metric tensor to be
\begin{eqnarray}
g_{MM}&=& \sqrt{6 \pi}G^2M (\frac{9}{4}\pi G^2J^2-\frac{2}{3}G^3M^3) (G^3M^3- \frac{27}{32}\pi G^2J^2)^{-3/2} \nonumber \\
g_{MJ}&=& -\frac{9}{8}\sqrt{6} \pi^{3/2} G^4JM^2 (G^3M^3- \frac{27}{32}\pi G^2J^2)^{-3/2} \nonumber \\
g_{JJ}&=& \frac{3}{4} \sqrt{6} \pi^{3/2} G^4M^3(G^3M^3- \frac{27}{32}\pi G^2J^2)^{-3/2}
\end{eqnarray}
Within this framework, we observe that there exists a very simple
description which divulges the geometric nature of the statistical
pair correlations. The fluctuating extremal Myers-Perry  black
holes may thus easily be determined in terms of the mass and
angular momentum of the underlying configurations. Moreover, it is
evident that the principal components of the statistical pair
correlations are positive definite for a range of the parameters
of concerned black holes, which physically signifies a certain
self-interaction of a fictitious particle moving on an intrinsic
surface $(M_2(R),g)$. In particular, it is clear in this case that
we have the following state-space metric constraints:
 \begin{eqnarray}
g_{MM} &>& 0, \ \ \forall \ (M,J) \mid \frac{27}{32}\pi J^2< GM^3 < \frac{27}{8}\pi J^2  \nonumber \\
g_{JJ} &>& 0, \ \ \forall \ admissible  \ (M,J)
\end{eqnarray}
Consequently, we may easily reveal that the common domain of the above state-space constraints defines
the range of physically sensible values of mass and angular momentum, such that the Myers-Perry
black holes may remain into certain locally stable statistical configurations.
We may also notice that the mass-mass component $g_{MM}$ of the state-space metric tensor is
asymmetrical, in contrast to $g_{MJ}$ and $g_{JJ}$.
This is physically be well-accepted, as the mass-mass component is somewhat
like the head on collision of two equal mass particles, which alternate more energy in contrast to the
other excitation, either involving a single rotating particle or a massless rotating (or spinning) particle.
It is worth to point out that the relative pair correlation function determines the rotation parameter
($a:= J/M$) of Myers-Perry black holes, which may be defined as the modulus of the ratio of mass-rotation
to rotation-rotation statistical correlations. In particular, we procure that the rotation parameter
thus apprised may be given as
\begin{eqnarray}
a: = \frac{3}{4} \vert \frac{g_{MJ}}{g_{JJ}}\vert.
\end{eqnarray}
Moreover, the stability of Myers-Perry statistical configurations may earnestly be analyzed
by computing the degeneracy of the associated two dimensional state-space manifold. In fact,
we can easily ascertain that the determinant of the state-space metric tensor may be given to be
\begin{equation}
g(M,J)= -3072 \pi^2 G^2M^4 \bigg(32 GM^3- 27\pi J^2\bigg)^{-2}
\end{equation}
The determinant of the metric tensor thus calculated is non-zero
for any set of given non-zero mass and angular momentum, and thus
it provides a non-degenerate state-space geometry for this
configuration. In turn, one may illustrate the order of
statistical correlations between the equilibrium microstates of
the Myers-Perry rotating black hole system. Besides the fact that
the principal components constraints $\lbrace g_{ii} > \ 0 \ \vert
\ i= M, J \rbrace$ imply that this system may accomplish certain
locally stable statistical configurations, however the negativity
of the determinant of the state-space metric tensor indicates that
the underlying systems may globally endure unstable, as well. This
connotes that there is no positive definite volume form on the
$(M_2,g)$, and thus one may conclude that this system might go to
some more stable brane configurations.

Furthermore, in order to examine certain global properties on
these black holes phase-space configurations, one is required to
determine the associated geometric invariants of the underlying
state-space manifold. For the Myers-Perry black holes, the
simplest invariant turns out to be the state-space scalar
curvature, which may easily be computed by using the intrinsic
geometric technology, defined as the negative Hessian matrix of the
entropy captured by the rotating contributions. Explicitly, we
discover that the state-space curvature scalar for the Myers-Perry
configurations may easily be depicted to be
\begin{equation}
R(M,J)= -\sqrt{\frac{27}{4\pi}} G^3 \bigg(32 GM^3- 27\pi J^2\bigg)^{-1/2}
\end{equation}
We recognize, in the framework of our state-space geometry, that the negative sign of the curvature scalar
signifies that this system is effectively an attractive configuration under the Gaussian fluctuation.
Thus, the Myers-Perry configurations are relatively less stable than the other configurations having
certain non-negative state-space scalar curvatures. Nevertheless, we observe that the curvature scalar thus
considered is inversely proportional to the determinant of the underlying state-space metric tensor,
In turn, we find that it remains a non-zero, finite, regular function of the mass and angular momentum
carried by the Myers-Perry black holes.

It is however important to mention that the state-space geometric
quantities may become ill-defined, if the state-space co-ordinates
being defined as the space-time parameters, jump from one existing
domain to the other domain of the ergo branch. This indicates that
the Myers-Perry black holes correspond to certain interacting
statistical configurations in a chosen ergo branch. In all
admissible domain of the physical parameters, we discover that the
rotating Myers-Perry black holes has no phase transition, and thus
the fundamental statistical configurations are completely free
from the critical phenomena.

Furthermore, we note, up to some easily appreciable constants $\alpha,\beta \in R$, that it is not
difficult to determine the constant entropy curve and constant state-space scalar curvature curve
for the rotating Myers-Perry black hole configurations. In particular, we may apparently figure out
that both the above possible curves, defined as real embedding functions, take a very simple form,
which may simultaneously be given as
\begin{equation}
M^3-\alpha J^2=\beta
\end{equation}
It thus turns out that the state-space quantities of the vacuum $M$-theory black holes,
particularly for the neutral Myers-Perry configurations, which are far from being supersymmetric,
may easily be analyzed by an appropriate combination of simple $D$-branes.
However, it is known since the study of Strominger and Vafa \cite{9601029v2} that the original string
theory black hole solutions are both charged and supersymmetric. In particular, they have pointed out that
(i) the charges of the black hole serve as tag marks which help to identify their microscopic constituents in
string theory and (ii) the supersymmetry requirement constrains the phase-space configuration of these black holes.
The next subsection would thus be devoted to investigate the notion of charges into our state-space geometry,
and there by we shall analyze the possible role of concerned charges in the veritable black hole configurations
in vacuum $M$-theory. In particular, we shall precisely compute the statistical pair correlations under this
consideration, which ascribes certain convincing physical meaning to the state-space quantities for the simplest
Kaluza-Klein configurations.
\subsection{Kaluza-Klein Black Holes}
In the present subsection, we shall study the state-space geometry arising from the two charged rotating
Kaluza-Klein black hole solutions. In particular, we shall devote our attention to the possible descriptions,
and their physical interpretations, of the state-space geometry arising from the vacuum $M$-theory black hole.
A similar two charge system arises from the $D_1$-$D_5$ configurations, which we shall analyze in section $3.4$.
In both cases, we shall discuss certain vital issues of the underlying equilibrium microstates, which describe
self and inter-brane statistical interactions, intertwined with the two charged rotating black holes. We find that
one may easily examine the nature of the associated correlation lengths for these simple $D$-brane configurations.

We begin our discussion by examining the state-space quantities of the Kaluza-Klein black holes, which
compound as well to the previously exposed properties of five-dimensional neutral Myers-Perry black holes.
It is worth to mention that one may easily consider certain $M$-theory vacuum solutions, which describe
the five-dimensional Kaluza-Klein black holes, and thus it appears interesting to find out their general
inter-relevance with those arising from the just encountered rotating Myers-Perry black holes.
Here, the present perspective of the state-space geometry may thus interestingly be divulged in the extremal limit,
which is akin to the decoupling limit of the underlying corresponding AdS/CFT correspondence.

We recommend \cite{gw} for detailed interpretations of these black hole solutions in ordinary four space-time dimensions.
In this description, our framework displays that the state-space geometry of general Kaluza-Klein black holes
may be characterized by the mass $M$, angular momentum $J$, electric charge $Q$, and magnetic charge $P$.
Here, we shall however focus our attention on the specific situations described in \cite{hep-th/9909102, hep-th/9505038},
and in particular an interesting case emerges when the concerned parameters of the solution appease an inequality,
\begin{eqnarray}
2G_4M \geq \bigg (Q^{2/3}+P^{2/3} \bigg)^{3/2}
\end{eqnarray}
Moreover, we may realize that the state-space geometry of these black hole solutions measures an
intriguing lower bound on the allowed angular momentum, than that arising from the limit of slow
rotations satisfying $G_4 J<PQ$. Our computation thus shows that these solutions are more strongly
saturated than the constrained solutions in the extremal limit. It is worth to note that the observation
of extremality and state-space conclusions are independent of the five dimensional rotations.
First, we would like to signal a certain interesting behavior of the pair correlation functions
for two charged black hole solutions in M theory, and then, from the view points of type IIA string theory,
our analysis indicates that the state-space correlations may easily be described in the dual description
of $D_0$ and $ D_6 $ brane cconfigurations.

Microscopically, the state-space geometry involves counting the microstates of a chosen configuration.
The case of Kaluza-Klein solutions may be realized by taking the product of the Kaluza-Klein circle $S^1$ with
$T^6$ with the volume $(2\pi)^6 V_6$. Thus, one may view them as vacuum solutions in M theory with
$S^1$ being the $M$-theory compactification circle.
Note, further, that our discussion may also begin by considering the usual coupling relation between the
$M$-theory and IIA string theory, which may in particular be ascribed by the relation $R_0=g l_s$.
Furthermore, the charges on the constituent $D_0$ and $D_6$-branes may be interpreted as the electric
and magnetic charges, which respectively correspond to the net numbers $N_0$ and $N_6$ of the basic branes.

We focus our attention on the state-space analysis of five-dimensional vacuum $M$-theory configurations
by considering the compact Kaluza-Klein circle of radius $R_0$, which is fibered over the two spheres
of spherical symmetry of the underlying four dimensional black hole solutions.
What follows here shows that the magnetic charges are quantized in terms of the Kaluza-Klein radius $R_0$,
and that the electric charge renders as the inverse of the radius $R_0$ of the Kaluza-Klein circle.
More precisely, the underlying electric and magnetic charges take integer values, in terms of the net
number of constituent $D_0$ and $D_6$ branes. In turn, one arrives at the simple quantization condition
that the existing charges may be inscribed as
\begin{eqnarray}
(Q, P)= \bigg(\frac{2G_4 N_0}{R_0}, \frac{N_6 R_0}{4}\bigg)
\end{eqnarray}
This is due to fact that the Kaluza-Klein circle bundles over $S^2$, which may in turn be labeled by
an integer. Thus, the corresponding electric charges are also quantized, since they correspond to
the momentum in the radial direction of the Kaluza-Klein circle.
It turns out that the Kaluza-Klein black holes are described in a microscopic framework, whose CFT
deals with $N_0$ number of $D_0$ branes and $N_6$ number of $D_6$ branes. There by, it is naturally
interesting to peruse the investigation of underlying statistical configurations in this perspective.
We therefore wish to analyze the state-space of two parameter extremal Kaluza-Klein black holes with
an angular momentum $J$. Such space-time solutions appear quite naturally in the string theory, see for
example \cite{larsen2}. In this case, one finds that the underlying horizon entropy may be given to be,
\begin{eqnarray}
S(N_0, N_6, J)&:=&\frac{A_{(4)}}{4G_4}= 2 \pi \sqrt{\frac{1}{4}N_0^2 N_6^2- J^2}
\end{eqnarray}
The state-apace geometry, constructed out of the equilibrium state of the
rotating two charged Kaluza-Klein black holes resulting from the entropy, may
now easily be computed, as earlier, from the negative Hessian matrix of the entropy,
with respect to the brane numbers and angular momentum $ \lbrace N_0, N_6, J \rbrace $,
which form the coordinates of the intrinsic state-space manifold.
Explicitly, we find that the components of the covariant metric tensor are given as,
\begin{eqnarray}
g_{N_0N_0}&=& 4 \pi N_6^2 J^2 (N_0^2 N_6^2- 4J^2)^{-3/2} \nonumber \\
g_{N_0N_6}&=& - \pi N_0 N_6 (N_0^2 N_6^2- 8J^2) (N_0^2 N_6^2- 4J^2)^{-3/2}  \nonumber \\
g_{N_0J}&=& -4 \pi JN_0 N_6^2 (N_0^2 N_6^2- 4J^2)^{-3/2} \nonumber \\
g_{N_6N_6}&=& 4 \pi N_0^2 J^2 (N_0^2 N_6^2- 4J^2)^{-3/2} \nonumber \\
g_{N_6J}&=& -4 \pi JN_0^2 N_6 (N_0^2 N_6^2- 4J^2)^{-3/2} \nonumber \\
g_{JJ}&=& 4 \pi N_0^2 N_6^2 (N_0^2 N_6^2- 4J^2)^{-3/2}
\end{eqnarray}
Incidentally, we observe, from the simple $D$-brane description,
that there exists an interesting brane interpretation, which
describes the state-space correlation formulae arising from the
corresponding microscopic entropy of the aforementioned non
supersymmetric extremal black hole solutions. Furthermore, our
state-space correlations turn out to be in a precise accordance
with the underlying attractor configuration being disclosed in the
special limit of Bekenstein-Hawking solutions. In the entropy
representation, it may thus be noticed that the Hessian matrix of
the entropy illustrates the nature of possible correlations
between the set of extensive variables, which in this case are
nothing more than the $D_6$ and $D_0$-brane charges and angular
momentum. As mentioned before, we may articulate in this case
again that, for all non-zero values of $N_0,N_6,J$, the principal
components of intrinsic state-space metric tensor satisfy
\begin{eqnarray}
g_{N_0N_0} &>& 0 \nonumber \\
g_{N_6N_6} &>& 0 \nonumber \\
g_{JJ} &>& 0
\end{eqnarray}
Substantially, the principal components of state-space metric
tensor signify heat capacities or the associated compressibility,
whose positivity indicates that the underlying statistical systems
are in local equilibrium, consisting of the $D_6$ and $D_0$-brane
configurations. Furthermore, we perceive that the ratios of
possible diagonal components vary as the inverse square, which
weakens faster, and thus relatively quickly comes into an
equilibrium configuration, than those involving the off diagonal
ones. We further observe that the ratios of non-diagonal components
vary inversely, and in turn remain comparable for a longer
domain of the parameters varying under the Gaussian fluctuations. In
particular, we easily inspect $ \forall i \neq j \in \lbrace 0,6
\rbrace $ that the relative correlation functions satisfy the
following simple relations:
\begin{eqnarray}
\frac{g_{N_iN_i}}{g_{N_jN_j}}&=& (\frac{N_j}{N_i})^2 \nonumber \\
\frac{g_{N_iN_i}}{g_{JJ}}&=& (\frac{J}{N_i})^2 \nonumber \\
\frac{g_{N_iJ}}{g_{N_jJ}}&=& \frac{N_j}{N_i} \nonumber \\
\frac{g_{N_iN_i}}{g_{N_iJ}}&=& -\frac{J}{N_i}
\end{eqnarray}
Moreover, one should note that the behavior of the brane-brane correlation defined as $g_{N_iN_j}$
is asymmetric, in comparison with all other pair correlation functions.
This may simply be understood by the fact that the brane-brane interaction imparts more energy
than either the self-interaction or that due to the rotation.
In particular, one discovers that the brane-brane relative pair correlation, with respect to the
rotation-rotation correlation, reads
\begin{eqnarray}
\frac{g_{N_0N_6}}{g_{JJ}}&=& -\frac{N_0N_6}{4}\bigg(1- 8(\frac{J}{N_0N_6})^2\bigg)
\end{eqnarray}
We thus deduce that the relative brane-brane correlation vanishes exactly at half angular momentum
of the vanishing entropy condition. This suggests that the brane-brane correlation is stable
under the Gaussian fluctuations, if the four dimensional angular momentum satisfies an inequality,
$ \vert J\vert< \frac{N_0N_6}{2\sqrt{2}}$.
Nevertheless, our state-space geometry enjoys a lower bound on the attainable angular momentum, than
the constraint arising from the acquainted slow rotations of the $M$-theory vacuum solutions.
Furthermore, it may be inferred from our analysis, independently of the five dimensional rotations,
that these solutions are more strongly saturated, than simply the condition of extremality, or
the decoupling limit concerning the AdS/CFT.

In order to investigate certain global properties of the
Kaluza-Klein configurations, we need to determine stability along
each direction, each intrinsic plane, as well as on the full
state-space manifold. In particular, for determining whether the
underlying configuration is locally stable on state-space planes,
one may compute the corresponding principal minor of the negative
Hessian matrix of the entropy. In this case, we may easily appraise
that the principal minor $p_2$, computed from the above state-space
metric, reduces to
\begin{eqnarray}
 p_2= \pi^2 N_0^2 N_6^2 \bigg(12J^2- N_0^2 N_6^2\bigg) \bigg(N_0^2 N_6^2- 4J^2\bigg)^{-2}
\end{eqnarray}
Thus, for all physically possible values of brane charges, the minor constraint $p_2>0$ results
into a restriction on the accessible values of the angular momentum.
%
In particular, we may easily ascertain the nature of the state-space geometry of Kaluza-Klein systems, i.e.
that the planar stability exists only in a limited domain of the configuration, whenever the angular momentum
picks up at least $ \vert J\vert= \frac{N_0N_6}{2\sqrt{3}}$, or alternatively the planar stability requires
that a given $D_0$-$D_6$-brane configuration is scarcely populated and thus the net brane charges are effectively
bounded by the four dimensional rotation.
Moreover, it is not difficult to investigate local stability in the full phase-space configuration,
which may easily be carried out by computing the determinant of the state-space metric tensor.
In this case, we observe that the determinant of the state-space metric tensor is
\begin{eqnarray}
g(N_0, N_6, J)= -4 \pi^3 N_0^4 N_6^4 \bigg(N_0^2 N_6^2- 4J^2\bigg)^{-5/2}
\end{eqnarray}
which in turn never vanishes for any given non-zero brane charges.
This non-degenerate metric tensor thus corresponds to a well-defined state-space,
which may in fact solely be parameterized in terms of the $D$-brane charges,
and an underlying angular momentum: $ \lbrace N_0, N_6, J \rbrace $.
We however emphasize that the determinant of the metric tensor does not take a positive definite
form, and thus there is no positive definite volume form on the defined state-space $(M_3,g)$.
It may thence be concluded that a Kaluza-Klein black hole, when considered as the bound state
of $D_0$-$D_6$-branes in the type-II string description, does not correspond to an intrinsically
stable statistical configuration.

Furthermore, in order to observe the nature of the global space-time properties,
one can construct certain local identifications in the five-dimensional black hole configurations.
In general, it turns out that the corresponding asymptotic spatial space-time geometry may globally
be determined to be an orbifold $\mathbf{R}^4/\mathbf{Z}_{N_6}$ \cite{itz}.
Thus, the only configuration with $N_6=1$ causes globally asymptotically flat solutions,
and all other possibilities with $N_6>1$ correspond to the black holes which sit at the tip of the conical space.
It is worth to note that the horizon topology changes whenever $P\neq 0$, and thus the magnetic charge
as an independent coordinate plays an important role in the realization of state-space geometry.
In particular, one finds for the $N_6=1$ that the Kaluza-Klein-circle $S^1$ and the four dimensional spherical
horizon $S^2$ combine into the topological $S^3$.
At this point, we however discover that nothing special happens either to the statistical pair correlations,
or to the determinant of the metric tensor, or to the state-space scalar curvature, as well for the choice of
$N_6=1$, except for the fact that all associated constraints are now applied on the value of $N_0$.
Furthermore, we may assert, from the duality invariance of the Kaluza-Klein entropy, that the analogous
state-space conclusions remain true under the exchange of brane numbers, {\it viz.}, $N_0 \leftrightarrow N_6$.

Physically, it is interesting to explore the five-dimensional interpretation of the Kaluza-Klein solutions.
In this case, it turns out, in the absence of a magnetic charge, that the horizon has a simple topology
$S^1 \times S^2$ , where $S^1$ is the usual Kaluza-Klein circle. Thus, the space-time solution of interest
turns out to be a boosted black string along the Kaluza-Klein direction.
On other hand, the vanishing of electric charge $Q=0$ and $J=0$ divulges the extremal limit. In this case
the considered space-time solution reduces to the Kaluza-Klein monopole, whose horizon geometry may in fact be
described as a cigar space-time, fibered over the spherical horizon topology of the four dimensional solution.
In the limits of either $\lbrace J=0, N_6=0 \rbrace$ or $\lbrace J=0, N_0=0 \rbrace$, we may easily apprise
that the state-space geometry becomes an ill-defined intrinsic Riemannian manifold, and there is no
interesting limit of the state-space scalar curvature.
At this point, we strongly feel that the Gaussian approximation alienates down and we thus need to go beyond
this consideration, in order to determine appropriately the long range behavior of the full statistical correlations.

However, we reveal, in the Gaussian approximation, that the statistical correlation volume, as a global invariant,
takes an interesting form for the extremal microscopic configurations.  Under these simplified identifications,
it follows that the state-space scalar curvature takes an intriguing form, for the Kaluza-Klein black holes.
In particular, it turns out that our geometric analysis does not entangle with the space-time orbifolds,
unless the entropy changes its duality invariant form. Correspondingly, one may easily notice, in this limit,
that the expression for the state-space scalar curvature may be given as
\begin{eqnarray}
 R(N_0, N_6, J)= -\frac{1}{\pi N_0^2 N_6^2} \bigg(N_0^2 N_6^2+ 6J^2\bigg) \bigg(N_0^2 N_6^2- 4J^2\bigg)^{-1/2}
\end{eqnarray}
The negative sign of the scalar curvature signifies that the underlying system is effectively
an attractive configuration, and thus it is a less stable than an arbitrary positively curved configuration.
Moreover, it is worth to mention that the scalar curvature thus determined never vanishes, for any
physically affirmed value of the $D$-brane charges and angular momentum, and it is an everywhere regular
function on the state-space manifold. Thus, for any non-zero brane charges, this system always
corresponds to a weakly interacting statistical configuration.

Here, we find that the regular state-space scalar curvature seems to be comprehensively
universal, for the given number of parameters of the configuration. In fact, the concerned
idea turns out to be related with the typical form of the state-space geometry arising from the
negative Hessian matrix of the duality invariant expression of the black hole entropy.
As the standard interpretation, the state-space scalar curvature describes the nature of
underlying statistical interactions of the possible microscopic configurations,
which in particular turn out to be non-zero for the Kaluza-Klein black holes.
Note that the absence of the divergences in the scalar curvature indicates that the present
black hole solution is an everywhere thermodynamically stable system on the state-space configuration.
Thus, it turns out that there are no any phase transitions or such critical phenomena
in the underlying state-space manifold of the $D_0$-$D_6$-brane configurations. In fact,
we may easily appreciate that the constant entropy curve is a standard curve, which may be given by
\begin{eqnarray}
N_0^2 N_6^2= J^2+ c,
\end{eqnarray}
where $c$ is some real constant, being determined from the given entropy expression.
This determines the $D_0$-$D_6$ Kaluza-Klein black brane embedding in the view-points of state-space geometry.
Moreover, we may also disclose in the present case that the curve of constant scalar curvature is given as,
\begin{eqnarray}
(N_0^2 N_6^2+ 6J^2)^2= k^2 N_0^4 N_6^4 (N_0^2 N_6^2- 4J^2),
\end{eqnarray}
where $ k $ is some real constant which may easily be fixed of the given value of the scalar curvature.
Moreover, it is not difficult to enunciate that the quantization condition existing on the electric
and magnetic charges signifies a general coordinate transformation on the state-space manifold,
which may be presented in terms of the net numbers of respective branes.
This in turn yields that the brane charges may be written as, $ Q= 2G_4 M_0 N_0 $ and $ P= 2G_4 M_6 N_6 $,
where $G_4$ is the usual four dimensional Newton's coupling constant given by $G_4=g^2 l_s^8/8V_6$.
In this concern, the article \cite{itz} implies that the masses of the individual $D_0$, $D_6$-branes
may be expressed as
\begin{eqnarray}
M_0 &=& {1\over g l_s  }  \nonumber \\
M_6 &=& {V_6\over g l_s^7 }
\end{eqnarray}
This suggests that the state-space fluctuations thus described may as well be expressed in terms of the
individual brane masses and the four dimensional rotation. In particular, it is interesting  to reveal
how our state-space geometry, expressed in terms of the $ \lbrace M_0, M_6, J \rbrace $, entangles with the
corresponding ADM mass, $ M= [(M_0 N_0)^{2/3} + (M_6 N_6)^{2/3}]^{3/2} $ of the extremal Kaluza-Klein black holes.
Thus, it may be plausible to describe possible energy excitations in the $D_0$-$D_6$ microscopic Kaluza-Klein
configurations.
More precisely, the four-dimensional mass is dominated by the mass of the Kaluza-Klein monopole
and, in this limit, the excitation energy above the Kaluza-Klein monopole is equal to the ADM mass
of the five-dimensional black hole.
Here, it may once again be seen from \cite{itz} that the electric-magnetic charges of the Kaluza-Klein
black holes may be determined by the five dimensional angular momenta $ \lbrace J_1, J_2 \rbrace $
and the four-dimensional angular momentum $J$ via the relations
\begin{eqnarray}
N_0(J_1, J_2, J)&:=&  4J^2 \frac{J_1+J_2}{(J_1-J_2)^2}  \nonumber \\
N_6(J_1, J_2, J)&:=& \frac{J_1-J_2}{2J}
\end{eqnarray}
This eases us to describe our state-space construction in term of the two five dimensional momenta
and a customary four dimensional angular momentum. Here, we reveal that the set $ \lbrace J_1, J_2, J \rbrace $
forms another coordinate chart for the state-space manifold of the rotating Kaluza-Klein black holes.
It has further been explained in \cite{itz} that, if the size of the black hole is much smaller than
the Kaluza-Klein compactification radius, then the finite-size effects become rather negligible, and
one recovers the neutral five-dimensional Myers-Perry black holes. Our investigations thus attribute
a unified framework, for analyzing the various possible representations and specifications of the
statistical fluctuation, underlying in the vacuum $M$-theory black hole configurations.

Furthermore, there might exist various possible microscopic descriptions of the state-space correlations,
what we have been invoking may efficiently be exploited from the view-points of counting the microstates
for supersymmetric four-dimensional black holes with a different choice of large charges \cite{Maldacena:1996gb}.
This observation is based on the related U-dualities, ensuring that the $D_3$-branes contain
precisely the right number of states to reproduce the leading order entropy of vacuum $M$-theory black holes.
This goes with the understanding that the macroscopic entropy of these black holes remains independent
of the moduli fields arising from the compactification. Here, it is worth to note that the corresponding
statistical pair correlations seem to be solely accomplished from those states that are associated with the
intersection point of the branes, see for further details \cite{Maldacena:1996gb}.

Moreover, Taylor has shown \cite{wati} that there exists certain interesting $D$-brane
configurations, for which the bound state of four $D_0$-branes and four $D_6$-branes may
be described by the gauge theory configuration living on the worldvolume of the $D_6$-brane.
Therefore, one may urge that the state-space geometric revelation of the possible configurations
lies on the fact that the T-duality along the three cycles of the torus makes this configuration
equivalent to that of the four $D_3$-branes. As invoked earlier, we may notice that the other string
duality relations may also be realized as the general coordinate transformations on the concerned
state-space manifold.

The other possibility materializing, in the present context of our state-space geometry, goes
with an understanding of the Kaluza-Klein black holes directly in the $M$-theory, possibly in
terms of gravitons and perhaps branes, or in terms of $D_6$-branes with a certain flux corresponding
to $D_0$-branes, rather than comporting directly with the $D_3$-brane configurations.
In particular, the quest of state-space manifold under consideration corresponds to an interesting
intrinsic geometric problem, against the counting of the number of certain instantons in the six-dimensional
Yang-Mills theory (see \cite{GopakumarGreen} for the D-instanton contributions to the graviton scattering
amplitudes and related $SU(N)$ Yang-Mills instanton corrections).
\subsection{Supersymmetric $AdS_5$ Black Holes}
In this subsection we analyze state-space manifold of supersymmetric $AdS_5$ black holes arising
in type-IIB string theory compactifications. In the framework of fractional-BPS supergravities,
we shall exploit the fact that the $\frac{1}{16}$-BPS solutions possess a characteristic property
that all the physical charges are not independent. We observe that the state-space geometry emerging
from the Gaussian fluctuations of the Bekenstein-Hawking entropy admits a remarkably simple expression,
in terms of the exclusive real charges and self dual angular momentum, which may intimate its microscopic
origin via the Cardy formula, or the universal Hardy-Ramanujan formula, determining the degeneracy of microstates.
We further point out that the physical properties of underlying state-space correlations achieve
notably illuminating forms, which simply emerge, due to fact that there are certain upper bounds
for the angular momenta. The concerned constraints may be furnished in terms of the electric charges
which describe the possible maximum rotation for the supersymmetric $AdS_5$ black hole configurations.

From the verge of supergravity considerations, it turns out that
the bosonic field contents of supergravity theories are
fundamentally the space-time metric tensor $g_{\mu\nu}$, the
graviphoton \cite{sherk1,sherk2, bf1, bf2} and the $n-1$ vector
fields, which may thence be composed together as $A^i$
($i=1,\cdots,n$), and the $n-1$ real scalars $\phi^a$
($a=1,\cdots,n-1$), see for details \cite{gu-sa}. It has
interestingly been visualized, in the case of the $U(1)^3 \subset
SO(6)$ truncation, that the $\mathcal N= 4$ gauged supergravity
may be embedded into type-IIB string theory \cite{cvetic}. In the
present case, one finds that the $AdS_5\times S^5$ configuration
thus specified has three vector fields, a tensor field $C_{123}$
and three constrained scalars $X^i$, which measure the squashing
of $S^5$ taking equal vacuum values, which in turn make the
internal $S^5$ to be a smooth round sphere. Furthermore, it is
possible to check that the insertions of these values admit an
intriguing physical appreciation for the supergravity black holes
in $AdS_5\times S^5$ solutions. Thus, one may easily divulge the
state-space manifold arising from the Bekenstein-Hawking entropy
of supersymmetric $AdS_5$ black holes.

As we have already mentioned the black holes of present interest carry two angular momenta in $AdS_5$,
{\it viz.}, $J_1$ and $J_2$, as well as they endure three $U(1)^3\subset SO(6)$ charges, say $Q_i$ ($i=1,2,3$).
However, there exists a unique feature of these supersymmetric $AdS_5$ black holes configuration, i.e. that
the associated physical charges are not all independent. Thus, these solutions may not always be regular,
see for example \cite{gu-re-1, cglp,kp}. However, similar notions hold for the general supersymmetric $AdS_n$
black holes in minimal gauged supergravities. Although there exists an ambiguity in the entropy expression,
it may however be defined in terms of the dependent physical charges. In order to compare it with
the microscopic descriptions, there indeed exist certain implicit prescriptions in the literature
\cite{kmmr,brs}. In turn, Kim et. al.
observed, in certain specific regimes of the parameters, that the underlying entropy for the symmetric
supergravity configurations admits a remarkably simple expression, in terms of the  reliant physical charges
\begin{eqnarray}
 S(Q_1,Q_2,Q_3, J)= 2 \pi \sqrt{Q_1 Q_2+ Q_2 Q_3+ Q_1 Q_3- N^2 J},
\end{eqnarray}
where $J=\frac{J_1+J_2}{2}$ is the self-dual part of the angular momenta, see for details \cite{hep-th/0607085}.
Thus, the intrinsic Riemannian geometry, as the equilibrium state-space configuration of the three charged
extremal rotating black hole resulting from the concerned fractional BPS-corrected entropy, may immediately
be computed, as earlier, from the negative Hessian matrix of the entropy. We find that the components of the
state-space metric tensor may easily be obtained, with respect to the underlying electric charges and the
self-dual angular momentum, as\\
\begin{eqnarray}
g_{Q_1Q_1}&=& \frac{\pi}{2}(Q_2+ Q_3)^2 (Q_1 Q_2+ Q_2 Q_3+ Q_1 Q_3- N^2 J)^{-3/2} \nonumber \\
g_{Q_1Q_2}&=& -\frac{\pi}{2}(Q_1 Q_2+ Q_2 Q_3+ Q_1 Q_3-Q_3^2- 2N^2 J) \nonumber \\
& & (Q_1 Q_2+ Q_2 Q_3+ Q_1 Q_3- N^2 J)^{-3/2} \nonumber \\
g_{Q_1Q_3}&=& -\frac{\pi}{2}(Q_1 Q_2+ Q_2 Q_3+ Q_1 Q_3- Q_2^2- 2N^2 J) \nonumber \\
& & (Q_1 Q_2+ Q_2 Q_3+ Q_1 Q_3- N^2 J)^{-3/2} \nonumber \\
g_{Q_1J}&=& -\frac{\pi}{2}N^2 (Q_2+ Q_3) (Q_1 Q_2+ Q_2 Q_3+ Q_1 Q_3- N^2 J)^{-3/2} \nonumber \\
g_{Q_2Q_2}&=& \frac{\pi}{2}(Q_1+ Q_3)^2 (Q_1 Q_2+ Q_2 Q_3+ Q_1 Q_3- N^2 J)^{-3/2} \nonumber \\
g_{Q_2Q_3}&=& -\frac{\pi}{2}(Q_1 Q_2+ Q_2 Q_3+ Q_1 Q_3- Q_1^2- 2N^2 J) \nonumber \\
& & (Q_1 Q_2+ Q_2 Q_3+ Q_1 Q_3- N^2 J)^{-3/2} \nonumber \\
g_{Q_2J}&=& -\frac{\pi}{2}N^2 (Q_1+ Q_3) (Q_1 Q_2+ Q_2 Q_3+ Q_1 Q_3- N^2 J)^{-3/2} \nonumber \\
g_{Q_3Q_3}&=& \frac{\pi}{2}(Q_1+ Q_2)^2 (Q_1 Q_2+ Q_2 Q_3+ Q_1 Q_3- N^2 J)^{-3/2} \nonumber \\
g_{Q_3J}&=& -\frac{\pi}{2}N^2 (Q_1+ Q_2) (Q_1 Q_2+ Q_2 Q_3+ Q_1 Q_3- N^2 J)^{-3/2} \nonumber \\
g_{JJ}&=& \frac{\pi}{2}N^4 (Q_1 Q_2+ Q_2 Q_3+ Q_1 Q_3- N^2 J)^{-3/2}
\end{eqnarray}
This framework thus affirms that there exists a lucid geometric
enumeration, which describes the possible statistical pair
correlations, determined in terms of the charges and angular
momentum of the supersymmetric $AdS_5$ black holes. Hitherto, it
is apparent that the principal components of the statistical pair
correlations are positive definite, for the whole allowed range of
concerned invariant parameters of the $AdS_5$ black holes. In
particular, it may easily be observed that the following
state-space metric constraints are satisfied
\begin{eqnarray}
g_{Q_1Q_1}&>& 0 \ \forall \ (Q_1,Q_2,Q_3, J) \mid Q_2 \neq -Q_3 \nonumber \\
g_{Q_2Q_2}&>& 0 \ \forall \ (Q_1,Q_2,Q_3, J) \mid Q_3 \neq -Q_1 \nonumber \\
g_{Q_3Q_3}&>& 0 \ \forall \ (Q_1,Q_2,Q_3, J) \mid Q_1 \neq -Q_2 \nonumber \\
g_{JJ}&>& 0 \ \forall \ admisible \ (Q_1,Q_2,Q_3, J) \mid N \neq 0
\end{eqnarray}
Physically, we may notice that the principal components of the
state-space metric tensor $\lbrace g_{Q_i Q_i}, g_{JJ} \ \vert \
i=1,2,3 \rbrace$ signify certain heat capacities (or the
associated compressibilities), whose positivity exhibits that the
underlying system is in local equilibrium configurations. Our
analysis further complies that the positivity of $g_{JJ}$ obliges
that the associated holographic conformal field theory living on
the boundary must prevail a non vanishing value of the central
charge.

Interestingly, it is worth to note that our geometric expressions arising from the black hole entropy
show that either one of the three electric charges may safely be turned off, say $Q_i= 0$, while having
a well-defined state-space geometry.
However, it is unfeasible to have an intrinsic $AdS_5$ black hole state-space configuration with only
a single electric charge, say $Q_i= Q_j= 0$, since the objects inside the square-root of the statistical
correlations cannot be any more well-defined real positive definite quantities.

Furthermore, the ratio of the principal components of statistical
pair correlations varies as an inverse square of the addition of
two other charges; while that of the other off diagonal
correlations only vary inversely. In particular, it is not
difficult to inspect, for non-identical $i,j,k \in \lbrace 1,2,3
\rbrace $, that the statistical pair correlations are consisting of
the following scaling properties:
\begin{eqnarray}
\frac{g_{Q_iQ_i}}{g_{Q_jQ_j}}&=& (\frac{Q_j+Q_k}{Q_i+Q_k})^2 \nonumber \\
\frac{g_{Q_iQ_i}}{g_{JJ}}&=& (\frac{Q_j+Q_k}{N^2})^2 \nonumber \\
\frac{g_{Q_iJ}}{g_{Q_jJ}}&=& \frac{Q_j+Q_k}{Q_i+Q_k} \nonumber \\
\frac{g_{Q_iQ_i}}{g_{Q_iJ}}&=& -\frac{Q_j+Q_k}{N^2} \nonumber \\
\frac{g_{Q_iQ_j}}{g_{Q_kJ}}&=& \frac{(Q_1 Q_2+ Q_2 Q_3+ Q_1 Q_3- 2N^2 J)- Q_k^2}{N^2(Q_i+Q_j)} \nonumber \\
\frac{g_{Q_iQ_j}}{g_{Q_{i,j}Q_k}}&=& \frac{(Q_1 Q_2+ Q_2 Q_3+ Q_1 Q_3- 2N^2 J)- Q_k^2}
{(Q_1 Q_2+ Q_2 Q_3+ Q_1 Q_3- 2N^2 J)- Q_{j,i}^2}
\end{eqnarray}
The concerned scaling investigations may further be strengthened by the fact that the self-dual
angular momentum $J$ of the $AdS_5$ black holes is constrained from above by the electric charges
$Q_i$ by the bound: $ J_1+J_2\leq \frac{2}{N^2}(Q_1Q_2+Q_2Q_3+Q_3Q_1) $, see for instance \cite{gu-re-2}.
Here, our analysis thus demonstrates that the positivity of state-space correlations among the non-identical
$U(1)$ charges $\lbrace Q_k \vert k= 1,2,3 \rbrace$ stipulates a modified lower bound for the self dual
angular momentum
\begin{eqnarray}
J\leq \frac{2}{N^2}(Q_1Q_2+Q_2Q_3+Q_1Q_3- Q_k^2)
\end{eqnarray}
This is because the brane-brane pair correlations involve the other remaining $U(1)$ charges
of the underlying configuration. Thus one establishes, for equal value of charges,
that the bound arising from the reality of the entropy is only three quarters of the stability
criteria of the pair correlations between the non-identical electric charges.

Apart from the positivity of principal components of the state-space
metric tensor, one demands that all the associated principal
minors should be positive definite, in order to accomplish the
local stability. It is nevertheless not difficult to compute the
principal minors of the associated Hessian matrix of the entropy
concerned with the spinning $AdS_5$ black holes. In fact, after
some simple manipulations one encounters that the local stability
conditions on the two dimensional surfaces and three dimensional
hypersurfaces of the state-space manifold may respectively be
measured by the following relations:
\begin{eqnarray}
 p_2 &=& \pi^2\bigg(N^2 J+ Q_3^2\bigg) \bigg(Q_1 Q_2+ Q_2 Q_3+ Q_1 Q_3- N^2 J\bigg)^{-2} \nonumber \\
 p_3 &=& 2 \pi^3 N^2 J \bigg(Q_1 Q_2+ Q_2 Q_3+ Q_1 Q_3- N^2 J\bigg)^{-5/2}
\end{eqnarray}
For all physically allowed values of invariant charges of the $AdS_5$ black holes,
we thus stipulate that the minor constraint $p_2>0$ obliges that the domain of the ascribed
angular momentum must respectively be greater than $J=-(Q_3/N)^2$, while the constraint
$p_3>0$ imposes that the angular momentum must be a positive definite real number.
In particular, we may easily inspect the nature of the state-space geometry for the spinning
$AdS_5$ black holes, i.e. that the planar and hyper-planar stabilities of the system together demand
just the existence of an arbitrary positive value of the self-dual angular momentum.

In addition, it is likewise evident that the local stability of the full phase-space configuration
may be determined by computing the determinant of the concerned state-space metric tensor.
Here, we may easily provide a compact formula, for the determinant of the metric tensor.
In particular, our intrinsic geometric analysis assigns the following expression,
for the determinant of the metric tensor, as the function of various possible values
of electric charges and the angular momentum:
\begin{eqnarray}
 g(Q_1,Q_2,Q_3, J)= -\pi N^4 \bigg(Q_1 Q_2+ Q_2 Q_3+ Q_1 Q_3- N^2 J\bigg)^{-3}
\end{eqnarray}
The determinant of the underlying metric tensor thus remains non-zero for the non-vanishing
central charge of the corresponding dual holographic $\mathcal N= 4 $ Yang-Mills theory.
This indicates that the state-space geometry is non degenerate at extremality, as a consequence of the higher
derivative central charge contributions. Note however that the determinant of the metric tensor does
not admit a positive definite form in the large charge limit, and thus there is no positive definite
volume form on the intertwined state-space $(M_4,g)$. This is indeed justifiable from the fact that
the equilibrium entropy approaches its maximum value, while the same may not remain valid in the
possible hyper-planes of the concerned state-space configurations. It may further be noted in this
description that the spinning $AdS_5$ black holes do not fabricate an intrinsically stable
statistical configuration. Thus, it is possible that the underlying microscopic states
may smoothly shift into a more stable ensemble of brane configurations.

Although the appraised expression of the entropy may not be unique, due to an existence of the relation
between the physical charges $Q_i$ and angular momentum $J$ for the allowed black hole solutions.
However, in the small black hole regime, where the solutions allow $Q_i \ll N^2$, the underlying entropy
can be transcribed into a perturbative form, for which we emphasize that there are indeed interesting
perspectives for the state-space geometry of the $AdS_5$ configurations.
Such a perturbative manifestation of the supersymetric black hole entropy has in particular been
advocated to meet the microscopic aspects of $AdS_5$ black holes, see for details \cite{kmmr}.
Eventually, our analysis discloses that the determinant of the metric tensor takes a small positive definite value.
Thus, one may realize an intrinsically stable statistical configuration, with the existence of the globally defined
positive definite volume form on the state-space manifold $(M_4,g)$.
Therefore, the state-space configuration, in the limit of vanishing angular momentum or infinite central
charge of the holographic CFT, inclines towards a degenerate intrinsic Riemannian manifold.

In order to conclusively analyze the nature of the interaction and the other concerned properties of the statistical
configurations, one needs to determine certain global invariants on the parametric state-space manifold $(M_4,g)$.
One may in fact ascertain that such a simplest invariant is the state-space scalar curvature, which may now be
computed in a straightforward fashion by applying our previously advertised intrinsic geometric technology.
The explicit expression for the scalar curvature is quite simple, and we notice that it may easily be expressed as
\begin{eqnarray}
 R(Q_1,Q_2,Q_3, J)= -\frac{9}{4\pi} \bigg(Q_1 Q_2+ Q_2 Q_3+ Q_1 Q_3- N^2 J\bigg)^{-1/2}
\end{eqnarray}
Here, it is worth to mention that the scalar curvature is a regular function of the electric
charges and angular momentum. In the large charge limit, in which the asymptotic expansion
of the concerned BPS-entropy calculation is valid, it inclines to a vanishingly small value.
The negative sign of the state-space scalar curvature discloses that the underlying configuration
is effectively an attractive system. Thus, the spinning $AdS_5$ black holes turn out to be less stable
than an arbitrary positively curved state-space configuration.
In addition, this is due to the fact that the phase-space configuration of supersymmetric solutions is
drastically constrained and is subject to powerful non-renormalization theorems and state-counting techniques.
Moreover, it is important to remember that the scalar curvature thus determined never vanishes, for any
physically affirmed value of the invariant charges and angular momentum, and is an everywhere regular
function on the state-space manifold, unless either one of the electric charge approaches infinity.
Thus, for any non-zero value of the charges and spinning angular momentum, the $AdS_5$ black hole always
corresponds to a weakly interacting statistical configuration. Furthermore, it is not difficult to check
that both the constant entropy and constant state-space scalar curvature curves take the same form
\begin{eqnarray}
Q_1 Q_2+ Q_2 Q_3+ Q_1 Q_3= N^2J+ b_i,
\end{eqnarray}
Such definite bound on a collection of states in itself is not surprising for the spinning black holes, as the
microscopic degeneracy depends on the ensemble of microstates, while the corresponding statistical correlations
depend on the state-space configuration. The concerned bound of the degeneracy of the microstates may in turn be
obtained from the central charge of an underlying four dimensional dual holographic conformal field theory.
Consequently, it is not difficult to manipulate that the constants $b_i$ may respectively be defined as
\begin{eqnarray}
b_1&:=& \frac{k_1^2}{4\pi^2}  \ for \ constant \ entropy \nonumber \\
b_2&:=& \frac{81}{16\pi^2k_2^2} \ for \ constant \ scalar \ curvature,
\end{eqnarray}
where $(k_1, k_2)$ are the chosen values of the entropy and scalar curvature for the state-space $(M_4,g)$.
In order to understand the origin of the statistical correlations underlying these objects,
it will also be interesting to consider an exclusive microscopic sector of these configurations
and then try to investigate the state-space quantities for the supersymmetric $AdS$-black holes.
The purpose of this investigation may thus be to explore the macroscopic and microscopic aspects
of the $\frac{1}{16}$-BPS objects, hoping that our geometric results would provide important
clues for future works giving an account on both sides: for instance, the state-space correlations
defining an intrinsic Riemannian geometry, arising from the Bekenstein-Hawking entropy formula,
and that in the counting degeneracy of the $AdS$-black hole microstates. For the other possible attempts
to analyze thermodynamics and the BPS aspects of the supersymmetric black holes, see \cite{kmmr,brs, silva}.

Decisively, our state-space geometric expressions which are exact in all regime of charges, exhibit certain
interesting comparisons and definite agreements with the expressions advocated in previous achievements.
The angular momentum factor originating from the central charge of the $\mathcal N=4$ Yang-Mills theory,
which is holographically dual to the string theory in $AdS_5\times S^5$, governs the nature of the correlation
length of the underlying statistical systems. In particular, it is worth to indicate that the prefactor,
$N^2=4c$, where $c$ signifies the central charge of the $\mathcal N=4$ Yang-Mills theory, determines
the typical characteristic of the fundamental microscopic theory. The symmetric supergravity condition
requires that $c$ should satisfy $c= \pi l^3/8G$. Here, the internal nature of the $AdS_5$ length parameter $l$
connotes the underlying microscopic objects, whether they are giant gravitons or wrapped branes \cite{gu-re-1}.
Our state-space expressions thus obtained are indeed remarkably simple and allude that there should exist
appropriate microscopic/macroscopic explanations for the statistical correlations for these configurations.
The observations may thus be a motivation factor to further investigate how the Cardy or Hardy-Ramanujan formula sets
down certain microscopic explanations for ascertaining similar examinations to the present, as well as to the other
possible supergravity solutions.

The points in question may nevertheless be addressed for the rotating objects in Minkowski space, as well.
In order to appraise this point more clearly, it turns out very interesting to consider well studied
$D_1$-$D_5$ configurations. In particular, we wish to visualize how the state-space correlators
behave when such an underlying BPS solution saturates the angular momentum bound.
For instance, the entropy of rotating $D_1$-$D_5$ on $T^5$ (or $K_3$) may be given by
$S= \gamma \pi\sqrt{Q_1Q_5-J}$, where $ \gamma $ depends on the chosen compactification.
Moreover, it turns out that the zero entropy limit of these configurations may be consigned
by the circular supertubes \cite{MT,pa-mar,bho,bhko}, or U-duals of them \cite{lu-ma,dghw,ch-oh,mnt}.
In the lieu of $AdS_5$ black holes state-space geometry, it may thus be intriguing to examine
the state-space correlators for a 2-charge black holes, whose degeneracy formula defines an
appropriate simple expression for the microscopic entropy, in large charge limit. 
Therefore, we shall devote the next subsection to a detailed analysis of these configurations.
In particular, we would like to allocate the state-space geometric meaning of the $N^2$ factor,
which makes the problem more challenging, from the view points of $D_1$-$D_5$ CFT.

\subsection{$D_1$-$D_5$ Configurations}
In this subsection, we shall analyze the state-space geometry of the most exhaustively contemplated
two charged black brane configurations, which naturally emerge from the equilibrium microstates of
various interesting bubbling supergravity solutions, see for details \cite{MT,pa-mar,bho,bhko}.
It turns our that one may easily ascribe the state-space definitions to the central charge
contributions associated with the rotating black holes in Minkowski space, and in turn
we may focus our attention on the possible U-dual configurations as well. The concerned
details of the intriguing solutions have been described in \cite{lu-ma,dghw,ch-oh,mnt}.
Here, the purpose is to exploit the state-space meanings of the $N^2$ factor arising from an elementary
four dimensional holographic conformal field theory living on the boundary \cite{hep-th/0607085}.
As we have encountered the state-space geometry of the $AdS_5$ black holes in the previous subsection,
this subsection analyzes the state-space fluctuations for the restricted $2$-charge $D_1$-$D_5$ rotating
solutions in the extremal limit, and there by examines the inclusion of a holographic central charge contribution.
It turns out that the involved entropy may be defined via an appropriate degeneracy formula to be
\begin{equation}
S(Q_1,Q_5,J):= 2\pi\sqrt{Q_1Q_5-N^2 J}
\end{equation}
The state-space geometry describing the correlations between the equilibrium microstates of the
two charged rotating extremal $D_1$-$D_5$ black holes, resulting from the degeneracy of the microstates
may easily be computed, as earlier, from the Hessian matrix of the entropy, with respect to the $D_1$, $D_5$
charges, {\it viz,}, $Q_1$,$Q_5$ and the angular momentum $J$. At this juncture, we obtain that the
components of the underlying state-space covariant metric tensor are given by
\begin{eqnarray}
g_{Q_1Q_1}&=& \frac{\pi}{2}Q_5^2 (Q_1Q_5-N^2 J)^{-3/2} \nonumber \\
g_{Q_1Q_5}&=& -\frac{\pi}{2}(Q_1 Q_5- 2N^2 J)(Q_1Q_5-N^2 J)^{-3/2} \nonumber \\
g_{Q_1J}&=& -\frac{\pi}{2}N^2 Q_5 (Q_1Q_5-N^2 J)^{-3/2} \nonumber \\
g_{Q_5Q_5}&=& \frac{\pi}{2}Q_1^2 (Q_1Q_5-N^2 J)^{-3/2} \nonumber \\
g_{Q_5J}&=& -\frac{\pi}{2}N^2 Q_1 (Q_1Q_5-N^2 J)^{-3/2} \nonumber \\
g_{JJ}&=& \frac{\pi}{2}N^4 (Q_1Q_5-N^2 J)^{-3/2}
\end{eqnarray}
We may thus stress out that the state-space geometry, materializing
from the Bekenstein-Hawking entropy of the $D_1$-$D_5$
configurations, admits remarkably simple expressions in terms of
physical charges and angular momentum. It may thence be expected
that the plausible microscopic preliminaries would be suggested
via the Cardy formula, or the associated general Hardy-Ramanujan
formula. As enumerated in the earlier sections, we may again
appreciate, for all non-zero values of the brane charges $Q_1,Q_5,J
$ and angular momentum $J$, that the principal components of
present state-space metric tensor satisfy
\begin{eqnarray}
g_{Q_1Q_1} &>& 0 \nonumber \\
g_{Q_5Q_5} &>& 0 \nonumber \\
g_{JJ} &>& 0
\end{eqnarray}
Physically, the principal components of the state-space metric tensor
signify heat capacities, or the relevant compressibilities, whose
positivity connote that the underlying system is in locally stable
equilibrium configurations of the $D_1$ and $D_5$-brane
microstates. Furthermore, it may easily be observed that the ratio
of diagonal components varies as the inverse square of the invariant
parameters which vary under the Gaussian fluctuations, whereas the
ratios involving off diagonal components vary only inversely in
the chosen parameters. This suggests that the diagonal components
weaken faster and relatively more quickly come into an equilibrium,
than the off diagonal components, which remain comparable for the
longer domain of invariant parameters defining the $D_1$-$D_5$
configurations. In particular, we may easily substantiate, for the
distinct $i, j \in \lbrace 1,5 \rbrace $, that the relative pair
correlation functions satisfy the following simple relations:
\begin{eqnarray}
\frac{g_{Q_iQ_i}}{g_{Q_jQ_j}}&=& (\frac{Q_j}{Q_i})^2 \nonumber \\
\frac{g_{Q_iQ_i}}{g_{JJ}}&=& (\frac{Q_j}{N^2})^2 \nonumber \\
\frac{g_{Q_iJ}}{g_{Q_jJ}}&=& \frac{Q_j}{Q_i} \nonumber \\
\frac{g_{Q_iQ_i}}{g_{Q_iJ}}&=& -\frac{Q_j}{N^2}
\end{eqnarray}
Moreover, it is very instructive to note that the behavior of brane-brane statistical pair
correlation, defined as $g_{Q_1Q_5}$, is asymmetric, in contrast to the other existing correlations.
In fact, one may understand it by arguing that the brane-brane interaction imparts more energy
than either the interaction corresponding to the self-interactions or that of the rotations.
In particular, our analysis proclaims that the relative pair correlation between the $D_1$-$D_5$
branes, with respect to rotation-rotation correlation turns out to be
\begin{eqnarray}
\frac{g_{Q_1Q_5}}{g_{JJ}}&=& -\frac{1}{N^4}\bigg(Q_1 Q_5- 2N^2 J \bigg)
\end{eqnarray}
This implies that the relative brane-brane statistical pair correlation function vanishes exactly
at the half value of the angular momentum, compared to that of the vanishing entropy condition.
Furthermore, it may easily be viewed, from the non-vanishing central charge, that the brane-brane
pair correlation is stable, only if the underlying angular momentum satisfies a lower bound
$ \vert J\vert< \frac{Q_1 Q_5}{2N^2}$. In order to entail definite global properties of the
concerned configurations, one may now be required to determine stabilities along each
direction and along each plane, as well as on the entire intrinsic state-space manifold.

It is worth to note that the appraisal of exhaustive state-space
stability demands that all the associated principal minors must be
positive definite, as the positivity of principal components of
metric tensor is defined only in the local neighborhood of a certain
chosen co-ordinate chart of the state-space manifold $(M_3,g)$. It
is nevertheless not difficult to compute the principal minor of
the associated Hessian matrix of the entropy, concerned with the
rotating two charged black holes. It is rather easy to explain the
physical picture of the solution set, and in fact, after some
simplifications, one discovers that the planar stability criteria
on the two dimensional surfaces of the state-space manifold may
simply be rendered to be
\begin{eqnarray}
p_2
= \pi^2 N^2 J \bigg(Q_1Q_5-N^2 J\bigg)^{-2}
\end{eqnarray}
In the view-points of the simplest $D_1$-$D_5$ solutions, it tuns out that the local
stability on the entire equilibrium phase-space configurations may clearly be determined
by computing the determinant of the underlying state-space metric tensor.
As in the previous examples, it is easy to observe that the state-space metric tensor
is a non-degenerate, and everywhere regular, function of the brane charges and angular momentum.
In particular, we may conveniently ascertain that the determinant of the metric tensor obtains a simple form
\begin{eqnarray}
g(Q_1,Q_5,J)= -\frac{1}{2}\pi^3 N^4 \bigg(Q_1Q_5-N^2 J\bigg)^{-5/2}
\end{eqnarray}
Here, we deduce that the determinant of the metric tensor does not take a well-defined
positive definite form, and thus there is no positive definite globally well-defined
volume form on the state-space manifold $(M_3,g)$, for the concerned systems.
Moreover, the non-zero state-space determinant $ g(Q_1,Q_5,J) $, thus constructed,
indicates that the extremal $D_1$-$D_5$-$J$ systems may decay into certain degenerate
vacuum state configurations, procuring the same corresponding microscopic entropy.
In turn, we may thus notify that the $D_1$-$D_5$-$J $ black holes, when considered as a bound state of
the $D_1$-$D_5$-brane microstates, do not correspond to an intrinsically stable statistical configuration.
It is worth to notice that the conclusions thus drawn remain independent of the microscopic type-II
string description, or the heterotic string description.

Furthermore, we may exhibit that the nature of the statistical interactions and the other global
properties of the $D_1$-$D_5$-$J$ configurations are indeed not really perplexing to anatomize.
In this regard, one only needs to compute certain global invariants of the state-space manifold
$(M_3,g)$, which can thus be determined in terms of the parameters of underlying brane configurations.
Here, we may work in the large charge limit, in which the asymptotic expansion of the entropy of the
two charge rotating $D_1$-$D_5$ system is valid. In particular, we notice that the state-space scalar
curvature may be appraised by the expression
\begin{eqnarray}
R(Q_1,Q_5,J)= -\frac{5}{4\pi} \bigg(Q_1Q_5-N^2 J\bigg)^{-1/2}
\end{eqnarray}
It is worth to mention that the scalar curvature thus determined never vanishes,
for all physically acceptable value of the brane charges and angular momentum, and
thus it is an everywhere non-vanishing function on the state-space manifold $(M_3,g)$.
The negative sign of the state-space scalar curvature signifies that the underlying statistical
system is effectively an attractive configuration, and thus it is less stable than the arbitrary
positively curved equilibrium intrinsic configurations. We may also notice, in the present case, that
the scalar curvature $ R(Q_1,Q_5,J) $ is a regular function of the $ \lbrace Q_1,Q_5,J \rbrace $
coordinatizing the underlying state-space configuration.
Thus, the underlying state-space configuration remains a non-degenerate, everywhere finitely curved,
intrinsic Riemannian manifold, which is free from quantum instabilities and vacuum phase transitions.
This makes sure, for all non-zero brane degeneracies, that the $D_1$-$D_5$ systems always correspond to
certain weakly interacting statistical configurations. It follows again that both the intriguing curves
of constant entropy $k_1$ and constant state-space scalar curvature $k_2$ take the same form
\begin{eqnarray}
Q_1 Q_5= N^2J+ c_i
\end{eqnarray}
This configuration thus defines an existence of veritable bounds on the collection of degenerate
correlated $D_1$-$D_5$ microstates, which is indeed not surprising for the two charged rotating black holes.
The concerned bounds may certainly be anticipated from the degeneracy of CFT microstates and from the derived
statistical correlations, which both depend on the corresponding phase-space configuration of the $D_1$-$D_5$
solutions. Such bounds, associated with the degeneracy of the microstates, may as well be obtained from
the central charge of an underlying four dimensional holographic dual conformal field theory. Furthermore,
it is easy to notice in the present case that the constants $c_i$ may respectively be defined as
\begin{eqnarray}
c_1&:=& \frac{k_1^2}{4\pi^2}  \ for \ constant \ entropy \nonumber \\
c_2&:=& \frac{25}{16\pi^2k_2^2} \ for \ constant \ scalar \ curvature
\end{eqnarray}
Moreover, we find an intriguing relation of the conclusions obtained from our state-space geometry
with the Mathur's fuzzball proposal of constructing microstates of the two charged rotating black holes.
As exercised in the connection with Mathur's fuzzball proposal, the typical microstates of an extremal black hole
cannot acquire a space-time singularity \cite{fuzzball}. Similarly, we may deduce from the present consideration
that even an intrinsic Riemannian manifold of the equilibrium microstates continues to remain non-singular.
In other words, the regularity of the state-space geometric curvature invariant exposes that the intrinsic
Riemannian geometry constructed out of the parameters of rotating two charge $D_1$-$D_5$ configurations
as extrema of the entropy, also remains regular. Consequently, the absence of divergences in the
state-space scalar curvature implies that the underlying statistical systems of the two charge rotating
black holes are thermodynamically stable under the Gaussian fluctuations, and thus there are no vacuum
phase transitions in the counterpoised microscopic configurations.

Very similar conclusions are also trivially accessible for the case of a more complicated
$U(1)^n$ supergravity theory. The general features of these solutions are rather involved, however
we find worth mentioning that they inspire some considerably interesting state-space characteristics.
In the next subsection, we shall generalize the above outlined state-space configuration of the two
charged $D_1$-$D_5$ solutions to the case of three charged $D_1$-$D_5$-$P$ configurations. In
particular, we may stimulate the present investigation for the spherical horizon topology BMPV black holes.
In order to adequately analyze these solutions, we shall put our emphasis on the standard $D_1$-$D_5$-$P$
frames and their dual $M$-theory backgrounds, to show that the state-space consideration pursued in this
subsection germinates an analogous bound on the admissible angular momentum, and that similar definite
conclusions arise for the three charge rotating configurations.
\subsection{BMPV Black Holes}
The present subsection examines the state-space configuration of the spherical topology black holes,
and experiments to extend our intrinsic geometric assessments, for the known supersymmetric black hole
solutions, in five-dimensional supergravity with a constant horizon topology, see for details \cite{bmpv}.
For the purpose of critical ratifications, we shall focus our attention on the BMPV black holes,
which may be obtained as the limiting case of supersymmetric black rings, when one takes zero-limit
of the ring radius, with the space-time coordinates and all other parameters of the solution held fixed.
The solution under interest may thus be ascertained by four parameters, {\it viz.}, $Q_i$ and $J$, where
the former retain their interpretation as conserved $M_2$-brane charges, while the latter determines
the angular momentum: $J:=J_\phi= J_\psi$ of the concerned BMPV black hole configurations.
It is known that the BMPV solutions are much more symmetrical than the supersymmetric black rings,
as they have non-trivial isometry group: ${\bf R} \times U(1) \times SU(2)$, see for details \cite{gmt}.
Here, the upshot is that we intend to discover the possible definitive conclusions emerging from
the event horizon, which has a topology of $S^3$, and sits at the origin of the space-time solution.

In order to explicate the state-space geometry of spherical BMPV black holes in the $D_1$-$D_5$-$P$ frame,
we may consider a five dimensional string in the maximally twisted sector with a linear momentum,
which is among one of the three conserved charges carried by both the bosonic and fermionic excitations
along the effective string. In particular, the possible fermionic excitations may also carry a polarization
in the transverse directions, which emerges due to the non-vanishing R-charges appearing in the associated UV-CFT.
Thus, for a given number of oscillators polarized in the same direction with a self-dual angular momentum,
the projection of oscillators onto the ascribed polarization, which restricts the phase-space configurations,
acquiesces our purpose for analyzing the nature of underlying statistical interactions.

Here, one may infer, via the Cardy formula, that the entropy of the supersymmetric BMPV rotating black holes
may easily be transcribed as a function of brane numbers, the self-dual angular momentum, and it does not
entangle with the Kaluza-Klein excitations or the $M_5$-brane charges, see for instance \cite{bmpv}.
The BMPV balck holes thus arise as an intriguing special case of the generic $M$-theory solutions,
when the dipole charges become redundant. In particular, one discovers that the horizon of these
solutions topologically reduces to the three sphere $S^3$. We thus observe that the analysis of
state-space geometry would present a detailed account of the statistical correlations and other
associated physical properties possessed by the spherical horizon BMPV black hole configurations.

For the purpose of our essential analysis, let us consider three independent asymptotic brane charges
$ \lbrace Q_1, Q_2, Q_3 \rbrace $ and an angular momentum $J$ describing the rotating BMPV configurations.
As just mentioned, the entropy of the spherical topology BMPV black holes, arising from the area of the
event horizon, may be disclosed in terms of the brane charges and an angular momentum \cite{bmpv}.
In particular, one finds that the entropy of the BMPV black holes divulges a simple expression
\begin{eqnarray}
 S(Q_1, Q_2, Q_3, J)= 2 \pi \sqrt{ \pi^2 Q_1 Q_2 Q_3- 16 G^2 J^2}
\end{eqnarray}
As proclaimed in the previous subsections, we notice, in this case also, that the state-space
geometry describing the nature of equilibrium brane microstates may be constructed out of the
three charges and angular momentum of the rotating spherical topology BMPV black holes.
As invoked earlier, the covariant metric tensor may immediately be computed from the negative Hessian
matrix of the concerned entropy, resulting from the underlying horizon area of the configration.
Thus, the asymptotic invariant charges and angular momentum: $ \lbrace Q_1, Q_2, Q_3, J \rbrace $
form the coordinate charts for the state-space manifold of our interest. Thus, with respect to these
parameters, we may describe the typical intrinsic geometric features of the BMPV black holes.
We thence notice that the components of the covariant metric tensor may easily be presented to be
\begin{eqnarray}
g_{Q_1 Q_1}&=& \frac{1}{2}\pi^5 Q_2^2 Q_3^2 (\pi^2 Q_1 Q_2 Q_3- 16 G^2 J^2)^{-3/2} \nonumber \\
g_{Q_1 Q_2}&=& -\frac{1}{2}\pi^3 Q_3 (\pi^2 Q_1 Q_2 Q_3- 32 G^2 J^2) (\pi^2 Q_1 Q_2 Q_3- 16 G^2 J^2)^{-3/2} \nonumber \\
g_{Q_1 Q_3}&=& -\frac{1}{2}\pi^3 Q_2 (\pi^2 Q_1 Q_2 Q_3- 32 G^2 J^2) (\pi^2 Q_1 Q_2 Q_3- 16 G^2 J^2)^{-3/2} \nonumber \\
g_{Q_1 J}&=& -16 \pi^3 G^2 J Q_2 Q_3 (\pi^2 Q_1 Q_2 Q_3- 16 G^2 J^2)^{-3/2} \nonumber \\
g_{Q_2 Q_2}&=& \frac{1}{2}\pi^5 Q_1^2 Q_3^2 (\pi^2 Q_1 Q_2 Q_3- 16 G^2 J^2)^{-3/2} \nonumber \\
g_{Q_2 Q_3}&=& -\frac{1}{2}\pi^3 Q_1 (\pi^2 Q_1 Q_2 Q_3- 32 G^2 J^2) (\pi^2 Q_1 Q_2 Q_3- 16 G^2 J^2)^{-3/2} \nonumber \\
g_{Q_2 J}&=& -16 \pi^3 G^2 J Q_1 Q_3 (\pi^2 Q_1 Q_2 Q_3- 16 G^2 J^2)^{-3/2} \nonumber \\
g_{Q_3 Q_3}&=& \frac{1}{2}\pi^5 Q_1^2 Q_2^2 (\pi^2 Q_1 Q_2 Q_3- 16 G^2 J^2)^{-3/2} \nonumber \\
g_{Q_3 J}&=& -16 \pi^3 G^2 J Q_1 Q_2 (\pi^2 Q_1 Q_2 Q_3- 16 G^2 J^2)^{-3/2} \nonumber \\
g_{J J}&=& 32\pi^3 G^2 Q_1 Q_2 Q_3 (\pi^2 Q_1 Q_2 Q_3- 16 G^2 J^2)^{-3/2}
\end{eqnarray}
Nevertheless, it is handy to observe that the statistical pair
correlations, thus ascertained, may in turn be accounted by a simple
microscopic description, expressed in terms of the brane charges
and angular momentum, connoting an ensemble of microstates for the
supersymmetric BMPV black holes. Furthermore, it is evident that
the principal components of the statistical pair correlations are
positive definite, for all allowed values of the asymptotic
invariant parameters of the black holes. In particular, we may
easily observe that the concerned state-space metric constraints
are
\begin{eqnarray}
g_{Q_1Q_1}&>& 0 \ \forall \ (Q_1,Q_2,Q_3, J)  \nonumber \\
g_{Q_2Q_2}&>& 0 \ \forall \ (Q_1,Q_2,Q_3, J)  \nonumber \\
g_{Q_3Q_3}&>& 0 \ \forall \ (Q_1,Q_2,Q_3, J)  \nonumber \\
g_{JJ}&>& 0 \ \forall \ admisible (Q_1,Q_2,Q_3, J) \mid Q_i \neq 0
\end{eqnarray}
Essentially, the principal components of the state-space metric tensor
$\lbrace g_{Q_i Q_i}, g_{JJ} \ \vert \ i=1,2,3 \rbrace$ signify a
set of definite heat capacities (or the related compressibilities),
whose positivity apprises that the BMPV black holes comply an
underlying locally equilibrium statistical configuration. It is
intriguing to note that the positivity of $g_{JJ}$ requires that
none of the brane charges of the associated $D_1$-$D_5$-$P$ CFT
(or $M_2$-brane charges in the dual description) should be zero.
This is clearly perceptible, because of the fact that the brane
configuration becomes unphysical.

It follows from the above expressions that the ratio of the
principal components of statistical pair correlations varies as an
inverse square of the asymptotic charges; in contrast, that of the off
diagonal correlations modulates only inversely. Interestingly, we
may easily visualize, for the distinct $i,j,k \in \lbrace 1,2,3
\rbrace $, that the statistical pair correlations, thus described,
are consisting of the following scaling properties:
\begin{eqnarray}
\frac{g_{Q_iQ_i}}{g_{Q_jQ_j}}&=& (\frac{Q_j}{Q_i})^2 \nonumber \\
\frac{g_{Q_iQ_i}}{g_{JJ}}&=& \frac{\pi^2}{64G^2}(\frac{Q_jQ_k}{Q_i}) \nonumber \\
\frac{g_{Q_iJ}}{g_{Q_jJ}}&=& \frac{Q_j}{Q_i} \nonumber \\
\frac{g_{Q_iQ_i}}{g_{Q_iJ}}&=& -\frac{\pi^2}{32G^2} (\frac{Q_jQ_k}{J}) \nonumber \\
\frac{g_{Q_iQ_j}}{g_{Q_kJ}}&=& \frac{1}{32G^2} (\frac{Q_k}{JQ_iQ_j})(\pi^2 Q_1 Q_2 Q_3- 32 G^2 J^2) \nonumber \\
\frac{g_{Q_iQ_j}}{g_{Q_{i,j}Q_k}}&=& \frac{Q_k}{Q_{j,i}}
\end{eqnarray}
As noticed in the previous configurations, it is not difficult to
analyze the local stability for the BMPV black holes, as well. In
particular, one may easily determine the principal minors
associated with the state-space metric tensor. Thus, we may demand
that all the principal minors must be positive definite. In this
case, we may adroitly compute the principal minors from the
Hessian matrix of the associated entropy concerning the three charge
rotating BMPV black holes. In fact, after some simple
manipulations we discover that the local stability criteria on the
three dimensional hyper-surfaces and the two dimensional surfaces
of the underlying state-space manifold may respectively be given
by the following relations:
\begin{eqnarray}
p_3 &=& 8 \pi^9  Q_1 Q_2 Q_3 \bigg(-\pi^2 Q_1 Q_2 Q_3+ 64 G^2 J^2\bigg)
\bigg(\pi^2 Q_1 Q_2 Q_3- 16 G^2 J^2\bigg)^{-5/2}  \nonumber \\
p_2 &=& 16 \pi^6(Q_3^2G^2J^2) \bigg(\pi^2 Q_1 Q_2 Q_3- 16 G^2 J^2\bigg)^{-2}
\end{eqnarray}
For all physically admitted values of invariant asymptotic charges of the BMPV black holes,
we may thus easily ascertain that the minor constraint, {\it viz.}, $p_2(Q_i, J)>0$ inhibits
the domain of assigned angular momentum, that must be a positive definite real number,
while the constraint $p_3(Q_i, J)>0$ imposes that the angular momentum must respectively be
greater than $J=\frac{\pi}{8G}(Q_1 Q_2 Q_3)^{1/2}$. In particular, these constraints enable
us to investigate the nature of the state-space geometry of the BMPV black holes. We may thus
observe that the presence of planar and hyper-planar instabilities is established for the spinning BMPV
black holes, which together demand for a restriction on the allowed value of the angular momentum.

Furthermore, we find that it is easy to determine the complete local stability on the full phase-space
configuration, which may be acclaimed by computing the determinant of the state-space metric tensor.
Nevertheless, it is not difficult to enumerate a compact formula for the determinant of the metric tensor.
For the different possible values of the charges, {\it viz.}, $\overrightarrow{Q}:=(Q_1,Q_2,Q_3)$ and
the angular momentum $J$, it may apparently be discovered from the present intrinsic geometric analysis,
that the following expression holds for the determinant of the metric tensor:
\begin{eqnarray}
g(\overrightarrow{Q}, J)= -16 \pi^{12} G^2 Q_1^2 Q_2^2 Q_3^2 \bigg(\pi^2 Q_1 Q_2 Q_3- 16 G^2 J^2\bigg)^{-3}
\end{eqnarray}
The determinant of the basic metric tensor is a small, non-zero quantity, in the large charge limit,
in which one acquires a non-vanishing central charge of the corresponding $D_1$-$D_5$-$P$ CFT,
or the associated worldvolume notion of the dual $M$-theory. Our analysis further discovers that
there exists a non-degenerate thermodynamic configuration with the extremal contributions.
However, it is worth to note that the determinant of the metric tensor does not take a positive
definite form, which thus shows that there is no positive definite volume form on the concerned
state-space manifold $(M_4,g)$ of the spinning BMPV black holes.
This is also intelligible from the fact that the responsible equilibrium entropy tends to its maximum value,
while the same may not remain valid on the chosen planes or hyper-planes of the entire state-space manifold.
It may in turn be envisaged, in either the $D_1$-$D_5$-$P$ description or $M$-theory description, that
the BMPV black holes do not correspond to an intrinsically stable statistical configuration. Thus,
it is very probable that the underlying ensemble of microstates may smoothly move into more stable
brane configurations.

Finally, in order to elucidate the universal nature of statistical interactions and the other properties
concerning the rotating BMPV black holes, one needs to determine certain global geometric invariants
on the state-space manifold $(M_4,g)$.
Here, we note that such a simplest invariant may be achieved by the state-space scalar curvature, which may
indeed be computed in the straightforward fashion by applying the formerly explained method of intrinsic geometry.
It turns out that the state-space configuration of the BMPV black holes is entirely simple, and in particular,
the explicit expression for the scalar curvature may be given to be
\begin{eqnarray}
R(\overrightarrow{Q}, J)=  \frac{3}{4\pi^3 Q_1 Q_2 Q_3} \bigg(\pi^2 Q_1 Q_2 Q_3- 64 G^2 J^2\bigg)
 \bigg(\pi^2 Q_1 Q_2 Q_3- 16 G^2 J^2\bigg)^{-1/2}
\end{eqnarray}
It may therefore be perceived, for the allowed value of conserved charges, that the state-space scalar curvature
remains a negative definite quantity. This makes perfect physical sense, i.e. that the underlying system is
effectively an attractive configuration. Here, the negative sign of the scalar curvature further signifies
that it is a less stable black hole system, than the definite statistical configurations possessing certain positive
state-space scalar curvature. Moreover, it is stimulating to expose that the scalar curvature thus demonstrated
is an everywhere regular function on the state-space manifold, and in particular, it never diverges for
the physically ratified values of $D_1$-$D_5$-$P$ or dual $M_2$-brane charges and the self angular momentum.
It may thus be indicated that the rotating BMPV black holes have no phase transitions and the underlying
statistical configuration is completely free from the critical phenomenon, in all tangibly admissible domain
of brane charges and angular momentum, describing the five dimensional minimal supergravity BMPV solutions.

In general, it may possibly be pronounced, for all non-zero brane charges, that the BMPV solutions render
to be weakly interacting statistical configurations, whenever there exists a non-vanishing state-space
scalar curvature. However, it turns out that the state-space scalar curvature vanishes for an absolute
value of the angular momentum. In this specific case, one may deduce that the underlying configuration
turns out to be a non-interacting statistical system and, thereafter, it changes the nature of the interactions.
In particular, we may simply celebrate that the changes in the nature of statistical interactions happen
precisely at the absolute value of angular momentum ascribed by
\begin{eqnarray}
 \vert J \vert=  \frac{\pi}{8 G} \sqrt{Q_1 Q_2 Q_3}
\end{eqnarray}
Furthermore, it is worth to  mention that the state-space quantities thus described entail
various intriguing generic intrinsic geometric properties, and in particular we may easily
notice, for the non-vanishing set of charges, that the determinant of the metric tensor remains non-zero,
while the scalar curvature clearly vanishes at the half value of the angular momentum, in
contrast to the zero entropy BMPV black holes.
It may further be envisaged, for the other possible supergravity solutions in five dimensions, that
they should indicate an interacting underlying statistical configuration, in accordance with the minimal
supergravity BMPV configurations. In these cases, it is interesting to note that the multifaceted contribution
of $M_5$-branes may be non-vanishing, and there would naturally be a higher number of charges. Thus, the underlying
state-space manifold would relatively be a higher dimensional intrinsic Riemannian manifold, than the
one analyzed for the restricted spherical horizon BMPV black holes.

We may however discover, for equal $M_2$ brane charges, {\it viz.}, $ Q_i= Q, \ i=1,2,3 $,
that the determinant of the state-space metric tensor and the associated scalar curvature
reduce to their corresponding easily determinable non-vanishing values.
It is worth mentioning that the indispensable contents of the state-space geometry, thus
elucidated, entangle with the notion of planar and hyper-planar stabilities. These constraints may
as well arise from the degeneracy of an ensemble of brane microstates constituting an equilibrium
statistical configuration. With a definite entropy of the spinning BMPV black holes, we may in fact
see that the constant entropy curve, for a given entropy $S_0$, take the form
\begin{eqnarray}
 Q_1 Q_2 Q_3- \alpha J^2= \beta
\end{eqnarray}
Here, it is obvious that the real constant $ \alpha,\beta $ may easily be defined to be
$\alpha:= 16G^2/\pi^2$ and $\beta:= S_0^2/4\pi^4$. Specifically, it is important to emphasize
that the state-space scalar curvature, as a function of the parameters, entails the very possible
existence of the phase-transitions, if any, underlying the spherical horizon BMPV configurations.
In particular, we may observe, for some given scalar curvature $k$, that the constant scalar
curvature curve turns out to be a simple curve which may rather be given as
\begin{eqnarray}
 9(Q_1 Q_2 Q_3- 4\alpha J^2)^2= 16 k^2 \pi^4 Q_1^2 Q_2^2 Q_3^2 ( Q_1 Q_2 Q_3- \alpha J^2)
\end{eqnarray}
It is worth to mention that there exists a remarkable aspect of the supergravity solutions, that is the way in which
they capture highly non-trivial constraints between the parameters that arise in their microscopic descriptions.
In this concern, it is known, for spherical horizon BMPV black holes, that the condition that the space-time
solutions be free from the causal pathologies, yields the correct upper bounds on the angular momentum required
by the microscopic $D_1$-$D_5$-$P$ CFT or worldvolume theory. Such supergravity constraints do also arise from the
state-space configuration, which may alternatively be apprised from the requirement that the solutions of interest
admit an intriguing thermal deformation.

Nevertheless, we may assert that our state-space geometry determines certain intricate physical properties of the
statistical pair correlation functions and correlation volume, which reveal the nature of associated parameters
prescribing an ensemble of microstates of the dual CFT living on the boundary of chosen black brane configurations.
Furthermore, we expect to consider it for a class of general higher dimensional black brane configurations,
where the propositions of the ordinary state-space geometric computations might not be very feasible with
multiple parameters. However, in such diverse situations one may possibly exhibit certain microscopic duality
acquisitions, with an appropriate comprehension of the required parameters defining the state-space coordinate
transformations on the concerned configurations.

In particular, our geometric concerns thus appraised may additionally be elucidated by the
parameters uprising in both the string theory or associated $M$-theory descriptions and their
gravity duals, from the very possible applications of an appropriate AdS/CFT correspondence.
This clearly demonstrates that the space of physically relevant $D_1$-$D_5$-$P$-solutions
is far from being completely mapped out, in our simple equilibrium state-space examinations.
We shall leave these issues for the future, for an appropriate explanation for their microscopic origins.
Further investigations are thus clearly needed, to establish a firm connection between the state-space,
worldvolume and supergravity constructions, for the two and three-charge configurations studied in this paper.
\section{Discussion of the Results}
In this Section we summarize each of the examples of geometries
considered in the paper, from the viewpoint of the corresponding
outlook and future investigations.
\subsection{Myers-Perry Black Holes}
The intrinsic Riemannian geometry of the rotating black brane
configurations entails an interesting feature and, in particular,
we deduce that the neutral Myers-Perry black holes admit a
non-degenerate, negatively curved state-space geometry.
Furthermore, we observe that the scalar curvature, as a function
of the mass and angular momentum, exhibits a non-vanishing and
regular structure over the entire state-space manifold. This
indicates that the underlying statistical configuration is
attractive and accompanies no phase transitions, under the
inclusion of leading order statistical correlations.

Nevertheless, we note that the macroscopic understanding of the
statistical interactions emerging from the attractor fixed point
analysis may be accomplished, by specifying the types of moduli
space geometry, while the corresponding large charged microscopic
properties may be appraised, with an appropriate consideration of
the $D$-brane or dual $M$-brane configurations. It is worth to
mention that the issues concerning the state-space geometry of the
Myers-Perry and Kaluza-Klein solutions may together be
categorized, in a unified perspective of vacuum $M$-theory black
holes. Thus, we wish to furnish the possible generalizations and
their interpretations in the next subsection.
\subsection{Kaluza-Klein Black Holes}
We find, in the case of extremal Kaluza-Klein black holes, that
the microscopic picture of state-space interactions may be
apprehended, in terms of the net number of the underlying $D_0$
and $D_6$ branes. In fact, this characterization may also be
exhibited, either by the individual brane masses, or in terms of
the five dimensional angular momenta describing the interested
Kaluza-Klein solutions. This is simply due to the fact that, when
taking the product with the flat $T^6$, one obtains a solution to
$M$-theory, whose IIA reduction acquires $D_0$ and $D_6$ charges.
Nevertheless, it is known that these black holes are not
supersymmetric, even in the extremal limit, which eludes nothing
more than the absence of supersymmetric bound states of the $D_0$
and $D_6$ branes, see for details \cite{KO}. However, there are
non-supersymmetric quadratically stable $D_0$-$D_6$ bound states
\cite{wati}, whose parameters intimate to serve a certain basis
towards the microscopic picture of state-space interactions. Our
analysis thus anticipates a simple stringy description to the
state-space interactions, which rely on an understanding of
microscopic entropy formula of the extremal black hole solutions.

More importantly, one can extend this argument to several other
$M$-theory black holes and, in particular, there exist certain
charged, nonsupersymmetric, extremal black branes \cite{nosusy},
for which our state-space descriptions may be envisaged to be
significant. It turns out that the state-space manifolds, thus
outlined, are well-defined on each hyper-plane, and everywhere
finitely curved on the full intrinsic configurations. However,
unlike this simple natural example, we are here interested in
describing the state-space manifolds of those higher dimensional
black holes, which correspond to the general vacuum solutions in
$M$-theory. In fact, the entropy of the pure black holes under
consideration can indeed be microscopically comprehended, in terms
of the Horowitz correspondence principal \cite{Horowitz:1996nw}.
Although such evaluations do not reproduce the precise coefficient
of the entropy, but up to a constant of proportionality, one may
easily ascertain the state-space quantities. Thus, it is immediate
to spell out the nature of Gaussian fluctuations, for these
configurations.

Furthermore, certain attempts can be made to describe the
statistical interactions geometrically, by different means,
concerning the earlier interests which try to capture an accurate
statistical entropy for the asymptotically flat black holes,
including the neutral Myers-Perry solutions, as well
\cite{ddbar,argurio}. This insinuates that our interpretations of
the metric tensor and scalar curvature are in perfect concordance
with the known statistical behavior of the associated $D_0$ and
$D_6$ black holes. In conclusion, the state-space geometry thus
accounted gives rise to a microscopic comprehension of the
fluctuating brane charge Kaluza-Klein black holes in string theory
or vacuum $M$-theory. Here, the brane charge quantizations and
energy excitation condition turn out to be certain general
co-ordinate transformation on the state-space manifold. Thus, the
state-space estimates may as well be transcribed in terms of these
parameters of the extremal black holes \cite{larsen2,sheinblatt}.

In the sequel, we out-line exotic interesting issues concerning
the state-space geometry and possible interpretations for the
neutral and charged $M$-theory black holes, which illustrate some
natural
generalizations in the other stimulating duality frames. 
%
%
We observe that the state-space geometry of the Kaluza-Klein black
holes is not only an interesting statistical configuration, but it
also ascribes an illuminating understanding of the phase
transition, if any, critical exponents and other physical
quantities, in terms of the parameters, either associated with the
microscopic $D$-brane/ $M$-brane configurations, or with the
macroscopic supergravity attractor solutions. It is important to
note that one can implicitly procure an idea of microscopic
descriptions of pair correlations and correlation volume, divulged
in the framework of our state-space manifold. Furthermore, the
state-space quantities thus obtained are indeed independent of the
Kaluza-Klein radius, and the physical nature thence endures to be
equivalent to that of the extremal Myers-Perry black holes,
arising in the limit when the limit of infinite radius has been
taken.

This is a strong indication that it would be feasible to determine
an intriguing microscopic counting argument for the statistical
pair correlations and correlation volume, which they may
respectively be associated with the covariant and invariant
quantities of the captivated state-space manifold. Our arguments
have, as well, been supported by the fact that the entropy of the
$M$-theory black holes depends on the corresponding integer
normalized net charges on the brane solutions, in a chosen duality
frame, see for further details \cite{larsen2,Dhar:1998ip}. In
order to comprehend the state-space geometry of the most general
Kaluza-Klein black holes, we may concentrate our attention on
various possible open questions and their interpretations. Some of
the issues in this concern may be summarized as follows.

Firstly, one should note that we have only examined the case of
extreme Kaluza-Klein black holes. However, in order to divulge the
full Kaluza-Klein configurations, one needs to comprehensively
consider the corresponding near-extremal or non-extremal
solutions, and then one would be obligated to count the
microstates. If it is possible to obtain the degeneracy formulae
in terms of the invariant parameters, then one can certainly take
account of the concerned intrinsic geometric quantities, and there
by our state-space analysis would explore the possible nature of
correlations in the vacuum $M$-theory equilibrium configurations.
These are precisely the ingredients that have also emerged from
recent developments in the study of finite-sized black-hole
microstates in the vacuum $M$-theory black holes \cite{larsen2}.
Our analysis would thus furnish a unified framework for
conceptualizing the underlying statistical fluctuations and their
limiting geometric models, which may explore the nature of the
general possible black brane configurations.

Interestingly, we identify, in these cases of more general
solutions, that the mass is only an additional parameter, which
gets added in the list of state-space variables. Thus, it may
easily be anticipated that the extremal state-space manifold gets
intrinsically embedded into an extra higher dimensional intrinsic
Riemannian manifold, which describes the entire Kaluza-Klein
configurations. In particular, it turns out, in the present case,
that the following embedding may be defined for the extremal
Kaluza-Klein black holes:
\begin{equation}
(M_3,g)\rightarrow (M_4,\tilde{g})
\end{equation}
Here, it is worth to mention that the extremal state-space
configuration may be examined as the BPS restriction to the full
counting entropy, with an intrinsic metric tensor,
$g:=\tilde{g}\vert_{M=M_0}$. It should be understood in this
expression that the restriction $M=M_0$ has been applied to the
assigned entropy of the non-extremal (or near extremal)
Kaluza-Klein solutions. It may thus be envisaged, from the
perspective of our state-space geometry, that an extremal
$M$-theory vacuum configuration, defined as an embedding $M_3
\hookrightarrow M_4$, would respectively be an interacting and
stable configuration on the hyper-spaces of concerned non-extremal
state-space configurations. Moreover, it is possible that the full
non-extremal Kaluza-Klein configurations might comprise certain
instabilities as well, whose signature would be encoded in the
invariant scalar curvature of the corresponding state-space
geometry.

Secondly, one can assimilate the state-space quantities of
Kaluza-Klein black holes, with globally asymptotically flat
solutions, simply by taking the limiting case of a single
$D_6$-brane in the corresponding counting problem. Although such
constructions are not yet rigorously exposed in the $M$-theory, as
they require at least four globally flat asymptotic solutions,
nevertheless it seems that the restricted two dimensional
state-space manifold may eventually investigate the peculiarities
of extremal Kaluza-Klein configurations. This is because an
assignment of the entropy may be justified to each brane
intersections, for a certain large number of such solutions. A
similar construction would also be admissible for the non-extremal
Kaluza-Klein configurations, for a single $D_6$-brane. In
particular, one finds that there exists a restricted three
dimensional state-space manifold, whose metric tensor and scalar
curvature might respectively provide the properties of the
statistical pair correlation functions and correlation volume of
the brane system, in the scenario of a large number of
intersecting brane solutions.

Thirdly, it may be expected that one can construct some natural
generalizations of the above state-space investigation on certain
general Calabi-Yau spaces, if it is still possible to ascribe a
counting argument to the corresponding microscopic degeneracy of
the brane microstates. Here, we would like to envisage in the
spirit of a well-known fact associated with the $T^6$ moduli, i.e.
that the corresponding mirror symmetry, defined as T-duality on a
$T^3$ might be perceived as a certain coordinate transformation on
the concerned state-space manifold. It is however exciting to
notice that the initial collection of $D_0$ and $D_6$-branes goes
over the collection of $D_3$-branes under this symmetry. Thus,
there may exist certain endurances for the very constructed
state-space correlations, as well. It would indeed be fascinating
to find a link between the microscopic configurations thus
originating, in the case of generalized Calabi-Yau spaces, and the
state-space signatures which have been explicitly considered.
Finally, we would like to mention that some of the observed
intrinsic geometric issues need to be carefully explored. In
particular, we intend that the approach being given in this paper
may favorably provide a key understanding of how the fluctuating
microstate geometries, in the semi-classical description, present
state-space correlations, in the regime of parameters where a
black hole exists.
\subsection{ Supersymmetric $AdS_5$ Black Holes}
The state-space geometry arising from the supersymmetric $AdS_5$
black holes exhibits an interesting set of scaling characteristics
for the statistical correlations. It is worth mentioning that the
normalization convention of the electric charges $Q_i$ depends on
the way in which one embeds the underlying gauged supergravity
into the higher dimensional string theory, or associated
$M$-theory configurations \cite{kmmr}. Furthermore, there exists a
natural normalization for the $AdS_5\times S^5$ background, such
that the conserved electric charges $Q_i$, in the supersymmetric
approximation, correspond to the internal Kaluza-Klein momenta
akin to the $U(1)^3\subset SO(6)$ isometry.

More specifically, the microscopic indications of our intrinsic
geometric estimations give an impression, i.e. that one may
potentially describe the basis notions, either in the framework of
dual giant gravitons, or in the construction of integral wrapped
brane solutions. Here, the brane charges or brane numbers play an
important role into the manifestly covariant computation of
correlations and critical phenomena, if any, in the underlying
statistical configurations. This is due to the fact that basic
electric charges or brane numbers are respectively associated,
either with the internal isometries of the theory, or with the
topological cycles of the compactifying internal manifold. In
particular, the charges $Q_i$ are integral for the consideration
of the $AdS_5\times S^5$ or $U(1)^3$, and thus the interested
state-space quantities arising from the concerned
Bekenstein-Hawking entropy exhibit a set of simple and exact
expressions, in the entire regime of the permissible physical
parameters. It is further instructive to notice that such a simple
explanation of the state-space analysis may shed light on the
corresponding fluctuations, comprising the microscopic degeneracy
and fundamental counting problem. It is thence expected that there
would probably exist a general Cardy or Hardy-Ramanujan formula,
which procures an account for its microscopic models, whose
equilibrium configurations may geometrically be exposed, in the
limit of Gaussian approximations.

Furthermore, we discover that the associated two point
off-diagonal pair correlation functions vary as the sum of the
inverse of two other brane charges, while the diagonal pair
correlations vary as the sum of the inverse square of two other
charges. We establish that the brane-brane pair correlation
functions satisfy a set of definite asymmetric relations, in
comparison with the other possible pair correlation functions, and
nonetheless vanish exactly at the half angular momentum of the
vanishing condition of the corresponding entropy. Here, one
achieves the scalar curvature signifying the correlation length,
which indicates that the associated microscopic configurations are
stable in all possible hyper planes. This also elucidates that the
corresponding interacting statistical systems are attractive
configurations, for all non-vanishing values of the entropy of
concerned black holes. Similar facts may further be enquired for
the case of the spinning black holes in the Minkowski space, as
well as in the richer isometric
solutions of $ \mathcal N= 2$ supergravity.


It appears interesting to explore the state-space geometry of the
$\frac{1}{16}$-BPS objects of type-IIB string theory, in
$AdS_5\times S^5$ configurations \cite{kmmr}. In particular, we
may investigate the statistical fluctuations, for the
supersymmetric black holes in gauged supergravity, and point out
that there exists simple expressions for the state-space
correlations arising from the Bekinstein-Hawking entropy, in terms
of physical charges and rotation. We may thus appraise, from the
simplicity of encompassed characteristics, that our analysis
suggests an existence of non singular correlations in the
underlying microscopic configurations, whose basis significance
may be defined in terms of the parameters of the BPS gravitons.

One of the intriguing features of our state-space covariant
objects, associated with the BPS black brane solutions assimilated
in this paper, is that they might be viewed as the correlators of
the parameters of the $\frac{1}{16}$-BPS deformations of either
the $\frac{1}{8}$-BPS giant gravitons of \cite{mikh}, or those of
the $\frac{1}{8}$-BPS dual giant gravitons of \cite{ma-sur}. It
would thus be exhilarating to contemplate our parametric geometric
quantities in these dual microscopic descriptions. These interests
emerge because the preferred solutions seem to be incorporated
into a single solution space of the $\frac{1}{16}$-BPS theory.
More strongly, we find at this point that if the entropy, as a
real embedding function of the invariant asymptotic charges,
endures fixed, then nothing much happens, except that there would
be an existence of certain general co-ordinate transformations on
the corresponding state-space manifold, which might change
completely one degenerate configuration into the others.

We stress that the state-space geometry of $AdS_5$ black holes
admit simple expressions not only for the $U(1)^3$ supergravity or
truncated $SO(6)$ gauged supergravity, but also for those obtained
by the very $S^5$ reduction of type-IIB string theory. The only
requirement for supergravity is that the scalars in the vector
multiplet should reside on a symmetric space, which may therefore
be accomplished by the $AdS_5$ supergravity with at least $16$
supersymmetries \cite{aw-to}. Moreover, we wish to further analyze
our state-space construction, for the $ \mathcal N= 2$ $AdS_5$
supergravity configurations, accomplished by compactifying
$M$-theory on a suitable 6-manifold. In this concern, there exists
an explicit example known as Maldacena-Nunez solution, see for
details \cite{ma-nu}.

It would thus be interesting to consider the state-space manifolds
of $AdS_5$ black holes in this background, carrying a certain
number of electric charges accomplishing from the membrane
wrapping an internal 2-cycle, as well as those coming from the
internal momenta, that are in turn analogous to $S^5$ Cartans of
the theory. Furthermore, it is known that the central charge
factor $c$ in front of the angular momentum $J$ is proportional to
$N^3$, where $N$ is the number of $M_5$ branes, on which the dual
superconformal field theory lives. In this concern, we would like
to account for the D-instanton contributions to the graviton
scattering amplitudes, and thus the corresponding $SU(N)$
Yang-Mills instanton corrections into the state-space
correlations, see for a definite consideration
\cite{GopakumarGreen}. Therefore, it would not be formidable to
analyze the coarse graining prelude of the state-space quantities,
with or without the instanton corrections, and the other
associated microscopic properties transpiring in the $AdS$-black
hole solutions.
\subsection{ $D_1$-$D_5$ Configurations}
With the most studied BPS black brane configurations, we notice
that the state-space manifolds associated with the $D_1$-$D_5$
solutions \cite{0706.3884v1} acquire the two striking microscopic
interpretations. In general, one arrives at the conclusion that an
akin representation exists in terms of either the $D_1$-$D_5$-$P$
CFT \cite{BK2}, or the four-dimensional black hole CFT
\cite{Bena:2005ni,CGMS}. On one hand, the first may be accounted
via the invariant parameters of the CFT microstates, the former
should be thought to be coming out of the ground states of
$D_1$-$D_5$-$P$ black branes, while the latter may be connoted in
the same way, as that of the parameters of the solutions of
\cite{lu-ma,LuninIZ}, describing the five-dimensional three-charge
black holes and the possible dual solutions of \cite{Bena2005ay},
which divulge the ground states of four dimensional four-charge
black holes. This shows that there exist many interesting
independent vacua of underlying $D_1$-$D_5$-$P$ CFTs. Thus, it is
beyond the present scope to explicitly establish a particular
microscopic composition, which corresponds to coarse grained
state-space geometries.

Based on the microscopic dispositions of \cite{lu-ma,BK2,LuninIZ},
we may envisage that our state-space constructions would
correspond, as well, to multiple brane configurations being dual
to the CFT states, with longer effective strings than the
solutions that come from only single brane. It may thus be pointed
out that an effective string picture stimulates an interesting
notion for the state-space pair correlation functions of well
established $D_1$-$D_5$ black hole configurations. Moreover, our
intrinsic Riemannian geometric analysis indicates that the maximum
correlation length, emerging from the scalar curvature, involves
the largest number of bubbles, which should correspond to the CFT
states with the longest tangible effective strings. Nevertheless,
it may be argued that the underlying statistical configurations of
bubbled black branes are free from the instabilities and/ or
vacuum phase transitions.

These inspirations stem naturally from the piercing mechanism of
state-space geometry to the regular $D_1$-$D_5$-$P$ system.
Moreover, the underlying physics is closely related to that of
other possible solutions containing the correct black hole/
string/ ring/ supertube vacuum configurations. As mentioned
earlier, the BPS black branes have two microscopic
interpretations: one in terms of the $D_1$-$D_5$-$P$ CFT
\cite{BK2}, the other in terms of the four dimensional black hole
CFT \cite{BK2,Bena2005ni,CGMS}. Thus, an investigation may as well
be explored for these solutions. Hence, our construction may be
disclosed, in terms of the parameters of the $D_1$-$D_5$-$P$
system, and those of the four dimensional black hole CFT, which
describes the BPS black branes. It may similarly be apprised that
both the urged descriptions should play an important role in the
determination of the correlated, interacting statistical basis,
for the general $D_1$-$D_5$-$P$ solutions.

Finally, the results thus obtained insinuate a number of
interesting consequences and intriguing relations for future
exploration. One of the most influential possibility is to start
from the microscopic $D_1$-$D_5$-$P$ description, and explore the
zero-entropy limit of the concerned BPS black brane
configurations. Here, the perceptible picture adduces that the
most generic bound states with definite branes, which could be
determined by a gas of positive and negative centers, with fluxes
threading the many non-trivial two-cycles of the GH base, having
no localized brane charges, are likewise possible to be elucidated
\cite{12oo,12ooo,BNTSBnext,12o3}. In turn, it is feasible to
demonstrate, with several dual $M_5$-brane contributions as well,
that there are no critical phenomenon on the non-degenerate
state-space manifolds, thus spanned by the conserved parameters,
in the vanishing entropy limit of the BPS solutions. Moreover,
very similar conclusions are encouraged for the state-space
geometric quantities defined for the other pertained definite
string theory, or M-theory black holes.
\subsection{BMPV Black Holes}
The state-space of three charged $D_1$-$D_5$-$P$-solutions shows
various interesting intrinsic Riemannian geometric properties, for
the spherical horizon topology black holes. One of the intriguing
features of the covariant structures, associated with the
supersymmetric configurations, has been based on the fact that
these solutions may uniquely be specified by their electric
charges, and definite angular momenta, which in the limit of
vanishing compactification radius lead to the rotating BMPV black
holes of the present inquisitiveness. In fact, there are no other
black hole solutions with a self dual angular momentum $J= J_\psi
= J_\phi$. Wherefore, the conserved electric charges and the
self-dual angular momentum of theory may always uniquely be chosen
for the BMPV black holes, which in particular form a coordinate
basis for the underlying intrinsic state-space manifold.
Significantly, the geometric nature of the Gaussian fluctuation,
thus celebrated, may properly be addressed, in terms of the
invariant physical parameters of the chosen solution. In turn, it
has been deduced that the state-space pair correlation functions
admit simple scaling relations, while the associated correlation
length of possible statistical configurations may as well be
ascertained, in terms of the concerned invariant parameters of the
theory.

On other hand, we have recently appraised various examples of a
class of generic state-space manifolds, whose microscopic
implications may be stimulated from the origins of coarse graining
entropy of extremal and non-extremal black brane solutions. It may
further be pointed out that the various statistical correlations,
based over a large number of equilibrium microstates, are possible
to be described in the framework of our state-space geometry, see
for further detail \cite{BNTBull}. In a close connection with the
exposed pursual, we pronounce that the ratio of various diagonal
components of the BMPV state-space configuration scale as inverse
square of the invariant asymptotic parameters varying under the
Gaussian fluctuations, whereas the ratios involving the diagonal
components transform only inversely in the chosen parameters. This
implies that the diagonal components weaken faster, and they
relatively more quickly attain an equilibrium statistical
configuration, than the underlying off-diagonal components. In
this case, we may further realize that the rotating BMPV black
holes procure a weakly interacting statistical configuration,
which has no phase transitions. The present analysis thus
instructs that the microscopic configurations of BMPV black holes
should be free from the critical phenomena.

Furthermore, it may be expected that alike statistical
illuminations would exist in the physically admissible domain of
brane charges and angular momenta, connoted by the miscellaneous
multi-charged rotating configurations. Here, our interest is in
particular on the very possible three-parameter supersymmetric
black brane solutions \cite{EEMR2,ElvangRT,Elvang:2005sa} of
minimal $D=5$ supergravities which, upon oxidation to the ten
dimensions, describe black supertubes carrying a definite number
of $D_1$-brane, $D_5$-brane and momentum charges. Nevertheless,
the concerned systems abiding equal number of $D_1$, $D_5$ branes
and Kaluza-Klein monopole with certain dipole moments, might
further be investigated from the perspective of our intrinsic
state-space geometry. It is tempting to annunciate that the
familiar state-space geometry of these configurations may in
general present many complexities, endorsed as the function of
various promising invariant parameters. Such an interesting
example may thus expound the case of black supertube solutions
entangling the three charges, three dipole moments and the radius
of ring. The concerned brane charges may as well be identified in
the M-theory dual configurations, so that they ascribe the $M_2$
brane charges, while the dipole moments account for the
$M_5$-charges.

Thus, the construction of our state-space geometry, enumerating
most general $D_1$-$D_5$-$P$ systems, contains several previously
divulged state-space geometries of \cite{BNTBull}. Hitherto, it
turns out, for the thrust of diversified analysis, that the
interesting family of the possible solutions may be summarized as
follows:
\begin{itemize}
\item Firstly, the state-space geometry thus developed would correspond to the one arising from the solution
of \cite{EEMR2,ElvangRT,Elvang:2005sa}, in the special case of
three equal charges and three equal dipoles.
\item Secondly, the state-space geometry may also expound to that of the two-charge supergravity
configurations of \cite{EMT}, when one of the invariant charges
and two dipoles vanish.
\item Thirdly, one should analyze the other solutions, in the zero-radius limit, which reduce
to the present solution having the four dimensional state-space
geometry described in terms of the parameters of the
supersymmetric spherical horizon BMPV black holes \cite{bmpv}.
\item Fourthly, one should compute the state-space geometric quantities of the $D_1$-$D_5$-$P$ solutions
in the infinite-radius limit, which would test for the six-charge
black strings of \cite{bena}.
\item Finally, one should examine the possible near-extremal and non-extremal cases of the full $D_1$-$D_5$-$P$
configurations. The concerned analysis would divulge a unified
framework, for the intrinsic geometric understanding of possible
statistical correlations. Thus, the nature of limiting microscopic
models may be explored, for a class of general fluctuating black
brane solutions.
\end{itemize}
\section{Conclusion and Outlook}
We studied state-space geometry of a class of higher dimensional
rotating black holes, obtained after an adequate compactification
of string theory and $M$-theory, and there by analyze the nature of
underlying statistical configurations, by incorporating
fluctuations into them. Mainly, the focus has been on the physical
parameters labeling the black brane configurations, which divulge
the microscopic and macroscopic acquisitions, and there by
critically examinining the possible implications arising from the
Gaussian fluctuations, exhibiting the symmetric character of
state-space metric tensor. The behavior of statistical
fluctuations has been explored from the perspective of an
intrinsic Riemannian geometry, emerging from the negative Hessian
of the coarse graining entropy defined over an ensemble of
microstates. It has been demonstrated that the principal components
of state-space metric tensor, which signify heat capacities or the
associated compressibilities are generically positive definite. It
is illuminating to notice that very similar state-space
geometric conclusions are accomplished for both neutral and
charged black hole solutions, being present in the vacuum
$M$-theory.

We manifestly show that the relative diagonal components weaken relatively faster, and thus they
more easily come into an equilibrium configuration, than the off diagonal components, which remain
comparable for a longer domain of the parameters varying under the Gaussian fluctuations.
In general, we observe that the brane-brane statistical pair correlation functions are asymmetrical
in nature, in contrast to the other possible statistical pair correlation functions.
This scaling property may naturally be expected from the fact that the brane-brane interactions
impart more energy than either the self-interactions or the energy due to the rotations.
This asymmetry generically intimates a lower bound on the angular momentum carried by the black holes.
We discover that the state-space geometry thus described is non-degenerate and finitely curved
for the neutral Myers-Perry black holes and charged Kaluza-Klein black holes.
This exhibits that the vacuum $M$-theory black holes are a statistical system in a local equilibrium.
The possible intrinsic geometric results obtained against the Myers-Perry solutions are quite illuminating.
The present consideration takes into account the fact that the underlying microscopic configurations
are consisting of a large number of brane-antibranes.

Furthermore, we deduce that the state-space invariant quantities thus acquired are negative, in general,
which connotes that the underlying black hole systems are effectively in an attractive configuration.
Importantly, it has been noted that the determined scalar curvatures never vanish, for a domain of the
physically justifiable value of invariant charges and angular momentum, for the favored black hole solutions.
This shows that these simple black holes correspond to an interacting statistical configuration, whose
state-space scalar curvatures rapture out, to be an everywhere regular function of the composing parameters.
For these natural systems, our results suggest that there exists an intriguing relationship between non-ideal
inter-particle (or more generally inter-brane) statistical interactions and thermodynamic phase transitions,
which may in particular be established from the perspective of an intrinsic Riemannian manifold, or vice-versa.
Our observations thus described are in turn consistent with the existing illustrations of the
microscopic CFT data of the corresponding black brane configurations.

In this paper, we have contemplated that the state-space geometry, arising from the fluctuating higher
dimensional rotating black brane configurations, may be exemplified by the string theory and $M$-theory.
It is significant to notice that our state-space investigations may be based on the coarse graining
phenomenon of the large number of degenerate CFT microstates describing an equilibrium statistical system,
and thus they require no direct understanding of the microscopic black brane configurations.
It has been discovered that the crucial ingredient, in analyzing the state-space manifold of the black
brane configurations, only depends on the degeneracy of micro-states defining the counting entropy.
Our conclusions are wherefore independent of the diverse microscopic properties of the elementary
conformal field theories.

Such an interesting supergravity system includes the case of supersymmetric $AdS_5$ black holes
and the accompanied $D_1D_5$ black brane configurations.
For the $AdS_5$ configurations, we focus our attention on the symmetric spaces. There by it
turns out that the statistical pair correlations comply truly captivating scaling properties.
The state-space geometry, thus defined, turns out to be stable only in the plane and possible hyper-planes
of the full state-space configuration, and the global parametric state-space is quadratically unstable.
The state-space scalar curvatures indicate that the underlying microscopic theory is free
from the possible phase transitions, and is a completely regular statistical configuration.
These notions are more direct to pronounce for the most prevalent $D_1$-$D_5$ black brane systems, and
here the supergravity conditions yield an intriguing stable and weakly interacting attractor solution.
Here, it may be appreciated that the holographic corrections may efficiently be accounted for, by
including the central charge term of the four dimensional boundary conformal field theory.
The modifications of the state-space geometry are thus possible, due to these contributions.
In particular, our conclusions show that the scalar curvatures, thus acquired, are only
marginally sensitive to the plausible CFT modifications.

As a final exercise, we explored the case of BMPV black holes obtained from the toroidal $M$-theory
compactification. Additionally, it is interesting to note that the dual $D_1$-$D_5$-$P$ description
provides an illuminating framework, in order to examine the nature of statistical fluctuations into this system.
The state-space geometry, thus intended, may indeed be shown to exhibit all the relevant properties
and, in particular, one may analyze the geometric nature of the correlation functions and correlation
volume of the concerned statistical configurations. Our computations present that the nature of statistical
fluctuations admits an analogous form, actualized for the $AdS_5$ black hole configurations.
It is worth to mention that the components of the state-space metric tensor, which comprise the two point
statistical correlation functions, appear to be intertwined with the vacuum fluctuations of the boundary
conformal field theory. This is because the required parameters of the black brane configurations
describing the microstates of dual conformal field theory, which exists on the boundary,
may be investigated by the AdS/CFT correspondence.

The geometric formalism thus described entails with an ensemble of
degenerate CFT ground states, which at an amusingly small constant
positive temperature exhibits the statistical correlations into
the basic equilibrium configurations. Furthermore, it is
interesting to note that the Gaussian fluctuations about the
equilibrium statistical configurations systematize the metric
tensor on the intrinsic state-space manifolds. The behavior of
brane-brane correlations in either cases turns out to be
asymmetrical, in comparison with the other statistical pair
correlation functions, and vanishes exactly at the half angular
momentum, of the one materializing from the vanishing entropy
condition. Our calculations display that the principal components
of the state-space metric tensor are positive definite, while the
non-identical components are not so, in general. Similarly, the
surface and hyper-surface principal minors are positive definite
functions of the invariant parameters, while for the full
state-space determinants they turn out to be negative definite
quantities.

The state-space geometries, thus exemplified for the higher dimensional rotating solutions, illustrate
that the potential statistical configurations are stable, only if at least one of the parameter remains fixed.
This is in perfect accordance with the well known fact that the only classical fluctuations
having a thermal origin divulge with the classical probability distribution, which defines
an intrinsic Riemannian metric tensor, over an ensemble of equilibrium microstates, and there by
render to a weakly interacting and attractive statistical configuration.
However, our state-space constructions, entangled with CFTs microstates, clearly celebrate that the
degeneracy and the signature of the state-space geometry is indefinite, and the positive definiteness
requirement is sensitive to the moduli space geometry and associated attractor solutions.
Moreover, the absence of divergences in the scalar curvatures imply that the considered solutions
are thermodynamically stable and there are no phase transitions in the underlying configurations.

The present investigation is thus motivated, as a prelude to the state-space geometry of an arbitrary
higher dimensional configuration. Particularly, we have shown that the examples thus explored have
an interesting set of state-space geometry, which describes the statistical fluctuations in the associated
rotating black brane configurations. The details of black hole solutions, concerning the revealed intrinsic
geometric enterprises, may further be observed. In general, we discover that the possible state-space geometry
may be characterized by the following cases:
(i) whenever there exists a positive definite determinant of the metric tensor, the underlying state-space
configurations are everywhere well-defined, and the concerned black holes realize stable statistical configurations.
(ii) whenever there exists a non-zero state-space scalar curvature, the underlying black hole configurations
correspond to an interacting statistical system. Nevertheless, the sign of the state-space curvature
characteristically plays an important role, in the determination of the nature of the interactions.

In this way, it is possible to expound, how the state-space pair correlation functions
characterize the behavior of fluctuating statistical configurations over an ensemble of
equilibrium microstates, defining the higher dimensional rotating black brane solutions.
In such cases, it has indeed been interesting to determine, whether these correlations, as
the function of all possible black brane charges, are positive definite and regular (or singular)
functions on the enticed state-space manifold.
Moreover, we have analyzed the stated black holes in the limit of vanishing state-space
scalar curvatures, which makes perfectly clear the nature of limiting statistical configurations.
Such limiting representations of the considered black hole solutions may as well be ascribed by
the definite constant entropy and scalar curvature curves, which in turn algebraically describe the
nature of corresponding state-space configuration.
Furthermore, we could easily reveal, whether there exist certain critical points, or critical lines,
and critical (hyper)-surfaces, in the fundamental configuration. In fact, if there exist any such
critical qualifiers, then they all would exhaustively be distinguished by the divergence structure
of the state-space scalar curvature.

The examples that we have accounted to analyze, for the rotating configurations, are neutral
Myers-Perry black holes, two charged Kaluza-Klein black holes, three charged $AdS_5$ black holes,
two charged $D_1D_5 $ configurations, and the associated spherical horizon BMPV black holes.
Moreover, we find yet instructive to further explore the state-space geometry, for the general
rotating black hole configurations emerging in the fundamental string theory and $M$-theory.
In the previous section, we presented various possible results and explained the encompassed motivations
for the specific cases of the extremal $M$-theory vacuum solutions, supersymmentric $D$-brane
solutions comprising the two or higher charged rotating black brane configurations.
%
%
In this respect, in the present investigation,
the main line of thought has been to develop an intrinsic Riemannian geometric notion to the
statistical fluctuations, which may be exercised between the most probable brane microstates of the
string theory or $M$-theory.

We thence vindicate the coarse graining origins of the state-space geometry, arising from the negative
Hessian matrix of the corresponding counting entropy, defined over an ensemble of  brane microstates
characterizing a chosen equilibrium statistical configuration.
Furthermore, we have exhibited, from this perspective, that the state-space geometry remains non-degenerate
and acquires a finite negative scalar curvature, for the exposed black holes at the attractor fixed points.
Interestingly, the example of $D_1$-$D_5P$ solutions may further be interpreted by considering a large number
of well separated Gibbon-Hawking charges, with a dissolving total Gibbon-Hawking charge.
In particular, one may interrogate the scaling behavior of statistical pair correlation functions,
describing the nature of various possible equilibrium microstates. In turn, our analysis reveals that
these may easily be assessed, by the conserved parameters of the underlying microscopic CFT configurations.
Finally, we are confronted with the general construction of our state-space geometry, and its potential role
to describe the issue of interactions in associated higher dimensional rotating black brane configurations.
In this concern, we have briefly addressed the result obtained for considered configurations, with numerous
open developments for the future investigations.

\vspace{.3cm}
\begin{Large} \noindent{\bf Appendix A} \end{Large}\\

We wish to explicitly give the involved intrinsic geometric quantities only for the case of two and
three dimensional state-space manifolds. Furthermore, we may also exhibit that similar outlines
hold for a class of general multi-parameter black brane configurations.
For the two dimensional state-space geometry parametrically defined by the two invariant parameters,
{\it viz.}, an electric charge $Q$ and angular momentum $J$, we expose that the components of the covariant
metric tensor may be given as
\begin{eqnarray}
g_{QQ}&=&- \frac{\partial^2 S}{\partial Q^2}, \nonumber \\
g_{QJ}&=&- \frac{\partial^2 S}{{\partial Q}{\partial J}},  \nonumber  \\
g_{JJ}&=&- \frac{\partial^2 S}{\partial J^2}
\end{eqnarray}
We may therefore notice that the components of the state-space
metric tensor are related to the statistical pair correlation
functions, which may as well be defined in terms of the parameters
describing the dual microscopic conformal field theory on the
boundary. This is because the underlying metric
tensor comprising Gaussian fluctuations of the entropy defines the
state-space manifold for the rotating black brane configuration.

Moreover, the positivity of the state-space metric tensor imposes
a stability condition on the Gaussian fluctuations of the
underlying statistical configuration, which requires that the
determinant and hyper-determinant of the metric tensor must be
positive definite. In fact, it is easy to express, in this
simplest case, that the determinant of the metric tensor turns out
to be
\begin{eqnarray}
\Vert g \Vert &= &S_{QQ}S_{JJ}- S_{QJ}^2
\end{eqnarray}

In order to have a positive definite metric tensor on the two dimensional state-space geometry,
one thus demands that the determinant of metric tensor must satisfy $\Vert g \Vert >0$,
which in turn defines a positive definite volume form on the concerned state-space manifold.
Furthermore, it is not difficult to calculate the Christoffel connection $\Gamma_{ijk}$,
Riemann curvature tensor $R_{ijkl}$, Ricci tensor $R_{ij}$, and the scalar curvature $ R $
for the two dimensional state-space intrinsic Reimannian manifold $(M_2,g)$.
In particular, it turns out that the above two dimensional state-space scalar curvature appears
as the inverse exponent of the inner product defining the pair correlation functions between
two arbitrary equilibrium microstates characterizing the black brane statistical configuration.
Explicitly, we find that the scalar curvature may be given by
\begin{eqnarray}
R&=& \frac{1}{2} (S_{QQ}S_{JJ}- S_{QJ}^2)^{-2}(S_{JJ}S_{QQQ}S_{QJJ}+ S_{QJ}S_{QQJ}S_{QJJ}\nonumber\\
&+& S_{QQ}S_{QQJ}S_{JJJ} -S_{QJ}S_{QQQ}S_{JJJ}- S_{QQ}S_{QJJ}^2- S_{JJ}S_{QQJ}^2 )
\end{eqnarray}
Notice that there exists an intriguing relation between the scalar curvature of the state-space
intrinsic Riemannian geometry characterized by the parameters of the equilibrium microstates,
with the correlation volume of the corresponding black brane phase-space configuration.
It is worth to mention that the state-space scalar curvature in general signifies certain possible
interactions in the underlying statistical configuration.
Following our previous observations \cite{BNTBull}, it is in particular evident that the scalar curvature
and corresponding Riemann curvature tensor of an arbitrary two dimensional intrinsic state-space manifold
$(M_2(R),g)$ may be given by
\begin{equation}
R=\frac{2}{\Vert g \Vert}R_{QJQJ}
\end{equation}
Furthermore, we find that the general coordinate transformations
on the state-space manifold, thus considered, may be expounded from
the perspective of fundamental string duality relations associated
with the invariant charges of the microscopic configurations.
Indeed, we show, from the perspective of an intrinsic Riemannian
geometry, that there exists an obvious mechanism on the black brane
side, i.e. that it is possible to illuminate the statistical notion of
associated correlations between the microstates of a brane
solution or vice-versa. It turns out that the state-space
constructions thus described clearly elucidate certain fundamental
issues, such as statistical interactions and stability of the
underlying brane configurations. In particular, it possible to
provide an intrinsic geometric realization of the equilibrium
statistical structures, which may be determined by the parameters
of the CFT microstates describing the rotating black brane
configurations.

Moreover, the state-space geometric structures thus disclosed may be expressed
for the case of the higher dimensional intrinsic Riemannian manifolds, as well.
In the case of the three dimensional state-space, we may easily define all the concerned quantities.
Specifically, the components of the metric associated with the rotating black brane configurations
carrying an electric charge $Q$, magnetic charge $P$, and an angular momentum $J$ are given by
\begin{eqnarray}
g_{QQ}&=& - \frac{\partial^2 S}{\partial Q^2} \nonumber \\
g_{QP}&=& - \frac{\partial^2 S}{\partial P \partial Q} \nonumber \\
g_{QJ}&=& - \frac{\partial^2 S}{\partial Q \partial J} \nonumber \\
g_{PP}&=& - \frac{\partial^2 S}{\partial P^2} \nonumber \\
g_{PJ}&=& - \frac{\partial^2 S}{\partial P \partial J} \nonumber \\
g_{JJ}&=& - \frac{\partial^2 S}{\partial J^2}
\end{eqnarray}

\vspace{.3cm}
\begin{Large} \noindent{\bf Appendix B} \end{Large}\\

In this appendix, we supply an explicit form of the most general state-space
scalar curvature describing the family of two charged rotating black holes.
Our analysis illustrates that the functional property of the specific scalar
curvatures may exactly be exploited by the definite behavior of $f(Q,P,J)$.
As we have accounted in section three, the various intriguing examples of
the concerned solutions include Kaluza-Klein black holes in vacuum $M$-theory
and the highly interesting $D_1$-$D_5$ systems. As mentioned in section $2$, we discover
that the $f(Q,P,J)$ indicates the possible nature of general three parameter state-space
configurations. In particular, one may notice, from the very definition of state-space
metric tensor, that the relevant function $f(Q,P,J)$ could be computed to be\\


\begin{eqnarray}
f(Q,P,J):= (f_1 -f_2) +2( f_3 -f_4) +3( f_5 -f_6) +2( 2f_7 +3 f_8)
\end{eqnarray}

where the $\{ f_i \}$ are given by the following expressions

\begin{eqnarray}
f_1&=& \frac{\partial^2 S}{\partial P^2} \frac{\partial^2
S}{\partial Q^2} \frac{\partial^3 S}{\partial Q^2
\partial P} \frac{\partial^3 S}{\partial P \partial J^2}
\frac{\partial^2 S}{\partial J^2} +\frac{\partial^2 S}{\partial
P^2} \frac{\partial^3 S}{\partial Q^3} \frac{\partial^2
S}{\partial Q \partial P} \frac{\partial^2 S}{\partial P \partial
J} \frac{\partial^3 S}{\partial J^3} \nonumber \\
&& +\frac{\partial^2 S}{\partial P^2} \frac{\partial^3 S}{\partial
P^2 \partial J} \frac{\partial^3 S}{\partial Q^2 \partial J}
\frac{\partial^2 S}{\partial Q^2} \frac{\partial^2 S}{\partial
J^2} +\frac{\partial^2 S}{\partial P^2} \frac{\partial^3
S}{\partial Q \partial P^2} \frac{\partial^3 S}{\partial Q
\partial J^2} \frac{\partial^2 S}{\partial Q^2} \frac{\partial^2
S}{\partial J^2} \nonumber \\ && +\frac{\partial^2 S}{\partial
P^2} (\frac{\partial^3 S}{\partial Q^2 \partial J})^2
(\frac{\partial^2 S}{\partial P \partial J})^2 +\frac{\partial^2
S}{\partial P^2} (\frac{\partial^3 S}{\partial Q \partial J^2})^2
(\frac{\partial^2 S}{\partial Q \partial P})^2 \nonumber \\
&& +\frac{\partial^3 S}{\partial P^3} \frac{\partial^2 S}{\partial
Q \partial P} \frac{\partial^2 S}{\partial Q \partial J}
\frac{\partial^3 S}{\partial J^3} \frac{\partial^2 S}{\partial
Q^2} +(\frac{\partial^3 S}{\partial Q^2 \partial P})^2
\frac{\partial^2 S}{\partial J^2} (\frac{\partial^2 S}{\partial P
\partial J})^2 \nonumber \\ && +(\frac{\partial^3 S}{\partial P^2 \partial J})^2
\frac{\partial^2 S}{\partial Q^2} (\frac{\partial^2 S}{\partial Q
\partial J})^2 +(\frac{\partial^3 S}{\partial P \partial J^2})^2
\frac{\partial^2 S}{\partial Q^2} (\frac{\partial^2 S}{\partial Q
\partial P})^2 \nonumber \\ && +(\frac{\partial^3 S}{\partial Q \partial P^2})^2
\frac{\partial^2 S}{\partial J^2} (\frac{\partial^2 S}{\partial Q
\partial J})^2 +\frac{\partial^3 S}{\partial P^3} \frac{\partial^2
S}{\partial P \partial J} \frac{\partial^2 S}{\partial Q \partial
J} \frac{\partial^3 S}{\partial Q^3} \frac{\partial^2 S}{\partial
J^2} \nonumber \\ && +\frac{\partial^3 S}{\partial P^2 \partial J}
(\frac{\partial^2 S}{\partial Q^2})^2 \frac{\partial^3 S}{\partial
P \partial J^2} \frac{\partial^2 S}{\partial P \partial J}
+\frac{\partial^3 S}{\partial Q \partial P^2} (\frac{\partial^2
S}{\partial J^2})^2 \frac{\partial^3 S}{\partial Q^2 \partial P}
\frac{\partial^2 S}{\partial Q \partial P} \nonumber \\
&& +\frac{\partial^3 S}{\partial P^3} \frac{\partial^3 S}{\partial
P \partial J^2} (\frac{\partial^2 S}{\partial Q^2})^2
\frac{\partial^2 S}{\partial J^2} +\frac{\partial^3 S}{\partial
P^3} \frac{\partial^3 S}{\partial Q^2 \partial P}
(\frac{\partial^2 S}{\partial J^2})^2 \frac{\partial^2 S}{\partial
Q^2} \nonumber \\ && +(\frac{\partial^2 S}{\partial P^2})^2
\frac{\partial^3 S}{\partial Q^2 \partial J} \frac{\partial^3
S}{\partial Q \partial J^2} \frac{\partial^2 S}{\partial Q
\partial J} +(\frac{\partial^2 S}{\partial P^2})^2
\frac{\partial^3 S}{\partial Q^3} \frac{\partial^3 S}{\partial Q
\partial J^2} \frac{\partial^2 S}{\partial J^2} \nonumber \\ && +\frac{\partial^2
S}{\partial P^2} \frac{\partial^3 S}{\partial P^2 \partial J}
(\frac{\partial^2 S}{\partial Q^2})^2 \frac{\partial^3 S}{\partial
J^3} +\frac{\partial^2 S}{\partial P^2} \frac{\partial^3
S}{\partial Q \partial P^2} (\frac{\partial^2 S}{\partial J^2})^2
\frac{\partial^3 S}{\partial Q^3} \nonumber \\
&& +(\frac{\partial^2 S}{\partial P^2})^2 \frac{\partial^2
S}{\partial Q^2} \frac{\partial^3 S}{\partial Q^2 \partial J}
\frac{\partial^3 S}{\partial J^3},
\nonumber \\
f_2&=& (\frac{\partial^3 S}{\partial P^2 \partial J})^2
(\frac{\partial^2 S}{\partial Q^2})^2 \frac{\partial^2 S}{\partial
J^2} +(\frac{\partial^3 S}{\partial Q \partial P^2})^2
(\frac{\partial^2 S}{\partial J^2})^2 \frac{\partial^2 S}{\partial
Q^2} \nonumber \\ && +\frac{\partial^2 S}{\partial P^2}
\frac{\partial^2 S}{\partial Q^2} \frac{\partial^3 S}{\partial Q^2
\partial P} \frac{\partial^2 S}{\partial P \partial J}
\frac{\partial^3 S}{\partial J^3} +\frac{\partial^2 S}{\partial
P^2} \frac{\partial^3 S}{\partial Q^2 \partial J} \frac{\partial^3
S}{\partial Q \partial J^2} \frac{\partial^2 S}{\partial Q
\partial P} \frac{\partial^2 S}{\partial P \partial J}
\nonumber \\ && +\frac{\partial^2 S}{\partial P^2}
\frac{\partial^3 S}{\partial Q^2 \partial P} \frac{\partial^2
S}{\partial P \partial J} \frac{\partial^2 S}{\partial Q \partial
J} \frac{\partial^3 S}{\partial Q \partial J^2} +\frac{\partial^2
S}{\partial P^2} \frac{\partial^3 S}{\partial Q^2 \partial J}
\frac{\partial^2 S}{\partial Q \partial P} \frac{\partial^2
S}{\partial Q \partial J} \frac{\partial^3 S}{\partial P \partial
J^2} \nonumber \\ && +\frac{\partial^2 S}{\partial P^2}
\frac{\partial^3 S}{\partial Q^3} \frac{\partial^2 S}{\partial Q
\partial P} \frac{\partial^3 S}{\partial P \partial J^2}
\frac{\partial^2 S}{\partial J^2} +\frac{\partial^2 S}{\partial
P^2} \frac{\partial^3 S}{\partial Q \partial P^2} \frac{\partial^2
S}{\partial Q \partial J} \frac{\partial^3 S}{\partial J^3}
\frac{\partial^2 S}{\partial Q^2} \nonumber \\ &&
+\frac{\partial^2 S}{\partial P^2} \frac{\partial^3 S}{\partial
P^2 \partial J} \frac{\partial^2 S}{\partial Q \partial J}
\frac{\partial^3 S}{\partial Q^3} \frac{\partial^2 S}{\partial
J^2} +\frac{\partial^3 S}{\partial Q \partial P^2}
\frac{\partial^2 S}{\partial J^2} \frac{\partial^3 S}{\partial Q^2
\partial P} \frac{\partial^2 S}{\partial P \partial J}
\frac{\partial^2 S}{\partial Q \partial J} \nonumber \\ &&
+\frac{\partial^3 S}{\partial P^2 \partial J} \frac{\partial^2
S}{\partial Q^2} \frac{\partial^3 S}{\partial P \partial J^2}
\frac{\partial^2 S}{\partial Q \partial P} \frac{\partial^2
S}{\partial Q \partial J} +\frac{\partial^3 S}{\partial Q \partial
P^2} \frac{\partial^2 S}{\partial P \partial J} \frac{\partial^2
S}{\partial Q \partial J} \frac{\partial^3 S}{\partial P \partial
J^2} \frac{\partial^2 S}{\partial Q^2} \nonumber \\ &&
+\frac{\partial^3 S}{\partial P^3} \frac{\partial^2 S}{\partial P
\partial J} \frac{\partial^3 S}{\partial Q^2 \partial J}
\frac{\partial^2 S}{\partial Q^2} \frac{\partial^2 S}{\partial
J^2} +\frac{\partial^3 S}{\partial P^3} \frac{\partial^2
S}{\partial Q \partial P} \frac{\partial^3 S}{\partial Q \partial
J^2} \frac{\partial^2 S}{\partial Q^2} \frac{\partial^2
S}{\partial J^2} \nonumber \\ && +\frac{\partial^3 S}{\partial P^2
\partial J} \frac{\partial^2 S}{\partial Q \partial J}
\frac{\partial^3 S}{\partial Q^2 \partial P} \frac{\partial^2
S}{\partial Q \partial P} \frac{\partial^2 S}{\partial J^2}
+\frac{\partial^3 S}{\partial P^2 \partial J} \frac{\partial^2
S}{\partial Q \partial P} \frac{\partial^2 S}{\partial P \partial
J} \frac{\partial^3 S}{\partial Q \partial J^2} \frac{\partial^2
S}{\partial Q^2} \nonumber \\ && +\frac{\partial^3 S}{\partial Q
\partial P^2} \frac{\partial^2 S}{\partial J^2} \frac{\partial^3
S}{\partial Q^2 \partial J} \frac{\partial^2 S}{\partial Q
\partial P} \frac{\partial^2 S}{\partial P \partial J}
+\frac{\partial^2 S}{\partial P^2} (\frac{\partial^3 S}{\partial Q
\partial P \partial J})^2 (\frac{\partial^2 S}{\partial Q
\partial J})^2 \nonumber \\ && +\frac{\partial^2 S}{\partial P^2}
(\frac{\partial^3 S}{\partial Q^2 \partial P})^2 (\frac{\partial^2
S}{\partial J^2})^2 +(\frac{\partial^2 S}{\partial P^2})^2
(\frac{\partial^3 S}{\partial Q^2 \partial J})^2 \frac{\partial^2
S}{\partial J^2} \nonumber \\ && +(\frac{\partial^2 S}{\partial
P^2})^2 \frac{\partial^2 S}{\partial Q^2} (\frac{\partial^3
S}{\partial Q \partial J^2})^2 +(\frac{\partial^3 S}{\partial Q
\partial P \partial J})^2 (\frac{\partial^2 S}{\partial Q \partial
P})^2 \frac{\partial^2 S}{\partial J^2} \nonumber \\ &&
+(\frac{\partial^3 S}{\partial Q \partial P \partial J})^2
(\frac{\partial^2 S}{\partial P \partial J})^2 \frac{\partial^2
S}{\partial Q^2} +\frac{\partial^2 S}{\partial P^2}
(\frac{\partial^3 S}{\partial P \partial J^2})^2
(\frac{\partial^2 S}{\partial Q^2})^2 \nonumber \\
&& +\frac{\partial^3 S}{\partial P^3} \frac{\partial^3 S}{\partial
P \partial J^2} \frac{\partial^2 S}{\partial Q^2}
(\frac{\partial^2 S}{\partial Q \partial J})^2 +\frac{\partial^3
S}{\partial Q \partial P^2} \frac{\partial^2 S}{\partial J^2}
\frac{\partial^3 S}{\partial Q^3} (\frac{\partial^2 S}{\partial P
\partial J})^2 \nonumber \\ && +\frac{\partial^3 S}{\partial P^3}
\frac{\partial^2 S}{\partial P \partial J} (\frac{\partial^2
S}{\partial Q^2})^2 \frac{\partial^3 S}{\partial J^3}
+\frac{\partial^3 S}{\partial P^3} \frac{\partial^3 S}{\partial
Q^2 \partial P} \frac{\partial^2 S}{\partial J^2}
(\frac{\partial^2 S}{\partial Q \partial J})^2 \nonumber \\ &&
+\frac{\partial^3 S}{\partial P^3} \frac{\partial^2 S}{\partial Q
\partial P} (\frac{\partial^2 S}{\partial J^2})^2 \frac{\partial^3
S}{\partial Q^3} +\frac{\partial^3 S}{\partial P^2 \partial J}
\frac{\partial^2 S}{\partial Q^2} \frac{\partial^3 S}{\partial
J^3} (\frac{\partial^2 S}{\partial Q \partial P})^2 \nonumber \\
&& +(\frac{\partial^2 S}{\partial P^2})^2 \frac{\partial^3
S}{\partial Q^3} \frac{\partial^2 S}{\partial Q
\partial J} \frac{\partial^3 S}{\partial J^3}
+\frac{\partial^2 S}{\partial P^2} \frac{\partial^3 S}{\partial
Q^2 \partial J} \frac{\partial^3 S}{\partial J^3}
(\frac{\partial^2 S}{\partial Q \partial P})^2 \nonumber \\
&& +\frac{\partial^2 S}{\partial P^2} \frac{\partial^3 S}{\partial
Q^3} \frac{\partial^3 S}{\partial Q \partial J^2}
(\frac{\partial^2
S}{\partial P \partial J})^2, \nonumber \\
f_3&=& \frac{\partial^2 S}{\partial P^2} \frac{\partial^2
S}{\partial Q^2} \frac{\partial^3 S}{\partial Q \partial J^2}
\frac{\partial^2 S}{\partial Q \partial P} \frac{\partial^3
S}{\partial P \partial J^2} + \frac{\partial^2 S}{\partial P^2}
\frac{\partial^3 S}{\partial Q^2 \partial P} \frac{\partial^2
S}{\partial P \partial J} \frac{\partial^3 S}{\partial Q^2
\partial J} \frac{\partial^2 S}{\partial J^2} \nonumber \\ &&
+ \frac{\partial^3 S}{\partial P^2 \partial J} \frac{\partial^3
S}{\partial Q \partial P^2} \frac{\partial^2 S}{\partial Q
\partial J} \frac{\partial^2 S}{\partial Q^2} \frac{\partial^2
S}{\partial J^2} + \frac{\partial^3 S}{\partial Q \partial P
\partial J} (\frac{\partial^2 S}{\partial Q \partial P})^3
\frac{\partial^3 S}{\partial J^3} \nonumber \\ && +
\frac{\partial^3 S}{\partial Q \partial P \partial J}
(\frac{\partial^2 S}{\partial P \partial J})^3 \frac{\partial^3
S}{\partial Q^3} + \frac{\partial^3 S}{\partial P^3}
\frac{\partial^3 S}{\partial Q \partial P \partial J}
(\frac{\partial^2 S}{\partial Q \partial J})^3 \nonumber \\ && +
\frac{\partial^3 S}{\partial Q \partial P^2} \frac{\partial^3
S}{\partial Q \partial J^2} (\frac{\partial^2 S}{\partial Q
\partial P})^2 \frac{\partial^2 S}{\partial J^2} +
\frac{\partial^3 S}{\partial P^2 \partial J} \frac{\partial^3
S}{\partial Q^2 \partial J} (\frac{\partial^2 S}{\partial P
\partial J})^2 \frac{\partial^2 S}{\partial Q^2}
\nonumber \\ && + \frac{\partial^2 S}{\partial Q^2}
\frac{\partial^3 S}{\partial Q^2 \partial P} (\frac{\partial^2
S}{\partial P \partial J})^2 \frac{\partial^3 S}{\partial P
\partial J^2} + \frac{\partial^2 S}{\partial P^2} \frac{\partial^3
S}{\partial P^2 \partial J} \frac{\partial^3 S}{\partial Q^2
\partial J} (\frac{\partial^2 S}{\partial Q \partial J})^2 \nonumber \\ && +
\frac{\partial^2 S}{\partial P^2} \frac{\partial^3 S}{\partial Q
\partial P^2} \frac{\partial^3 S}{\partial Q \partial J^2}
(\frac{\partial^2 S}{\partial Q \partial J})^2 + \frac{\partial^3
S}{\partial Q^2 \partial P} \frac{\partial^3 S}{\partial P
\partial J^2} (\frac{\partial^2 S}{\partial Q \partial P})^2
\frac{\partial^2 S}{\partial J^2}, \nonumber \\
f_4&=& \frac{\partial^2 S}{\partial P^2} \frac{\partial^2
S}{\partial Q^2} \frac{\partial^3 S}{\partial Q \partial P
\partial J} \frac{\partial^2 S}{\partial Q \partial P}
\frac{\partial^3 S}{\partial J^3} + \frac{\partial^2 S}{\partial
P^2} \frac{\partial^3 S}{\partial Q \partial P \partial J}
\frac{\partial^2 S}{\partial Q \partial P} \frac{\partial^2
S}{\partial Q \partial J} \frac{\partial^3 S}{\partial Q \partial
J^2} \nonumber \\ && + \frac{\partial^2 S}{\partial P^2}
\frac{\partial^3 S}{\partial Q^2 \partial J} \frac{\partial^2
S}{\partial P \partial J} \frac{\partial^3 S}{\partial Q \partial
P \partial J} \frac{\partial^2 S}{\partial Q \partial J} +
\frac{\partial^3 S}{\partial Q^2 \partial P} \frac{\partial^2
S}{\partial J^2} \frac{\partial^3 S}{\partial Q \partial P
\partial J} \frac{\partial^2 S}{\partial Q \partial P}
\frac{\partial^2 S}{\partial P \partial J} \nonumber \\ && +
\frac{\partial^3 S}{\partial Q \partial P \partial J}
\frac{\partial^2 S}{\partial Q \partial P} \frac{\partial^2
S}{\partial P \partial J} \frac{\partial^3 S}{\partial P \partial
J^2} \frac{\partial^2 S}{\partial Q^2} + \frac{\partial^3
S}{\partial Q^2 \partial P} \frac{\partial^3 S}{\partial P
\partial J^2} \frac{\partial^2 S}{\partial Q \partial P}
\frac{\partial^2 S}{\partial P \partial J} \frac{\partial^2
S}{\partial Q \partial J} \nonumber \\ && + \frac{\partial^3
S}{\partial P^3} \frac{\partial^3 S}{\partial Q \partial P
\partial J} \frac{\partial^2 S}{\partial Q \partial J}
\frac{\partial^2 S}{\partial Q^2} \frac{\partial^2 S}{\partial
J^2} + \frac{\partial^3 S}{\partial P^2 \partial J}
\frac{\partial^2 S}{\partial Q \partial J} \frac{\partial^3
S}{\partial Q \partial P \partial J} \frac{\partial^2 S}{\partial
P \partial J} \frac{\partial^2 S}{\partial Q^2} \nonumber \\ && +
\frac{\partial^3 S}{\partial Q \partial P^2} \frac{\partial^2
S}{\partial J^2} \frac{\partial^3 S}{\partial Q \partial P
\partial J} \frac{\partial^2 S}{\partial Q \partial P}
\frac{\partial^2 S}{\partial Q \partial J} + \frac{\partial^3
S}{\partial P^2 \partial J} \frac{\partial^3 S}{\partial Q^2
\partial J} \frac{\partial^2 S}{\partial Q \partial P}
\frac{\partial^2 S}{\partial P \partial J} \frac{\partial^2
S}{\partial Q \partial J} \nonumber \\ && + \frac{\partial^3
S}{\partial Q \partial P^2} \frac{\partial^3 S}{\partial Q
\partial J^2} \frac{\partial^2 S}{\partial Q \partial P}
\frac{\partial^2 S}{\partial P \partial J} \frac{\partial^2
S}{\partial Q \partial J} + \frac{\partial^3 S}{\partial Q
\partial J^2} (\frac{\partial^2 S}{\partial Q \partial P})^3
\frac{\partial^3 S}{\partial P \partial J^2} \nonumber \\ && +
\frac{\partial^3 S}{\partial Q^2 \partial P} (\frac{\partial^2
S}{\partial P \partial J})^3 \frac{\partial^3 S}{\partial Q^2
\partial J} + \frac{\partial^3 S}{\partial P^2 \partial J} \frac{\partial^3
S}{\partial Q \partial P^2} (\frac{\partial^2 S}{\partial Q
\partial J})^3 \nonumber \\ && + \frac{\partial^3 S}{\partial Q \partial P \partial J}
\frac{\partial^2 S}{\partial Q \partial J} \frac{\partial^3
S}{\partial P \partial J^2} (\frac{\partial^2 S}{\partial Q
\partial P})^2 + \frac{\partial^3 S}{\partial Q^2 \partial P} (\frac{\partial^2
S}{\partial P \partial J})^2 \frac{\partial^3 S}{\partial Q
\partial P \partial J} \frac{\partial^2 S}{\partial Q \partial J}
\nonumber \\ && + \frac{\partial^3 S}{\partial P^3}
\frac{\partial^2 S}{\partial P \partial J} \frac{\partial^3
S}{\partial Q^2 \partial J} (\frac{\partial^2 S}{\partial Q
\partial J})^2 + \frac{\partial^3 S}{\partial P^3}
\frac{\partial^2 S}{\partial Q \partial P} \frac{\partial^3
S}{\partial Q \partial J^2} (\frac{\partial^2 S}{\partial Q
\partial J})^2 \nonumber \\ && + \frac{\partial^3 S}{\partial P^2
\partial J} \frac{\partial^2 S}{\partial Q \partial J}
\frac{\partial^3 S}{\partial Q^3} (\frac{\partial^2 S}{\partial P
\partial J})^2 + \frac{\partial^3 S}{\partial P^2
\partial J} (\frac{\partial^2 S}{\partial Q \partial J})^2
\frac{\partial^3 S}{\partial Q \partial P \partial J}
\frac{\partial^2 S}{\partial Q \partial P} \nonumber \\ && +
\frac{\partial^3 S}{\partial Q \partial P \partial J}
\frac{\partial^3 S}{\partial Q^2 \partial J} (\frac{\partial^2
S}{\partial P \partial J})^2 \frac{\partial^2 S}{\partial Q
\partial P} + \frac{\partial^3 S}{\partial Q^2 \partial P}
\frac{\partial^2 S}{\partial P \partial J} \frac{\partial^3
S}{\partial J^3} (\frac{\partial^2 S}{\partial Q \partial P})^2
\nonumber \\ && + \frac{\partial^3 S}{\partial Q \partial P
\partial J} \frac{\partial^2 S}{\partial P \partial J}
\frac{\partial^3 S}{\partial Q \partial J^2} (\frac{\partial^2
S}{\partial Q \partial P})^2 + \frac{\partial^3 S}{\partial Q
\partial P^2} (\frac{\partial^2 S}{\partial Q \partial J})^2
\frac{\partial^3 S}{\partial Q \partial P \partial J}
\frac{\partial^2 S}{\partial P \partial J} \nonumber \\ && +
\frac{\partial^3 S}{\partial Q^3} \frac{\partial^2 S}{\partial Q
\partial P} \frac{\partial^3 S}{\partial P \partial J^2}
(\frac{\partial^2 S}{\partial P \partial J})^2 + \frac{\partial^3
S}{\partial Q \partial P^2} \frac{\partial^2 S}{\partial Q
\partial J} \frac{\partial^3 S}{\partial J^3} (\frac{\partial^2
S}{\partial Q \partial P})^2 \nonumber \\ && + \frac{\partial^2
S}{\partial P^2} \frac{\partial^3 S}{\partial Q \partial P
\partial J} \frac{\partial^2 S}{\partial P \partial J}
\frac{\partial^3 S}{\partial Q^3} \frac{\partial^2 S}{\partial J^2}, \nonumber \\
f_5&=& \frac{\partial^2 S}{\partial P^2} \frac{\partial^3
S}{\partial Q^3} \frac{\partial^2 S}{\partial P \partial J}
\frac{\partial^2 S}{\partial Q \partial J} \frac{\partial^3
S}{\partial P \partial J^2} + \frac{\partial^2 S}{\partial P^2}
\frac{\partial^3 S}{\partial Q^2 \partial P} \frac{\partial^2
S}{\partial Q \partial P} \frac{\partial^2 S}{\partial Q \partial
J} \frac{\partial^3 S}{\partial J^3} \nonumber \\ && +
\frac{\partial^3 S}{\partial P^3} \frac{\partial^2 S}{\partial P
\partial J} \frac{\partial^2 S}{\partial Q \partial J}
\frac{\partial^3 S}{\partial Q \partial J^2} \frac{\partial^2
S}{\partial Q^2} + \frac{\partial^3 S}{\partial P^3}
\frac{\partial^2 S}{\partial Q \partial P} \frac{\partial^2
S}{\partial J^2} \frac{\partial^3 S}{\partial Q^2 \partial J}
\frac{\partial^2 S}{\partial Q \partial J} \nonumber \\ && +
\frac{\partial^3 S}{\partial P^2 \partial J} \frac{\partial^2
S}{\partial Q \partial P} \frac{\partial^2 S}{\partial P \partial
J} \frac{\partial^3 S}{\partial Q^3} \frac{\partial^2 S}{\partial
J^2} + \frac{\partial^3 S}{\partial Q \partial P^2}
\frac{\partial^23 S}{\partial Q \partial P} \frac{\partial^2
S}{\partial P \partial J} \frac{\partial^3 S}{\partial J^3}
\frac{\partial^2 S}{\partial Q^2}, \nonumber \\
f_6&=& \frac{\partial^2 S}{\partial P^2} \frac{\partial^2
S}{\partial Q^2} \frac{\partial^3 S}{\partial P \partial J^2}
\frac{\partial^3 S}{\partial Q^2 \partial J} \frac{\partial^2
S}{\partial P \partial J} + \frac{\partial^2 S}{\partial P^2}
\frac{\partial^3 S}{\partial Q^2 \partial P} \frac{\partial^2
S}{\partial Q \partial P} \frac{\partial^3 S}{\partial Q \partial
J^2} \frac{\partial^2 S}{\partial J^2} \nonumber \\ && +
\frac{\partial^2 S}{\partial P^2} \frac{\partial^3 S}{\partial P^2
\partial J} \frac{\partial^2 S}{\partial Q \partial J}
\frac{\partial^3 S}{\partial Q \partial J^2} \frac{\partial^2
S}{\partial Q^2} + \frac{\partial^2 S}{\partial P^2}
\frac{\partial^3 S}{\partial Q \partial P^2} \frac{\partial^2
S}{\partial J^2} \frac{\partial^3 S}{\partial Q^2 \partial J}
\frac{\partial^2 S}{\partial Q \partial J} \nonumber \\ && +
\frac{\partial^3 S}{\partial P^2 \partial J} \frac{\partial^3
S}{\partial Q^2 \partial P} \frac{\partial^2 S}{\partial P
\partial J} \frac{\partial^2 S}{\partial Q^2} \frac{\partial^2
S}{\partial J^2} + \frac{\partial^3 S}{\partial Q \partial P^2}
\frac{\partial^2 S}{\partial Q \partial P} \frac{\partial^3
S}{\partial P \partial J^2} \frac{\partial^2 S}{\partial Q^2}
\frac{\partial^2 S}{\partial J^2} \nonumber \\ && +
\frac{\partial^3 S}{\partial Q \partial P^2} \frac{\partial^3
S}{\partial Q \partial J^2} (\frac{\partial^2 S}{\partial P
\partial J})^2 \frac{\partial^2 S}{\partial Q^2} +
\frac{\partial^3 S}{\partial P^2 \partial J} \frac{\partial^3
S}{\partial Q^2 \partial J} (\frac{\partial^2 S}{\partial Q
\partial P})^2 \frac{\partial^2 S}{\partial J^2} \nonumber \\ && +
\frac{\partial^2 S}{\partial P^2} \frac{\partial^3 S}{\partial Q^2
\partial P} \frac{\partial^3 S}{\partial P \partial J^2}
(\frac{\partial^2 S}{\partial Q \partial J})^2 + \frac{\partial^2
S}{\partial P^2} (\frac{\partial^3 S}{\partial Q \partial P
\partial J})^2 \frac{\partial^2 S}{\partial Q^2} \frac{\partial^2
S}{\partial J^2}, \nonumber \\
f_7&=& \frac{\partial^2 S}{\partial P^2} \frac{\partial^2
S}{\partial Q^2} \frac{\partial^3 S}{\partial Q \partial P
\partial J} \frac{\partial^3 S}{\partial Q \partial J^2}
\frac{\partial^2 S}{\partial P \partial J} + \frac{\partial^2
S}{\partial P^2} \frac{\partial^3 S}{\partial Q^2
\partial J} \frac{\partial^2 S}{\partial Q \partial P}
\frac{\partial^3 S}{\partial Q \partial P \partial J}
\frac{\partial^2 S}{\partial J^2} \nonumber \\ && +
\frac{\partial^2 S(}{\partial P^2} \frac{\partial^3 S}{\partial Q
\partial P \partial J} \frac{\partial^2 S}{\partial Q \partial J}
\frac{\partial^3 S}{\partial P \partial J^2} \frac{\partial^2
S}{\partial Q^2} + \frac{\partial^2 S}{\partial P^2}
\frac{\partial^3 S}{\partial Q^2 \partial P} \frac{\partial^2
S}{\partial J^2} \frac{\partial^3 S}{\partial Q \partial P
\partial J} \frac{\partial^2 S}{\partial Q \partial J}
\nonumber \\ && + \frac{\partial^3 S}{\partial P^2 \partial J}
\frac{\partial^2 S}{\partial Q \partial P} \frac{\partial^3
S}{\partial Q \partial P \partial J} \frac{\partial^2 S}{\partial
Q^2} \frac{\partial^2 S}{\partial J^2} + \frac{\partial^3
S}{\partial Q \partial P^2} \frac{\partial^2 S}{\partial P
\partial J} \frac{\partial^3 S}{\partial Q \partial P \partial J}
\frac{\partial^2 S}{\partial Q^2} \frac{\partial^2 S}{\partial
J^2} \nonumber \\ && + \frac{\partial^3 S}{\partial Q \partial
P^2} \frac{\partial^2 S}{\partial Q \partial P} \frac{\partial^3
S}{\partial P \partial J^2} (\frac{\partial^2 S}{\partial Q
\partial J})^2 + \frac{\partial^3 S}{\partial P^2 \partial J}
\frac{\partial^3 S}{\partial Q^2 \partial P} \frac{\partial^2
S}{\partial P \partial J} (\frac{\partial^2 S}{\partial Q \partial
J})^2 \nonumber \\ && + \frac{\partial^3 S}{\partial P^2
\partial J} \frac{\partial^2 S}{\partial Q \partial J}
\frac{\partial^3 S}{\partial Q \partial J^2} (\frac{\partial^2
S}{\partial Q \partial P})^2 + \frac{\partial^3 S}{\partial Q
\partial P^2} (\frac{\partial^2 S}{\partial P \partial J})^2
\frac{\partial^3 S}{\partial Q^2 \partial J} \frac{\partial^2
S}{\partial Q \partial J} \nonumber \\ && + \frac{\partial^3
S}{\partial Q^2 \partial P} \frac{\partial^2 S}{\partial Q
\partial P} \frac{\partial^3 S}{\partial Q \partial J^2}
(\frac{\partial^2 S}{\partial P \partial J})^2  + \frac{\partial^3
S}{\partial Q^2 \partial J} \frac{\partial^2 S}{\partial P
\partial J} \frac{\partial^3 S}{\partial P \partial
J^2} (\frac{\partial^2 S}{\partial Q \partial P})^2, \nonumber \\
f_8&=& (\frac{\partial^3 S}{\partial Q \partial P \partial J})^2
\frac{\partial^2 S}{\partial Q \partial P} \frac{\partial^2
S}{\partial P \partial J} \frac{\partial^2 S}{\partial Q \partial
J} \end{eqnarray}

\begin{Large} \noindent{\bf Acknowledgement} \end{Large}\\

This work has been supported in part by the European Research
Council grant n.~226455, ``SUPERSYMMETRY, QUANTUM GRAVITY AND
GAUGE FIELDS (SUPERFIELDS)''.
B. N. T. would like to thank Prof. A. Sen and Prof. R. Emparan and Prof. S. Minwalla for necessary discussions
on stability of the rotating black hole configurations and possible phase transions during the
\textit{``Spring School on Superstring Theory and Related Topics-2008, ICTP Trieste, Italy''};
Prof. R. Gopakumar for valuable explanation on black brane configurations and their degeneracy of microstates during the
\textit{``National Conference on New Trends in Field Theories, Nov. 2008, Banaras Hindu University, Varanasi, India''};
Prof. V. Ravishankar, Prof. P. Jain, Prof. U. B. Tewari, Prof M. K. Harbola, and Prof. R. K. Thareja
for providing their viable support during the preparation of this manuscript;
Vinod Chandra for ample discussions on intrinsic geometric character of Hot QCD and related configurations.
B. N. T. further acknowledges  CSIR, New Delhi (India) for the financial support under the research grant
\textit{``CSIR-SRF-9/92(343)/2004-EMR-I''}.

\end{document}